\documentclass[11pt,a4paper]{article}
\usepackage{jcappub}
\usepackage{color}
\usepackage{mathtools}
\usepackage{diagbox}
\usepackage{appendix}
\usepackage{amsmath,amssymb,amsfonts,dcolumn,color,latexsym,epsfig}
\usepackage{graphicx}
\usepackage{caption}
\usepackage{subcaption}
\graphicspath{ {./plots/} }
\usepackage{placeins}
\usepackage[x11names]{xcolor}
\usepackage{hyperref}
\usepackage{natbib}
\usepackage{float}
\usepackage{booktabs}
\usepackage{comment}
\usepackage{lipsum}
\usepackage{tikz}
\usepackage{multirow}
\usepackage{makecell}

\usepackage{pifont}
\newcommand{\cmark}{\ding{51}}%
\newcommand{\xmark}{\ding{55}}%
\begin{document}
	\title{Shadows of generalised Hayward spacetimes : in vacuum and with plasma}

 \author[a]{Suvikranth Gera}
  \author[a,b]{Saurabh Kumar}
  \author[b]{Poulami Dutta Roy}
  \author[a]{Sayan Chakrabarti}

\affiliation[a]{Department of Physics, IIT Guwahati, Guwahati, Assam, India.}
 \affiliation[b]{Chennai Mathematical Institute, Siruseri 603103, Tamil Nadu, India}

\emailAdd{suvikranthg@gmail.com}
 \emailAdd{ksaurabhkumar.712@gmail.com}
\emailAdd{poulami@cmi.ac.in}
\emailAdd{sayan.chakrabarti@iitg.ac.in}

\abstract{
 We investigate the shadow properties of a wide class of spacetimes arising from different parameter regimes of the generalized Hayward metric, characterized by two independent parameters $(\sigma, \kappa)$ (Phys. Rev. D 106, 044028).  This metric extends the original Hayward regular black hole solution by introducing distinct mass functions in the $g_{tt}$ and $g_{rr}$ components, giving rise to four types of wormholes ( which include multi-peak effective potentials), a regular black hole, and a singular black hole solutions allowing for a unified treatment of black hole mimickers. We compute the shadow radii for all spacetimes in vacuum and in the presence of plasma, using both homogeneous and non-homogeneous plasma profiles. Our results show that certain wormhole solutions particularly the Hayward-Damour-Solodukhin class—can exhibit multiple photon spheres, leading to shadow features that differ significantly from the Schwarzschild black hole. When these results are compared with Event Horizon Telescope observations of Sgr~$A^\star$, we find that regular black holes remain observationally viable but only within a narrow parameter space. In contrast, wormhole solutions with multi-peak effective potentials are more consistent with shadow constraints than those with single peaks. This contrasts with quasinormal mode studies, which favored single-barrier potentials, and may imply detectable late-time echoes in gravitational wave signals.
 
}

\maketitle
\newpage

\section{Introduction}

Black holes (BH) are ubiquitous in physics, and at the same time they are the simplest objects in General Relativity(GR). For this reason they are often termed as the hydrogen atom of GR. Their extreme gravitational fields make them ideal candidates for testing GR in the strong-field regime.  The onset of testing GR in the strong-field regmie began with the direct observations of gravitational waves by the LIGO-Virgo-KAGRA collaboration \cite{LIGOScientific:2016aoc}, which were closely followed by the Event Horizon Telescope's (EHT) first observations via the Very Long Baseline Interferometry (VLBI) of the supermassive BHs at the centers of the galaxies M87 \cite{EventHorizonTelescope:2019dse} and Sgr $\rm A^*$ \cite{EventHorizonTelescope:2022wkp}. It is to be noted that the EHT observations have achieved the required frequency to resolve the horizon-scale emission from the vicinity of the supermassive BHs at the center of the galaxies. As a result, the observed light from these extreme compact objects includes photons whose trajectories near the horizon of the BHs are deviated from their actual travel paths and describes nearly bound spherical orbits around the BH. This humongous achievement was made possible by a global network of radio telescopes working in parallel to create a virtual earth-sized observatory. By capturing and analyzing the radio waves emitted by the surrounding accretion disk, the EHT has once again provided a confirmation of the theoretical predictions of photon paths within the theory of GR. 

%Although GR has achieved significant successes, it also faces severe challenges on the theoretical front.  One of the most problematic areas in GR is the presence of the singularities at the core of the BH spacetimes, which raises serious question about the credibility of the theory.

Despite GR's remarkable observational successes, it faces significant theoretical challenges, which include but are not limited to the complete understanding of curvature singularity and a comprehensive handle over the quantization of gravity. The issue that GR is ultraviolet (UV)-incomplete is related to this fact. It is therefore proposed that GR requires modifications in the regions where the spacetime curvature becomes high. Such modifications are necessary if one requires  a theory of gravity which is UV complete. 

A potential approach to address the curvature singularity issue is to consider specific types of matter distributions, which will lead to singularity-free solutions. These regular solutions can be broadly divided into spacetimes with and without a horizon, like the regular BHs and traversable wormholes, respectively. Chronologically, Bardeen first proposed the metric for such a nonsingular BH geometry \cite{bardeen_1968}. This was followed up by works on coupling GR with nonlinear electrodynamics (NLED) resulting in a plethora of regular BH solutions with electric and magnetic charges \cite{Ayon-Beato:1998hmi, Ayon-Beato:1999qin, Ayon-Beato:1999kuh, ayon-Beato_2000, ayon-Beato_2004, Dymnikova20044417, Bronnikov2001, Shankaranarayanan20041095}. One of the seminal works in the direction of obtaining a regular blackhole solution was by  Hayward, who proposed a static spherically symmetric BH solution for which all possible curvature scalars are finite \cite{hayward_2006}. This metric describes an isolated regular four dimensional spherically symmetric spacetime and contains a length scale $\ell$, apart from which, there is only one parameter: the mass of the BH. At large distances the metric takes the Schwarzschild form, while at the origin it is regular and has de Sitter form. The distinguishing feature of such a metric is its scaling behaviour, i.e. the scaling transformation of the coordinates, parameters and the metric preserves the form of the metric.  

However, a closer examination reveals that such ad hoc constructions often trade curvatures singularities for new issues. For instance, traversable wormholes typically require matter that violates energy conditions \cite{morris_1988,thorne_1988}, while regular black holes are prone to mass-inflation instabilities \cite{ori_1991,poisson_1989,carballo_2018,carballo_2021} at least in the context of GR. However, beyond GR several modified gravity theories admit wormhole solutions without invoking exotic matter \cite{lobo_2008,varieschi_2015, kord_2015, ovgun_2019, zubair_2017, lobo_2009, boehmer_2012, shaikh_2016,kanti_2012,mehdizadeh_2015, maeda_2008, kanti_2011,shaikh_2015}, and recent work has explored ways to mitigate the mass-inflation instability in regular black holes \cite{Carballo-Rubio:2022kad,bonanno_2021}. Nevertheless, these issues remain far from conclusively resolved.

The literature related to shadows of different extreme compact objects including BHs is very rich and because of the present interest in EHT collaboration's result, there has been a flurry of works in this direction. Here, we briefly mention some of the important works towards the development of the subject. The shadow of the Schwarzschild black hole was first discussed in \cite{Synge:1966mon, Luminet:1979nyg}. Bardeen \cite{bardeen_1973}, again, was the first to describe the nature of the shadow cast by a rotating Kerr BH, consequently it was generalised in the case of a Kerr-Newman BH in \cite{deVries:1999tiy}. The shadow of the Kerr BH as well as a Kerr naked singularity was discussed in \cite{PhysRevD.80.024042} by introducing two new observables. Going beyond the standard GR solutions of Einstein's equations, the study of shadows in BHs in other theories had been discussed extensively in the literature, viz. Einstein-Maxwell-Dilaton-Axion BH \cite{Shao-Wen_Wei_2013}, Kerr-Taub-NUT BH \cite{Abdujabbarov:2012bn}, shadows of static spherically symmetric and rotating Kerr BHs in modified gravity (MOG) theory \cite{Moffat:2015kva}, rotating braneworld BH \cite{PhysRevD.85.064019}, rotating non-Kerr type BHs \cite{PhysRevD.88.064004}, BHs in an expanding Universe \cite{Roy:2020dyy}, BHs in dynamical Chern-Simons modified gravity \cite{Rodriguez:2024ijx} to name a few. For more detailed on the methods and results in calculating shadows, we refer the readers to the existing reviews \cite{Cunha:2018acu,Perlick:2021aok,Lupsasca:2024wkp}. On the other hand, the study of shadows in regular BH backgrounds has gained considerable interests in recent years. There has been a lot of works done in this direction, see \cite{Li:2013jra,Abdujabbarov:2016hnw,Stuchlik:2019uvf,Dymnikova:2019vuz,GHOSH2020115088,Uniyal:2023ahv,KumarWalia:2024yxn} for more details. 

%In our work, we take a pragmatic approach and address the observational aspects of a class of regular spacetime solutions called the generalized Hayward metric recently developed in Dutta Roy and Kar \cite{DuttaRoy:2022ytr}, which describes {\color{red}traversable} wormholes, regular BH spacetimes along with the singular BH depending on the choice of the metric parameters. {\color{red} Note that the singularity or a lack thereof is defined with respect to curvature scalars. The generalized Hayward metric is a perfect testing ground for comparing the shadows of the regular spacetimes to those of the singular BHs. {\color{red}To our knowledge, this is the first study incorporating compact objects like regular BHs, wormholes, and singular BHs into a single framework}. Even though studies on shadows of compact objects  (see \cite{Kumar:2023wfp} for details on wormhole shadows and for more references on the topic) have been performed individually, the present approach to study a spacetime solution that changes its characteristics based on the parameter values has not been done before.  As a part of this study, we also discovered that some of these cases possess an anti-photon ring along with the usual photon ring. We explore in detail the occurrence of photon/anti-photon rings in all classes of geometries originating from the generalized Hayward metric and probe the capability of the shadow observations to distinguish the regular spacetimes from the singular BH. }

In our work, we take a pragmatic approach and address the observational aspects of a class of regular spacetime solutions called the generalized Hayward metric recently developed in Dutta Roy and Kar \cite{DuttaRoy:2022ytr}, which describes traversable wormholes, regular BH spacetimes along with the singular BH depending on the choice of the metric parameters.  Note that the singularity or a lack thereof is defined with respect to curvature scalars. The generalized Hayward metric is a perfect testing ground for comparing the shadows of the regular spacetimes to that of the singular BHs.To our knowledge, this is the first study incorporating compact objects like regular BHs, wormholes, and singular BHs into a single framework. Even though studies on  shadows of compact objects  (see \cite{Kumar:2023wfp} for details on wormhole shadows and for more references on the topic) have been performed individually, the present approach to study a spacetime solution that changes its characteristics based on the parameter values has not been done before.  As a part of this study, we also discovered that some of these cases possess an anti-photon ring along with the usual photon ring. We explore in detail the occurrence of photon/anti-photon rings in all classes of geometries originating from the generalized Hayward metric and probe the capability of the shadow observations to distinguish the regular spacetimes from the singular BH.

It is known that the shadow of a compact object is associated with the photons scattered from the innermost unstable circular null orbits and, hence, is a characteristic of the compact object. However, this holds true assuming the trajectories of these photons are unaffected throughout their journey from the point of emission to the typical asymptotic observer. In a realistic scenario, compact objects are always surrounded by an interstellar medium, which has a non-trivial effect on the photon trajectories. Hence, it is natural to incorporate the effects of the environment in the computations of shadow. As a first step to address these issues, we analytically handle the computations of shadow in the presence of a plasma surrounding the compact object. Although our analysis corresponds to a simplified scenario which are different from a real astrophysical environment, these analytic calculations allow us to understand the explicit dependencies of the shadow profile on the metric and plasma parameters in contrast to numerical simulations. 

To this extent, we follow the techniques described in \cite{Perlick:2015prd} and restrict to non-magnetized cold plasma consisting of two fluids, which are modeled by ions and electrons. The Hamiltonian for the photon trajectories in the presence of plasma in a curved spacetime background has been done at various levels of sophistication. For example, the rigorous derivation of Hamiltonian for magnetized plasma was achieved by Breur and Ehlers in \cite{Breuer:1980proc,Breuer:1981proc}. For a much simpler case of non-magnetized plasma fluids that we are interested in, the same analysis was carried out in \cite{Perlick:2000book,Synge:1960book}. These results were then applied to the photon trajectories in the equatorial plane of Kerr spacetime by Perlick and others in \cite{Perlick:2000book,Bisnovatyi:2008grav,Bisnovatyi:2010mon,Tsupko:2013prd} and in a general setup in \cite{Morozova:2013}. 

Finally, we address these predictions in the context of recent observational constraints from BH shadows of Sgr $A^\star$ from the EHT. Our results show that plasma effects, especially in non-homogeneous configurations, significantly constrain the viable parameter spaces of these regular spacetimes.

The rest of the paper is organized as follows. In section \ref{sec:1}, we briefly discuss the generalized Hayward metric and the spacetimes arising for different metric parameter values. Section \ref{sec:2} consists of a brief derivation of the shadow radius and other important quantities for the case of a static spherically symmetric solution in both vacuum and in the presence of a plasma medium.  Section \ref{sec:3}  consists of the shadow, with and without plasma, for the different spacetimes originating from the generalized Hayward metric. In section \ref{sec:4}, we obtain additional constraints on the spacetime parameters and plasma parameter using EHT data for Sgr $A^\star$. We summarise the conclusions in Section \ref{sec:5}.
Unless otherwise stated we have used geometrised units $G = c = 1$.

%%%%%%%%%

\section{Review of spacetimes originating from generalised Hayward metric}\label{sec:1}
In this section, we will discuss the structure of the generalized Hayward metric and give a brief recap about the different spacetimes that originate for different parameter values. 

	\begin{spreadlines}{-0.4em}
		\begin{figure}[ht]
			\centering
			\includegraphics[width= 0.8\textwidth ]{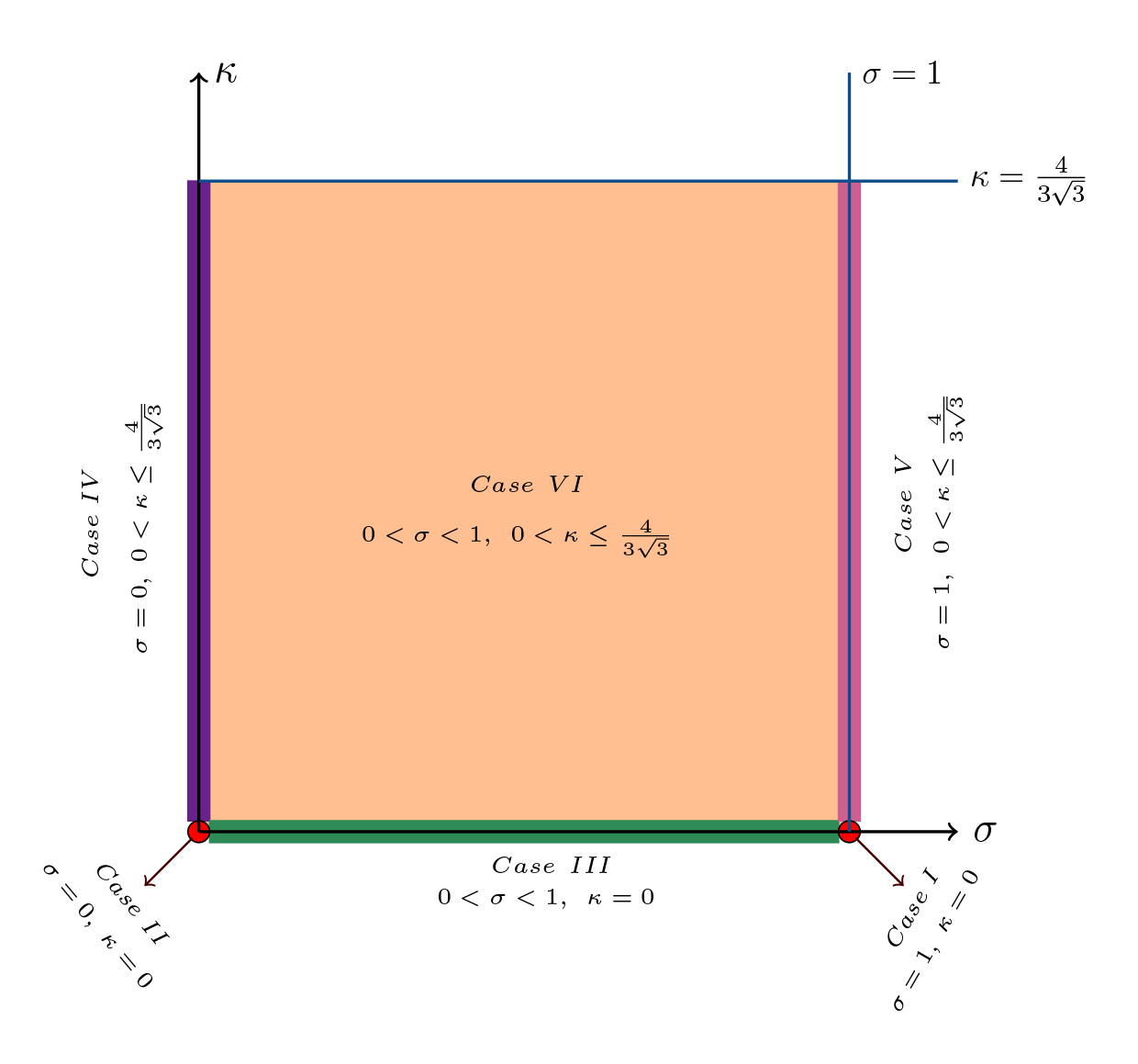}
			\caption{Parameter space of ($\sigma$, $\kappa$) showing different spacetimes corresponding to different ranges of the parameters.}
			\label{fig:parameter_space}
		\end{figure}
	\end{spreadlines}	
%%%%%%%%%%%%%%%%%%%%%%%%%%%%%%%%%%%%%%%%%%%%%

The spherically symmetric, static generalized Hayward metric, introduced in \cite{DuttaRoy:2022ytr} is of the form
	\begin{equation}
		ds^2 = - f_1(r) dt^2 + \frac{dr^2}{f(r)} + r^2 (d\theta^2 + \sin^2 \theta d\phi^2),
		\label{eq:metric}
	\end{equation}
	where $f_1(r)$ and $f(r)$ are chosen respectively as 
	\begin{equation}
		f_1(r) = 1- \frac{2 M_1 r^2}{ r^3 + 2 M_1 \ell^2}; \hspace{0.2in}  f(r) = 1- \frac{2 M r^2}{ r^3 + 2 M \ell^2}.
	\end{equation}
It is important to note that for $M_1 =M$, this metric reduces to the Hayward regular BH, with the square of the parameter $\ell$ being identified with the inverse of the cosmological constant in the small $r$ limit. In fact, the construction of the generalized Hayward metric is motivated from the Hayward BH case by considering two different mass parameters in the $g_{tt}$ and $g_{rr}$ components. The incorporation of a different mass parameter \( M_1 \) in the \( g_{tt} \) component of the metric, such that \( M_1 \neq M \), is inspired by the Damour-Solodukhin wormhole construction based on the Schwarzschild metric \cite{damour_2007}. By introducing a different mass parameter into the metric of the Schwarzschild BH, the well-studied Damour-Solodukhin wormhole is obtained, while a procedure on the same vein on the Hayward regular BH resulted in the generalized Hayward metric as shown in eq.(\ref{eq:metric}). Following \cite{DuttaRoy:2022ytr}, we re-write eq.(\ref{eq:metric}) as,
	\begin{eqnarray}
		ds^2 = - \Big( 1- \frac{2 \sigma x^2}{x^3+ 2 \sigma \kappa^2}\Big) dt^2 + \frac{dr^2}{\Big( 1- \frac{2 x^2}{x^3+ 2 \kappa^2}\Big)} + r^2 (d\theta^2 + \sin^2 \theta d\phi^2),\label{eq:dimensionless_metric}
	\end{eqnarray}
where the parameters $M,M_1,\ell$ are written in a dimensionless form using $\sigma=\frac{M_1}{M}$, $\kappa=\frac{\ell}{M}$ with the radial coordinate $x=\frac{r}{M}$. We will use the metric of the form given in eq.(\ref{eq:dimensionless_metric}) for rest of the discussions. For different values of $(\sigma,\kappa)$, the spacetimes are classified into wormholes/BHs depending on the presence of throat/event horizon. 
	
Fig.(\ref{fig:parameter_space}) shows the six different classes of spacetimes corresponding to different values of $(\sigma,\kappa)$ whose properties are summarized in Table \ref{tab:classification}.
    \begin{table}[h]
		\centering
		\begin{tabular}{|c|c|c|c|c|}
			\hline
			& \textbf{Spacetime} &  $\sigma$ &  $\kappa$ & \textbf{Curvature Singularity} \\ 
			\hline 
			\hline
			Case I & Schwarzschild BH & 1 & 0 & \cmark \\
			\hline
			Case II & Schwarzschild Wormhole & 0 & 0 & \xmark \\
			\hline
			Case III & Damour-Solodukhin Wormhole & (0,1) & 0 & \xmark \\
			\hline
			Case IV & Hayward Wormhole & 0 & $(0,\frac{4}{3\sqrt{3}}]$ & \xmark\\
			\hline
			Case V & Hayward BH & 1 & $(0,\frac{4}{3\sqrt{3}}]$ &\xmark\\
			\hline
			Case VI & Hayward-Damour-Solodukhin Wormhole & (0,1) & $(0,\frac{4}{3\sqrt{3}}]$ & \xmark\\
			\hline
		\end{tabular}
		\caption{Table shows the classification of spacetimes into BHs and wormholes depending on the values of ($\sigma,\kappa$). The only spacetime with curvature singularity is the Schwarzschild BH while all others are regular.}
		\label{tab:classification}
	\end{table}

\section{Null Geodesics and Shadow}\label{sec:2}
 
 The `shadow' in the context of BHs and other astrophysical compact objects  is defined as a dark silhouette against a bright background which is formed due to the gravitational bending of light. Historically this has been called by various names and has various definition. A comprehensive list of these have been discussed in \cite{Perlick:2021aok}. In the context of this paper we define shadow as the two dimensional closed curve on a celestial sphere that separates  the captured and scattered orbits \cite{Gralla:2019xty,Vazquez:2003zm,Luminet:1979nyg}. These celestial coordinates can be expressed in terms of the observer's position relative to the BH and the tangents to the photon geodesics \cite{Cunha:2018acu,Vazquez:2003zm,Bardeen:1973tla,deVries:1999tiy}
 \begin{align} \label{eq:celestial_coordiantes}
 	\alpha & = -r_{o}^{2} \sin \theta_{o} \left.\frac{d \phi}{dr}\right\rvert_{r_{o}}, \\ \nonumber
 	\beta &= \left. r_{o}^{2} \frac{d\theta}{dr}\right\rvert_{r_o},
 \end{align}
where $\alpha$ and $\beta$ are celestial coordinates. In case of static spherically symmetric metric the closed curve is a circle and the radius of this circle is known as the shadow radius and is determined by
\begin{equation}\label{eq:shadow_celestial}
	r_{sh}^{2}= \alpha^{2} +\beta^{2}.
\end{equation}

Hence calculating shadow radius essentially reduces to finding the tangents to the null geodesics and plugging that information into the above equations. In principle there are various approaches for this calculation. In the remainder of this section we reproduce the known results and set the conventions to be used in the later sections of this work based on the Hamilton-Jacobi approach. In the process, we also define other useful quantities, such as the impact parameter and the photon sphere radius. We  also restrict this analysis  to the static spherically symmetric solutions with the metric ansatz being  %(A much more general approach to this derivation is discussed in \cite{Vazquez:2003zm}):

\begin{equation}
	ds^{2}=-f_{1}(r) dt^{2} + \frac{dr^{2}}{f(r)}+r^{2}d\Omega^{2},
\end{equation}
the choice of which is motivated by the form of the generalized Hayward metric, given in eq.(\ref{eq:dimensionless_metric}), which is the focus of this work. The effective Lagrangian for the geodesic equations in the above spacetime is given by
\begin{equation*}
	\mathcal{L}= -f_{1}(r) \dot{t}^{2} + \frac{\dot{r}^{2}}{f(r)}+r^{2}\dot{\theta}^{2}+ r^{2}\sin\theta^{2} \dot{\phi}^{2}, 
\end{equation*}
whose canonical momentum is defined as
\begin{equation}
	\mathit{p}_{\alpha} = \frac{1}{2} \frac{\partial \mathcal{L}}{\partial \dot{x}^{\alpha}},
\end{equation}
with the dot denoting derivative with respect to the affine parameter. For the static spherically symmetric metric considered, the conserved momenta $(\mathit{p}_{\phi},\mathit{p}_{t})$, corresponding to the two Killing vectors, gives $r^{2}\sin^{2}\theta \dot{\phi}\equiv L$ and $-f_{1}(r)\dot{t}\equiv -E$, respectively. The total angular momentum per unit mass and total energy per unit mass is denoted by $L$ and $E$, respectively.

The Hamilton-Jacobi action given by,
\begin{equation*}
	\mathcal{S}=\frac{1}{2}\tilde{\mu}^{2}\lambda -E t + L \phi + S_{r}(r) + S_{\theta}(\theta),
\end{equation*}
% the relevant equation of motion is 
% \begin{equation}
% 	\frac{\partial \mathcal{S}}{\partial \lambda}= -\frac{1}{2}g^{\mu\nu}\frac{\partial \mathcal{S}}{\partial x^{\mu}}\frac{\partial \mathcal{S}}{\partial x^{\nu}}
% \end{equation}
which gives rise to the equation of motion the photons follow:
%For the case of photons that follow null geodesics, the above equation reduces to 
\begin{equation}
	0=  \frac{E^{2}}{f_{1}(r)}-f(r)\left( \frac{\partial \mathcal{S}_{r}}{\partial r} \right)^{2}-\frac{1}{r^{2}}\left( \frac{\partial \mathcal{S}_{\theta}}{\partial \theta} \right)^{2}-\frac{1}{r^{2}\sin^{2}\theta}L^{2}. \label{eq:eqm}
\end{equation}
The above equation can be decoupled into radial and angular parts as follows
\begin{align}
	p_{\theta}^{2}&\equiv \left( \frac{\partial \mathcal{S}}{\partial \theta} \right)^{2} = \mathcal{K}-L^{2}\csc^{2}\theta, \label{eq:p_theta}\\ 
	p_{r}^{2}&\equiv \left( \frac{\partial \mathcal{S}}{\partial r} \right)^{2}= \frac{1}{f(r)}\left(\frac{E^{2}}{f_{1}(r)}-\frac{\mathcal{K}}{r^{2}}\right), \label{eq:p_r}
\end{align} 
with $ \mathcal{K} $ being the Carter constant.%We now introduce the impact  parameter  $ \ell = \frac{L}{E} $.
 Without loss of generality, we can restrict our analysis to the equatorial plane ($ \theta = \frac{\pi}{2} $) due to the spherical symmetry of the spacetime. This would entail $ \frac{\partial \theta}{\partial \lambda}=0 $ and hence one gets $\mathcal{K} = L^{2}$ in such cases of equatorial motion, as is evident from eq.(\ref{eq:p_theta}). 
The radial equation simplifies to
\begin{align}
	\frac{1}{f(r)}\dot{r}= \sqrt{\frac{1}{f(r)f_{1}(r)}\left( E^{2}-\frac{f_{1}(r)\mathcal{K}}{r^{2}} \right)}
	\implies \dot{r} = L\sqrt{\frac{f(r)}{f_{1}(r)}\left( \frac{1}{b^2}-\frac{f_{1}(r)}{r^{2}} \right)},
	%\implies \dot{r}= E\sqrt{V_{eff}}
\end{align}
with $b=\frac{L}{E}$ being the impact parameter.
The geodesic equation of motion $r(\phi)$ is thus given by
\begin{equation}
	\left( \frac{p_r}{p_\phi} \right)^{2}= \left( \frac{d r}{d \phi} \right)^{2} = r^{4}\frac{f(r)}{f_{1}(r)}\left( \frac{1}{b^2} - V_{\rm eff}\right) ,
\end{equation}
with the effective potential being of the form 
\begin{equation}
    V_{\rm eff} (r)  =  \frac{f_1(r)}{r^2}.
    \label{eq:potential}
\end{equation}
For circular geodesics, it is known that $ \dot{r} = \ddot{r}=0 $. 
Choosing $M=1$ makes the coordinate $x$, introduced in eq.(\ref{eq:dimensionless_metric}), equivalent to radial coordinate $r$. However, to avoid confusion we will use $x$ in the following sections when discussing equations specific to the generalised Hayward metric.
%In the case of a generalised Hayward metric, it is much more convenient to use a new coordinate $ x = \frac{r}{M} $ instead of $ r $. As in most cases, the parameter $ M $ is set to unity for all the computational purposes, hence both these coordinates become the same, however we will continue to use the coordinate $ x $. 
%For the generalised Hayward metric this translates to the following set of equations:
Hence, from eq.(\ref{eq:p_r}) and its derivative, we get the following set of equations,
\begin{align}
	\dot{x}&= \frac{f(x)}{f_{1}(x)}\left( \frac{1}{b^{2}}-\frac{f_{1}(x)}{x^{2}} \right)=0, \label{eq:V}\\
	\ddot{x}&=\frac{\left( f_{1}(x) f'(x)-f(r) f_{1}'(x) \right)}{f_{1}^{2}(x)}\left( \frac{1}{b^{2}}-\frac{f_{1}(x)}{x^{2}} \right)+ \frac{ f(x)}{x^{2} f_{1}(x)}\left( -f_{1}'(x)+\frac{2 f_{1}(x)}{x} \right)=0,\label{eq:V2}
\end{align}
solving which enables us to determine  the photon sphere radius $ x_{\rm{ph}} $ and the corresponding  impact parameter $b $. The critical orbits are unstable if  $ \dddot{x}>0 $ and stable if $ \dddot{x}<0 $. Note that the unstable null circular orbits are the photon spheres and the stable null circular orbits are known as anti-photon spheres.

 Using the above information and plugging them into eq.\eqref{eq:celestial_coordiantes} and eq.\eqref{eq:shadow_celestial} we obtain:
 \begin{equation}
	x_{sh}^{2}= \frac{b^{2}}{\frac{f(x_{o})}{f_{1}(x_{o})}\left(1-\frac{f_{1}(x_{o})}{x_{o}^{2}}b^{2} \right)}.
\end{equation}
Note that we have defined $x_{sh} = \frac{r_{sh}}{M}$. However as emphasised earlier we are setting $M=1$, which makes both these definitions equivalent.

For an observer at asymptotic infinity, due to the asymptotic flat nature of the spacetime i.e $ x_{o}\rightarrow \infty $, the shadow radius reduces to  $x_{sh}= b$.
%We will  case by case for the class of generalised Hayward spacetimes.

\subsection{Shadows with plasma medium}
In reality, astrophysical objects are surrounded by an interstellar medium, generally considered to be plasma in the near horizon region. When photons pass through this plasma, they interact with it, altering the trajectory of the photons. Consequently, this interaction affects the size of the shadow of the central object as observed by a distant observer. Our current analysis focuses on homogeneous and non-homogeneous plasma profiles and observing their effect on the shadows. In this section, we will briefly discuss the impact of the plasma medium on the shadow radius following \cite{Perlick:2015prd}. These findings will be used to analyze the effects of the above mentioned plasma profiles on generalized Hayward spacetimes.

We would restrict our analysis to the non-magnetized cold plasma with the plasma frequency of the form,
\begin{equation}
	\omega_{p}(r)=\frac{4\pi e^{2}}{m}N(r),
\end{equation}
where $ e $ is electron charge,  $ m $ the electron mass and $ N(r) $ is the number density of the electrons. The plasma medium acts as an effective refractive index for the photons propagating in this medium, and the effective refractive index is dependent on the radial coordinate $ r $, as well as on the frequency of the photon $ \omega $, and is given by
\begin{equation}
	n^{2}(r)=1-\frac{\omega_{p}^{2}(r)}{\omega^{2}}.
	\label{eq:ref_ind}
\end{equation}
In the presence of surrounding plasma, the null geodesic equation of motion, given in eq.(\ref{eq:eqm}), gets modified to 
\begin{equation*}
	0=  \frac{E^{2}}{f_{1}(r)}-f(r)\left( \frac{\partial \mathcal{S}_{r}}{\partial r} \right)^{2}-\frac{1}{r^{2}}\left( \frac{\partial \mathcal{S}_{\theta}}{\partial \theta} \right)^{2}-\frac{1}{r^{2}\sin^{2}\theta}L^{2}-\omega_{p}^{2}(r).
\end{equation*}
Since the plasma medium does not depend on the angular coordinates, the same technique of variable separation, as done for the vacuum case, works for this case also leads to
\begin{align}
	p_{\theta}^{2}&\equiv \left( \frac{\partial \mathcal{S}}{\partial \theta} \right)^{2} = \mathcal{K}-L^{2}\csc^{2}\theta\\
	p_{r}^{2}&\equiv \left( \frac{\partial \mathcal{S}}{\partial r} \right)^{2}= \frac{1}{f(r)}\left(\frac{E^{2}}{f_{1}(r)}-\frac{\mathcal{K}}{r^{2}}-\omega_{p}^{2}(r)\right).
\end{align} 
We get $ \mathcal{K} =L^{2} $ by restricting the analysis to equatorial plane, since spherical symmetry still holds, similar to the vacuum case. Assuming that the frequency of photon observed by the asymptotic observer as $ \omega_{0}\equiv E $, for the case of asymptotically flat spacetime, the frequency $ \omega $ observed by a static observer as a function of $ r $ is given by
\begin{equation}
	\omega(r)=\frac{\omega_{0}}{\sqrt{f_{1}(r)}},
\end{equation} 
using which the refractive index, in presence of plasma, can be written as
\begin{equation}
	n^{2}=1-\frac{\omega_{p}^{2}}{\frac{\omega_{0}^{2}}{f_{1}(r)}}.
\end{equation}
In order to reach the asymptotic observer, the photons must satisfy $ n^{2}>0$ which implies $\frac{\omega_{0}^{2}}{f_{1}(r)}> \omega_{p}^{2}(r)$.

Similar to the vacuum case, we translate to $ x$ coordinate and also introduce an new parameter $ \Omega(x)= \frac{\omega_{p}^{2}}{E^{2}} $. Hence, the equivalent condition to eq.(\ref{eq:V}) and (\ref{eq:V2}) for the circular orbits in presence of plasma is,
\begin{align}
	\dot{x}&= \frac{f(x)}{f_{1}(x)}\left( \frac{1}{b^{2}}- \frac{f_{1}(x)}{x^{2}}-\frac{\Omega(x)f_{1}(x)}{b^{2}} \right)=0\\
	\ddot{x}&= \left(\frac{f(x)}{f_{1}(x)}\right)'\left( \frac{1}{b^{2}}- \frac{f_{1}(x)}{x^{2}}-\frac{\Omega(x)f_{1}(x)}{b^{2}} \right)+  \frac{f(x)}{f_{1}(x)}\left( -\frac{f_{1}'(x)}{x^{2}} + \frac{2f_{1}(x)}{x^{3}}-\frac{(\Omega(x)f_{1}(x)')}{b^{2}} \right)=0
\end{align}
Using the above equations, the impact parameter can be written in terms of the photon sphere radius $x_{\rm ph}$ as,
\begin{equation}
	b^{2} = \frac{x^{2}}{f_{1}}\left( 1-\Omega f_{1} \right)\Big\vert_{x=x_{\rm ph}} \label{eq:impact_plasma}.
\end{equation}
The $ x_{\rm ph} $ has two possible branches of solutions given by
\begin{align}
& f(x) = 0 \label{eq:Rph_P1},\\
	&\frac{1}{f_{1}(x)}(1-\Omega(x) f_{1}(x))\Big(-f_{1}'(x)+\frac{2 f_{1}(x)}{x}\Big)-(\Omega(x) f_{1}(x))'=0.
	\label{eq:Rph_P2}
\end{align}
The shadow radius observed by an observer at $ x_{o} $ is given by
\begin{align}
	x_{\rm sh}^{2}&=\frac{\frac{x_{\rm ph}^{2}}{f_{1}(x_{\rm ph})}\left( 1-\Omega(x_{\rm ph})f_{1}(x_{\rm ph}) \right)}{\frac{f(x_{o})}{f_{1}(x_{o})}\left(1-\frac{b^{2}}{x^{2}_{o}}f_{1}(x_{o})-\Omega(x_{o})f_{1}(x_{o})\right)}.
\end{align}
For an asymptotic observer at $ x_{o}\rightarrow \infty $, the shadow radius simplifies to
\begin{equation}
	x_{\rm sh}^{2}\Big\vert_{x_{o}\rightarrow\infty}=\frac{\frac{x_{\rm ph}^{2}}{f_{1}(x_{\rm ph})}\left( 1-\Omega(x_{\rm ph})f_{1}(x_{\rm ph}) \right)}{\left(1-\Omega(\infty)\right)}.
	\label{eq:Rsh_P}
\end{equation}
Note that the numerator in the above equation is required to be positive as it represents the impact parameter, as shown in eq.(\ref{eq:impact_plasma}), which is a physical quantity and hence cannot be negative. 

Now let us apply the above results to the two plasma profiles we are interested. We begin by considering the homogeneous plasma profile given by $ \Omega = k_{0} $.  Solving equations \eqref{eq:Rph_P1} and \eqref{eq:Rph_P2} with $ \Omega = k_{0} $ gives the impact parameter 

% \iffalse

% For this case, the possible solutions for the photon sphere are given by:
% \begin{align}
% 	f(x_{\rm{ph}})=0 \label{homogeneousplasmarpheq1}\\
% 	-f_{1}(x_{\rm{ph}}) + \frac{2 f_{1}(x_{\rm{ph}})}{x_{\rm{ph}}}-2 \kappa_{0} \frac{f_{1}^{2}(x_{\rm{ph}})}{x_{\rm{ph}}}=0 \label{homogeneousplasmarpheq2}
% \end{align}
% \fi
\begin{equation}
	b^{2}=\frac{x_{\rm ph}^{2}}{f_{1}(x_{\rm ph})}\left(1-k_{0}f_{1}(x_{\rm ph})\right) \label{eq:homogeneousplasmaimpactparameter}
\end{equation}
with the shadow radius obtained from eq.\eqref{eq:Rsh_P}
\begin{equation}
	x_{\rm sh}^{2}=\frac{\frac{x_{\rm ph}^{2}}{f_{1}(x_{\rm ph})}\left( 1- k_{0}f_{1}(x_{\rm ph}) \right)}{\left(1-k_{0}\right)}.
	\label{eq:homogeneousplasmashadow}
\end{equation}
Since the impact parameter, given by  eq.\eqref{eq:homogeneousplasmaimpactparameter}, is a physical quantity which needs to be positive, the denominator of eq.\eqref{eq:homogeneousplasmashadow} also needs to be positive, hence giving $ 0\leq \kappa_{0}<1 $. Further constraints on this parameter would depend on the refractive index condition and will be studied for each spacetime from the generalised Hayward metric.

For the non-homogeneous case, $ \Omega =  \frac{k_x}{x} $.  Similar to the homogeneous profile, solving equations \eqref{eq:Rph_P1} and \eqref{eq:Rph_P2} we get the impact parameter to be
\begin{equation}
	b^{2}=\frac{x_{\rm ph}^{2}}{f_{1}(x_{\rm ph})}\left(1-\frac{k_{x}}{x_{\rm ph}}f_{1}(x_{\rm ph})\right).\label{eq:nonhomogeneousplasmaimpactparameter}
\end{equation}
Note that as $ x \rightarrow \infty $, the plasma profile $ \Omega \rightarrow 0 $, hence the shadow radius in eq.\eqref{eq:Rsh_P} reduces to
\begin{equation}
	x_{\rm sh}^{2}= b^{2}=\frac{x_{\rm ph}^{2}}{f_{1}(x_{\rm ph})}\left(1-\frac{k_{x}}{x_{\rm ph}}f_{1}(x_{\rm ph})\right).	\label{eq:nonhomogeneousplasmashadow}
\end{equation}
In this case, the constraints on the plasma profile parameter are obtained from the impact parameter and the refractive index, which need to be analyzed for each spacetime. No broad statements can be made at this point for the general function.

\section{Shadow of Generalised Hayward metric}\label{sec:3}
In this section, we will apply the above results for the various spacetimes of the generalized Hayward metric given in eq.\eqref{eq:dimensionless_metric}. and summarized in Table \ref{tab:classification}.

\subsection{ Schwarzschild black hole  $\sigma =1$, $ \kappa=0 $:}
\begin{figure}[h]
	\makebox[0.95\paperwidth][c]{
		\hspace*{-3.5cm}
		\begin{subfigure}[t]{0.4\paperwidth}
			\includegraphics[width=0.75\textwidth]{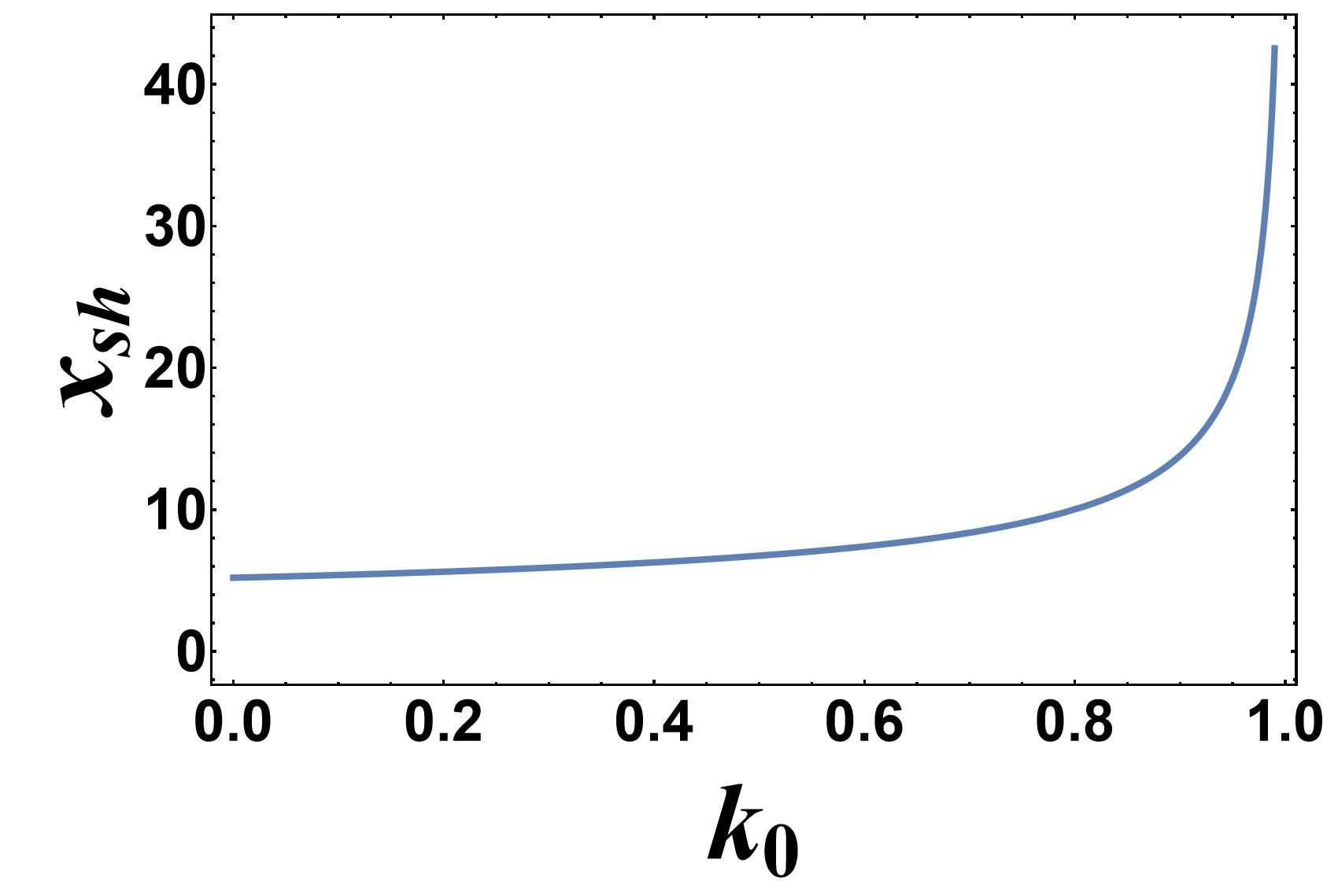}
			\caption{$ \Omega= k_{0} $}\label{fig:SBH_homo_shadow}
		\end{subfigure}
		\begin{subfigure}[t]{0.4\paperwidth}
			\centering
			\includegraphics[width=0.75\textwidth]{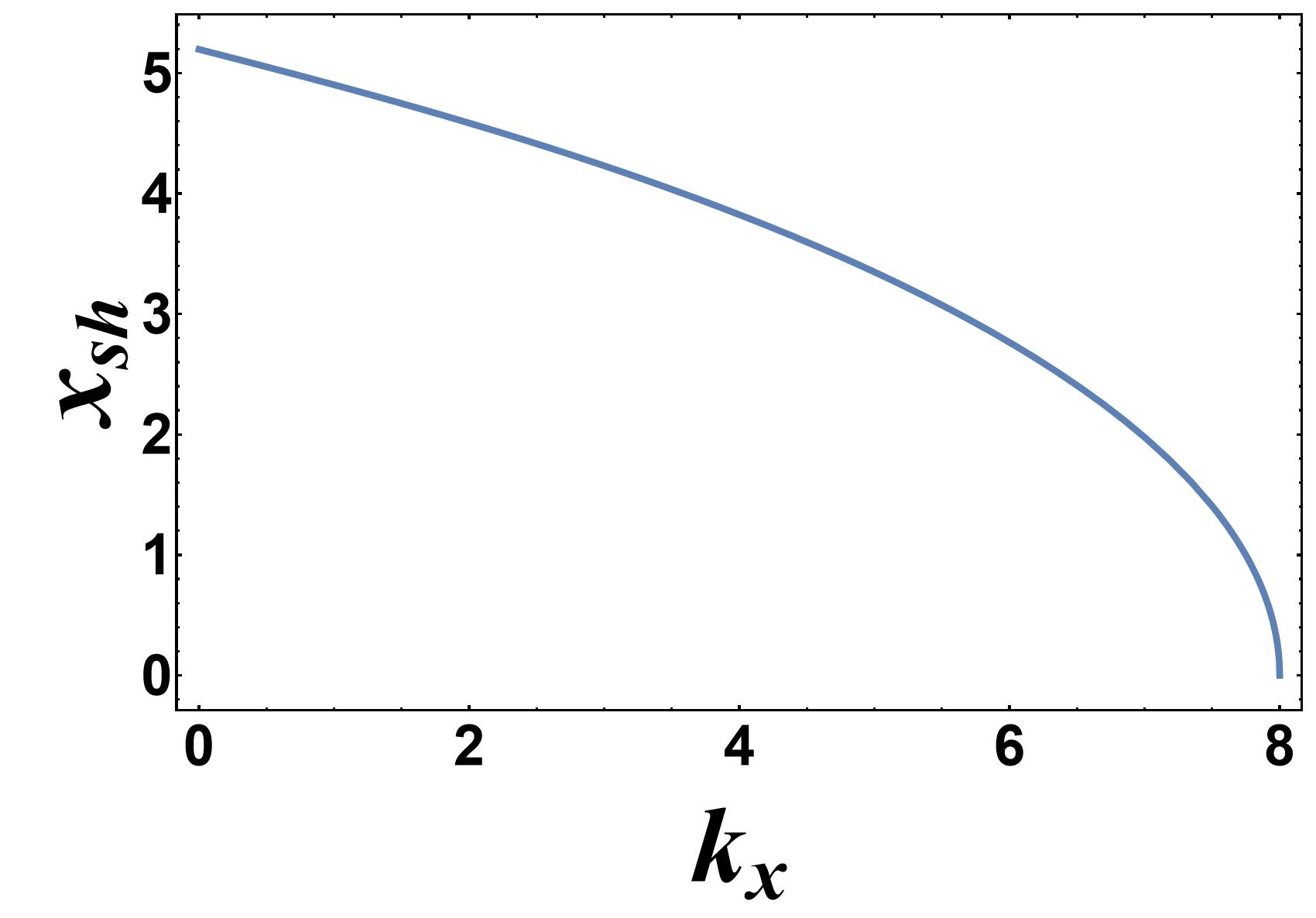}
			\caption{$ \Omega=\frac{ k_{x}}{x} $}\label{fig:SBH_nonhomo_shadow}
		\end{subfigure}
	}
	\caption{Plots showing variation of the shadow radius of the Schwarzschild BH as a function of the plasma parameter. The left and right panels denote the effect of the homogeneous and non-homogeneous plasma profiles on the shadow radius, respectively. } 
\end{figure}
By setting $ \sigma =1$, $ \kappa=0 $ in eq.\eqref{eq:dimensionless_metric}, we recover the standard Schwarzschild solution. The shadow radius of Schwarzschild BH was first analyzed by \cite{Synge:1966mon}. This is one of the most well-known results in the literature, and we include these results for completeness. %These results are further extended by considering various additional complexities, such as the presence of plasma medium or dark matter background.
Solving eqns.\eqref{eq:V} and \eqref{eq:V2}, we get $x_{\rm ph}= 3$, $b^2 =27$ and 
the shadow radius being $x_{\rm sh}= 3\sqrt{3}$. 
%The instability of the photon orbit is well studied and understood in various reviews and references. For example, the original works of  \cite{Synge:1966mon} discuss them in detail, and we will not be discussing the same.

We now focus on the influence of the plasma on the shadow of Schwarzschild BH. For the case of homogeneous plasma, the eq.\eqref{eq:Rph_P1} does not lead to any solutions. However, eq.\eqref{eq:Rph_P2} reduces to
\begin{equation}
	(1-k_{0})x^{2}+(4k_{0}-3)x-4k_0=0.
\end{equation}
Solving this quadratic equation we find that the $ x_{\rm ph}>x_{\rm h} $ for all the range of $ k_{0} $ where $x_{\rm h}$ is the horizon radius. Note that the shadow radius diverges at $ k_{0}\rightarrow 1 $, which is observed in the Fig.(\ref{fig:SBH_homo_shadow}). 

Similarly, for the case of non-homogeneous plasma, the eq.\eqref{eq:Rph_P2} reduces to
\begin{equation}
	2 x^3+(-k_{x}-6) x^2+4 k_{x} x-4 k_{x}=0.
\end{equation}
Substituting the solution of the above equation in eq.\eqref{eq:nonhomogeneousplasmashadow}, it can be checked that the shadow radius becomes zero at $ k_{x}=8 $, beyond which the photon sphere does not exist. The shadow radius decreases monotonically as a function of the plasma parameter $k_x$ as shown in Fig.(\ref{fig:SBH_nonhomo_shadow}).

%\FloatBarrier

\subsection{ Schwarzschild Wormhole $ \sigma=0,\kappa=0 $:\label{sec:SWH}}
\begin{figure}[h]
	\centering
	\includegraphics[width=0.45\textwidth]{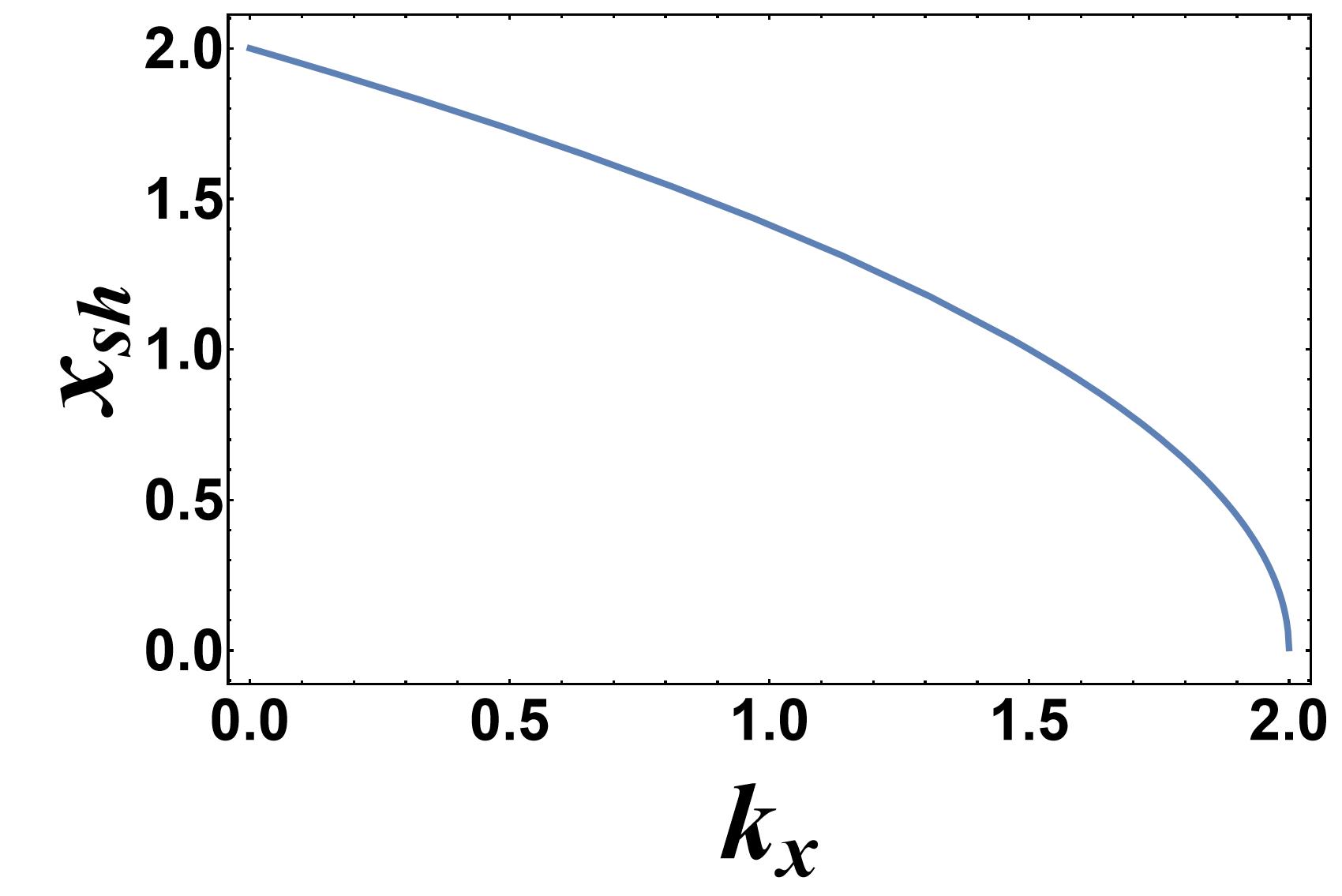}
	\caption{Plot shows the monotonic decrease of the shadow radius for the Schwarzschild Wormhole with increasing non-homogeneous plasma parameter where the plasma profile is given as $\Omega= \frac{k_{x}}{x}$. Physical solutions exist till $k_x =2$.}\label{fig:SWH_nonhomo_shadow}
\end{figure}

For $\sigma =\kappa =0$, the metric functions in eq.(\ref{eq:dimensionless_metric}) become $ f_{1}(x) =1$ and $ f(x)= 1-\frac{2}{x} $.
Solving eq.\eqref{eq:V},\eqref{eq:V2} we get $x_{\rm ph}=2, \, b^{2}=x_{\rm ph}^{2}=2, \,x_{\rm sh}=\sqrt{b^{2}}=2$.

Now we look at the influence of plasma on the shadow of Schwarzschild wormhole spacetime. For the case of homogeneous plasma profile, we observe that the photon sphere is still located at the throat $ x= 2 $ obtained from solving $f(x)=0$. Since $ f_{1}(x)=1 $, the shadow radius from eq.\eqref{eq:homogeneousplasmashadow} becomes independent of the plasma parameter $k_0$. Hence it is identical to the vacuum case. 

Moving on to the case of non-homogeneous plasma profile, the photon sphere equations \eqref{eq:Rph_P1} and \eqref{eq:Rph_P2} provide us with the solutions $x_{\rm ph}=2$ and $x_{\rm ph}=2-k_{x}$. The shadow radius corresponding to the two photon radius is thus $x_{\rm sh}^2 = 2(2-k_x)$ and $x_{\rm sh}^{2}=-\frac{k_{x}}{4}$ respectively. Of the above solutions, only the first one is physically reasonable. Hence, from $x_{\rm sh}^2 = 2(2-k_x)$, we can deduce that the shadow radius decreases monotonically with increasing plasma parameter $k_x$. At $k_x=2$, the shadow radius vanishes and becomes unphysical beyond this point as shown in Fig.(\ref{fig:SWH_nonhomo_shadow}).

\begin{figure}[h]
	\makebox[0.95\paperwidth][c]{
		\hspace*{-3.5cm}
		\begin{subfigure}[t]{0.4\paperwidth}
			\includegraphics[width=0.95\textwidth]{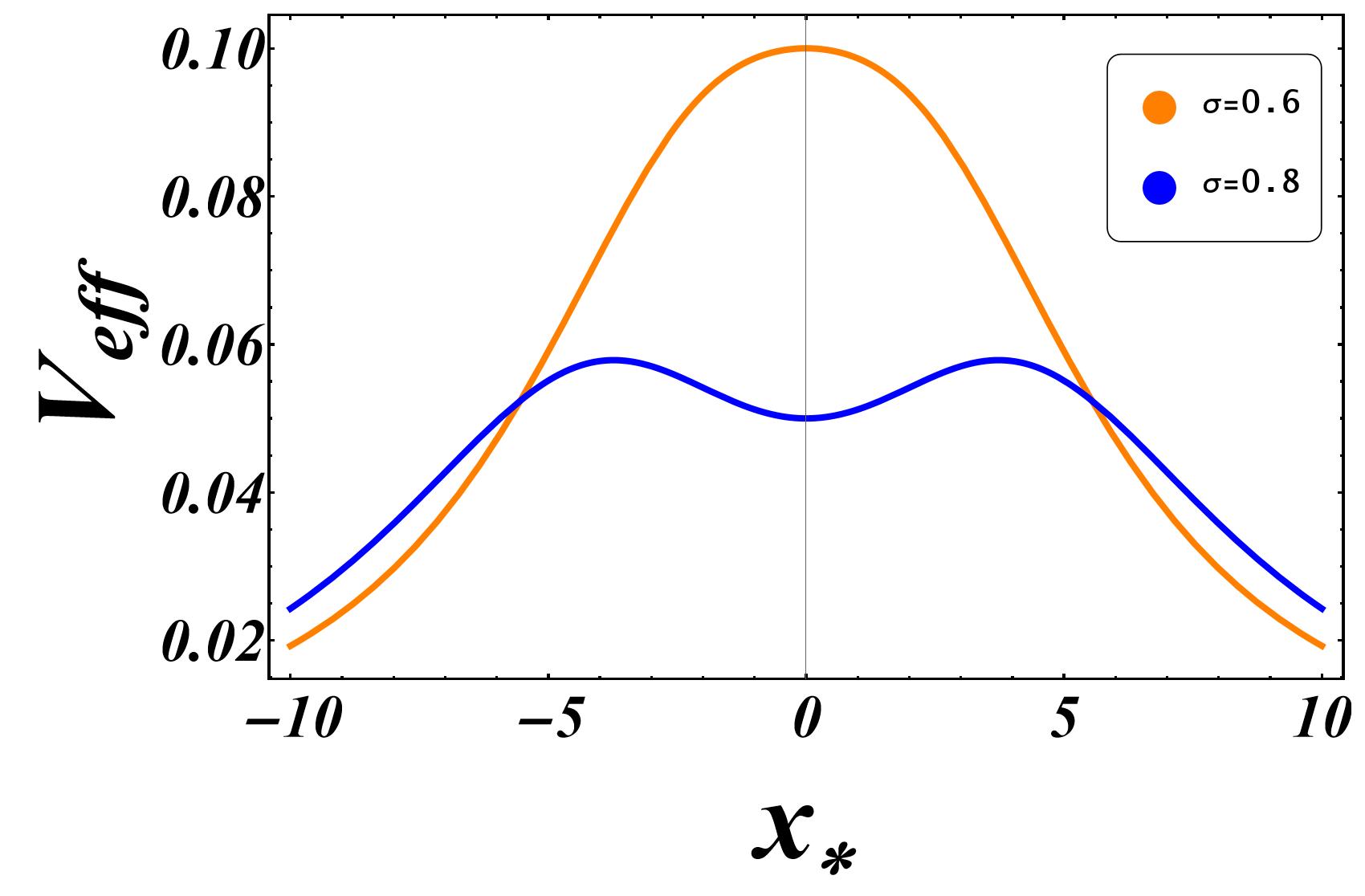}
			\caption{}%Effective Potential}
            \label{fig:DSWH_pot}
		\end{subfigure}
		\begin{subfigure}[t]{0.45\paperwidth}
			\includegraphics[width=0.95\textwidth]{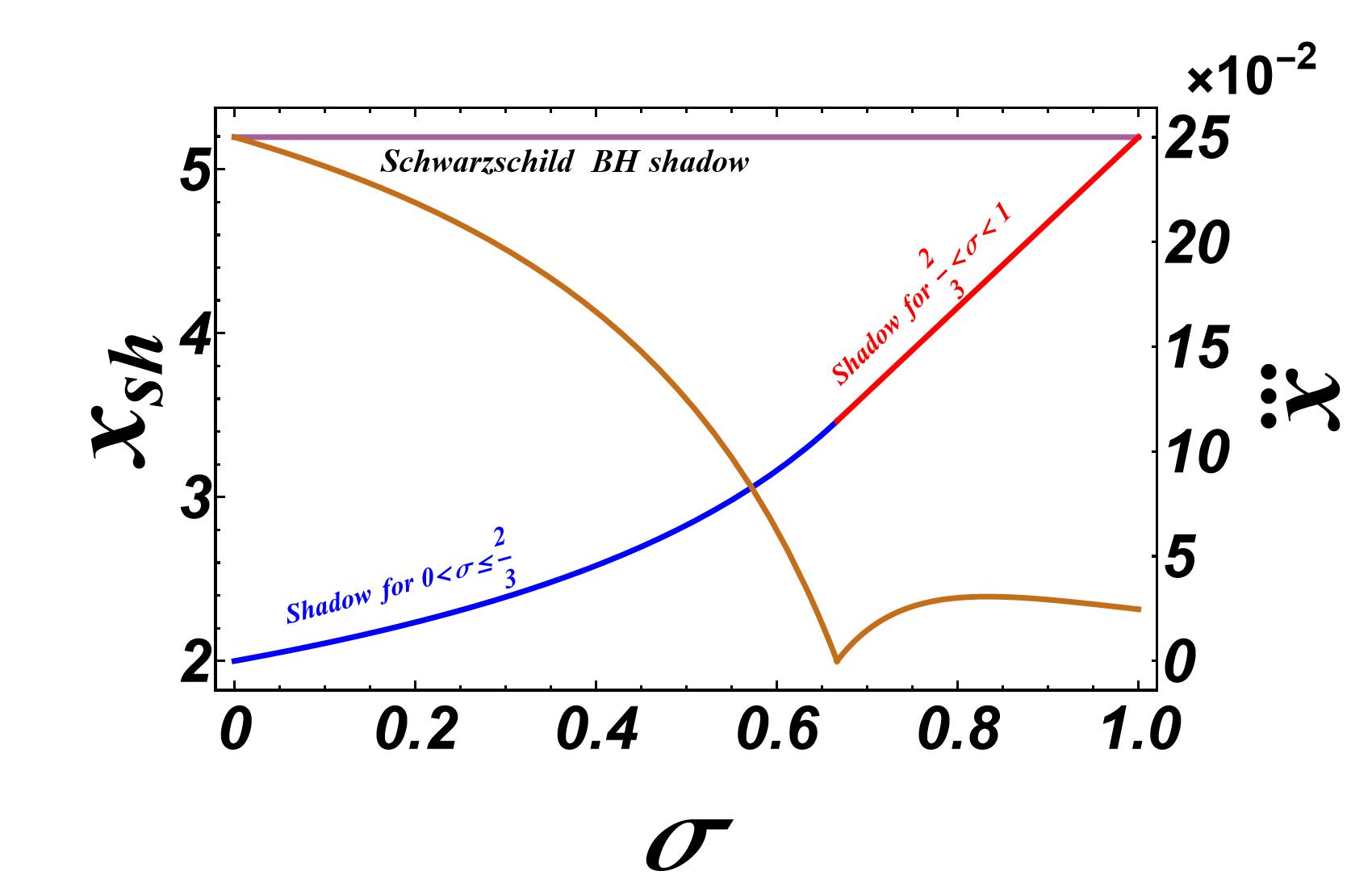}
			\caption{}%Shadow Radius }
   \label{fig:DSWH_shadow}
		\end{subfigure}
%		\begin{subfigure}[t]{0.3\paperwidth}
%			\centering
%			\includegraphics[width=0.95\textwidth]{plots/nonplasma/Damour-Solodukin_Stability.jpg}
%			\caption{Stability}
%		\end{subfigure}
	}
\caption{(a) The plot shows the variation of the effective potential for the Damour-Solodukhin wormhole as a function of tortoise coordinate $x_*$. The nature of the effective potential depends on the parameter $\sigma$. (b) The plot shows the shadow radius for the complete range of the parameter $\sigma$ which are compared with the Schwarzschild BH shadow denoted by the violet line. The brown line shows the variation of $\dddot{x}$ which is positive for the complete range indicating that the photon spheres are unstable.}
\end{figure}

%\FloatBarrier

\subsection{Damour-Solodukhin Wormhole $ 0<\sigma<1 , \kappa=0 $ :  \label{DSW}} 

For the Damour-Solodukhin wormhole, the metric components simplify to $f_{1}(x)=1-\frac{2\sigma}{x}$ and $f(x)=1-\frac{2}{x}$.
Unlike the previous cases, for this wormhole we have two possible solutions to eq.\eqref{eq:V}, which will be discussed in detail one by one. The first set of solutions corresponds to $f(x)=0$, which gives $x_{\rm ph}=2$. The critical impact parameter for this is obtained by solving eq.\eqref{eq:V2}, which gives $b^{2}= \frac{4}{1-\sigma}$. It is evident that these circular geodesics are located at the throat. Note that the photons will be observable to an asymptotic observer only if the photon orbits are unstable which is determined by the condition $ \dddot{x}\vert_{x=x_{\rm ph}}>0 $. Hence we check the stability of the null geodesics by computing $\dddot{x}\vert_{x=x_{\rm ph}}$ which for this first set of solutions is given by
\begin{equation}
	\dddot{x}\vert_{x=x_{\rm ph}} = \frac{2-3\sigma}{8(1-\sigma)}.
\end{equation}
From the above expression we obtain the following condition on $\sigma$,
\begin{align}
	\dddot{x}\vert_{x=x_{\rm ph}} &>0\quad \implies 0<\sigma < \frac{2}{3} \implies \text{photon sphere}\\
	\dddot{x}\vert_{x=x_{\rm ph}} &<0\quad \implies \frac{2}{3}<\sigma<1 \implies \text{anti-photon sphere}
\end{align}
The shadow radius for the range of parameter $0<\sigma < \frac{2}{3}$ is thus obtained from this set of solution i.e. $x_{\rm ph}=2$ and $b^{2}= \frac{4}{1-\sigma}$ as shown by the blue curve in Fig.(\ref{fig:DSWH_shadow}). For $\sigma = \frac{2}{3}$, the $\dddot{x}$ is identically 0 and higher order derivative tests are performed to ensure that the unstable photon sphere exists at $x_{\rm}=2$.
	\begin{figure}[h]
		\makebox[0.95\paperwidth][c]{
			\hspace*{-3.5cm}
			\begin{subfigure}[t]{0.4\paperwidth}
				\includegraphics[width=0.75\textwidth]{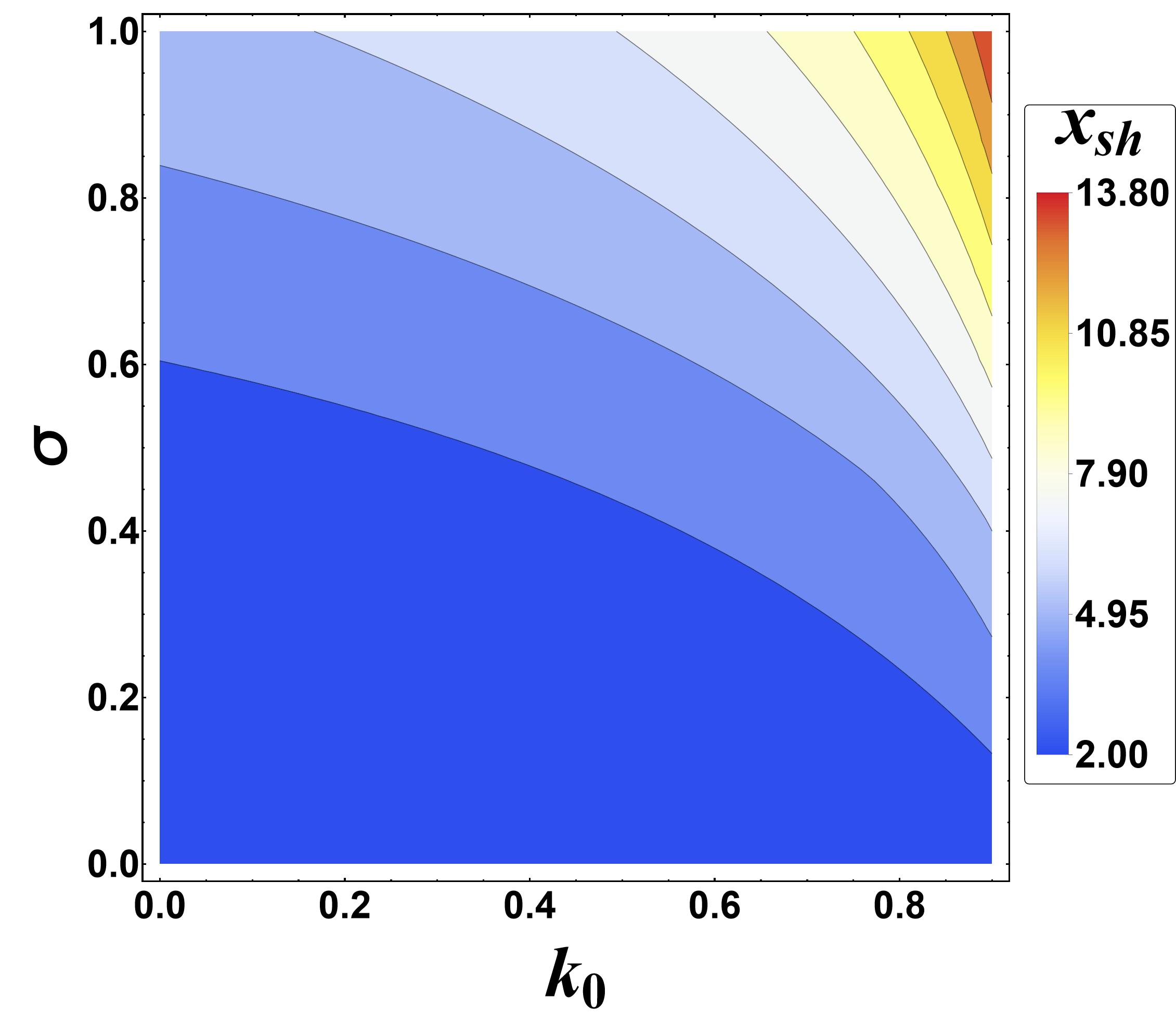}
					\caption{}%Shadow Plot}
     \label{fig:DSWH_homo_shadow}
			\end{subfigure}
			\begin{subfigure}[t]{0.4\paperwidth}
				\centering
				\includegraphics[width=0.75\textwidth]{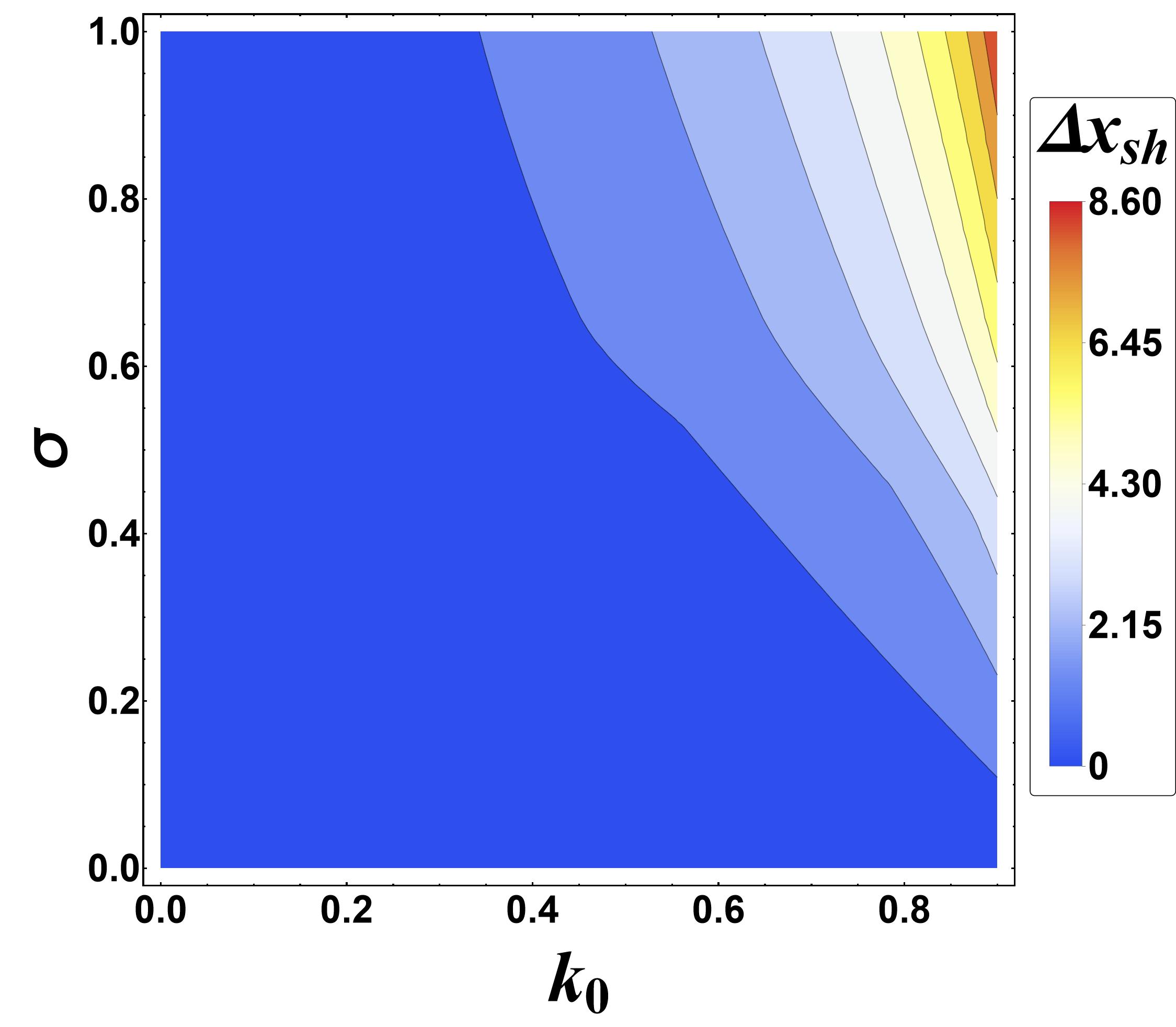}
			\caption{}%$ \Delta x_{sh} = x_{sh} -x_{sh}\vert_{k_{0}=0} $}
   \label{fig:DSWH_homo_diff}
			\end{subfigure}
		}
\caption{Figure shows the effect of homogeneous plasma profile given by $ \Omega=k_{0}$ on the shadow radius of the Damour-Solodukhin wormhole. (a) The plot shows the variation of the shadow radius as a function of $\sigma$ and $k_0$. Larger shadow radius is observed for higher values of both $\sigma$ and $k_0$. (b) The plot shows the variation of the difference $\Delta x_{\rm sh} =x_{\rm sh} - x_{\rm sh}\vert_{k_0=0}$ between the shadow radius with and without plasma, as a function of $\sigma$ and $k_0$. }
	\end{figure}
  \begin{figure}[h]
		\makebox[0.95\paperwidth][c]{
			\hspace*{-3.5cm}
			\begin{subfigure}[t]{0.4\paperwidth}
				\includegraphics[width=0.75\textwidth]{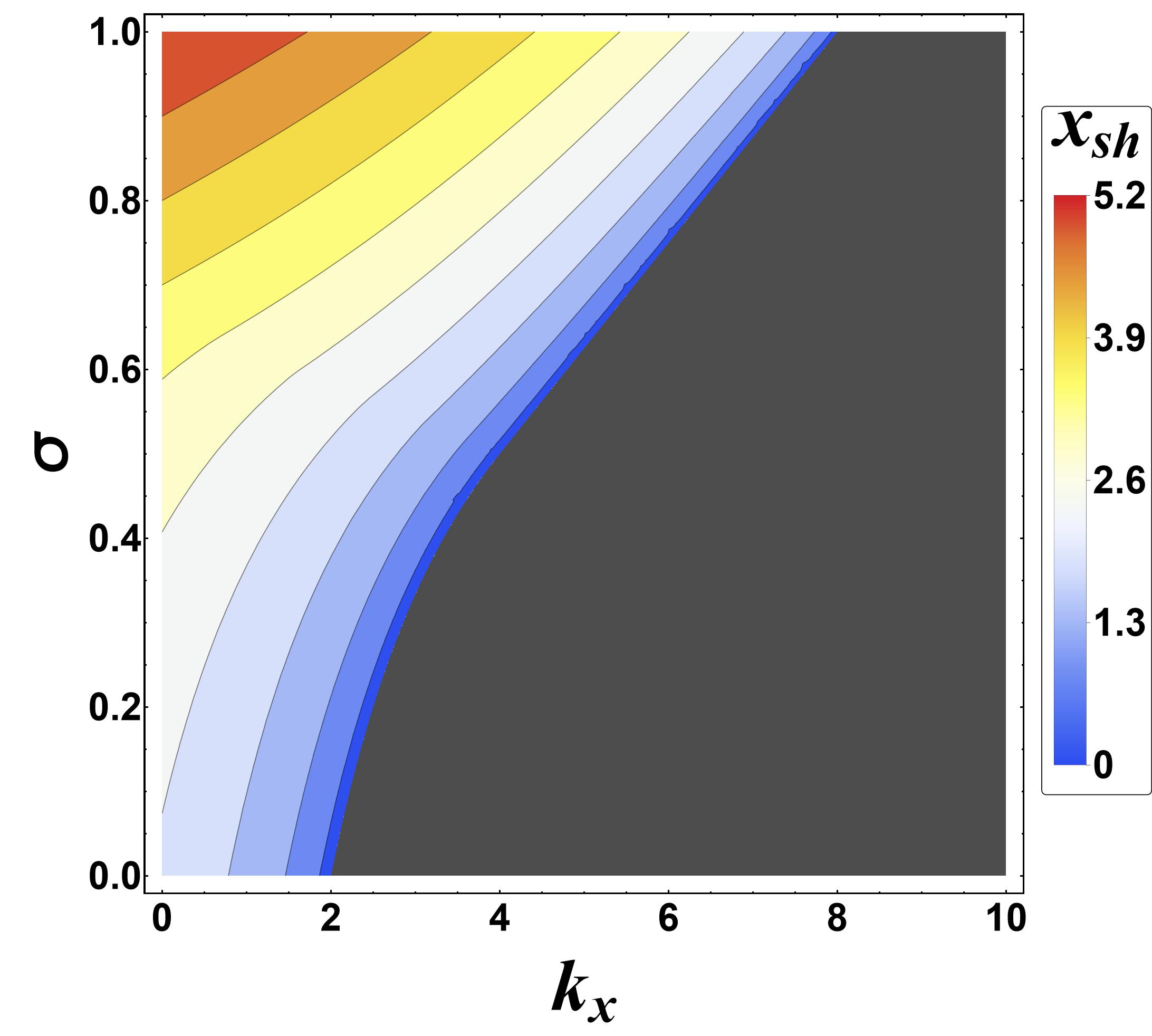}
				\caption{}%Shadow Plot}
    \label{fig:DSWH_nonhom_shadow}
			\end{subfigure}
			\begin{subfigure}[t]{0.4\paperwidth}
				\centering
				\includegraphics[width=0.75\textwidth]{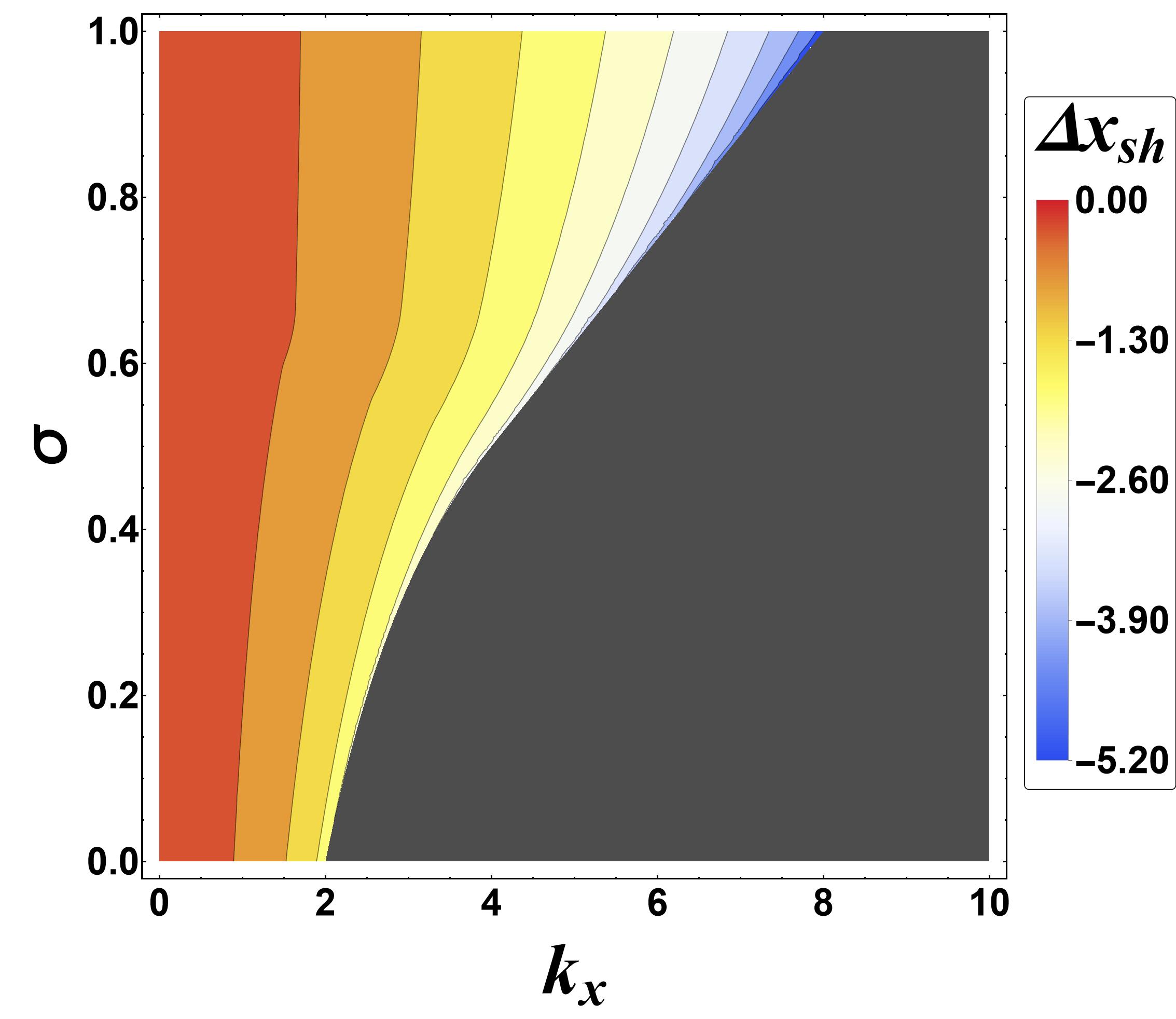}
				\caption{}%$ \Delta x_{sh} = x_{sh} -x_{sh}\vert_{k_{r}=0} $}
    \label{fig:DSWH_nonhom_diff}
			\end{subfigure}
		}
		\caption{Figure shows the effect of non-homogeneous plasma profile given by $ \Omega=\frac{k_{x}}{x}$ on the shadow radius of the Damour-Solodukhin wormhole. The grey region represents the values of parameters for which there is no shadow. (a)The plot shows the variation of the shadow radius as a function of $\sigma$ and $k_x$. Larger shadow radius is observed for higher values of $\sigma$ and smaller values of $k_x$, unlike the homogeneous plasma case. (b) The plot shows the variation of the difference $\Delta x_{\rm sh}= x_{\rm sh} - x_{\rm sh}\vert_{k_x=0}$ between the shadow radius with and without plasma, as a function of $\sigma$ and $k_x$.}
	\end{figure}
The second set of solutions to eq.\eqref{eq:V}, along with eq.\eqref{eq:V2}, gives $x_{\rm ph} = 3\sigma$ and $b^{2} = 27 \sigma^{2}$.
This set of solutions are only valid when $ \sigma\geq\frac{2}{3} $ since below this value, the photon sphere radius would be located inside the throat radius and thus will not be part of the manifold. The stability condition for this case reduces to
\begin{equation}
	\dddot{x}\vert_{x=x_{\rm ph}} = \frac{2 (3 \sigma -2)}{81 \sigma ^5},
\end{equation}
which implies
\begin{align}
	\dddot{x}\vert_{x=x_{\rm ph}} &>0\quad \implies \frac{2}{3} <\sigma<1 \implies \text{photon sphere}.
\end{align}
Thus the shadow radius for parameter range $\frac{2}{3}<\sigma<1$ is obtained using $x_{\rm ph} = 3\sigma$ and $b^{2} = 27 \sigma^{2}$ as shown by the red curve in Fig.(\ref{fig:DSWH_shadow}). As $\sigma \rightarrow 0$, the shadow of the wormhole approaches that of the Schwarzschild BH denoted by the violet line in Fig.(\ref{fig:DSWH_shadow}) for comparison. The $\dddot{x}$ is positive for all values of $x_{\rm sh}$ and $\sigma$ denoted by the brown curve in Fig.(\ref{fig:DSWH_shadow}) which indicates their instability.

The difference in the two branches of the solutions of eq.\eqref{eq:V}, as mentioned above, can be clearly seen from the effective potential plots shown in  Fig.(\ref{fig:DSWH_pot}), where the nature of the potential changes from a single to a double peak with the critical point being $\sigma=\frac{2}{3}$. The potential for the wormhole is plotted as a function of tortoise coordinate $x_*$ which is defined as $d x_*^2 = \frac{dx^2}{f_1(x) f(x)}$ and ranges from $(-\infty, \infty)$.

%Hence we conclude that the photon sphere, which determines the shadow radius, is located at the throat $x=2$ for $ \sigma<\frac{2}{3} $ and at $ x= 3\sigma $ for $ \sigma>\frac{2}{3} $. The corresponding shadow radius can be computed from the above information and is shown in Fig.\ref{fig:DS_shadow}.

%In the plot of the shadow radius above, we can see that as the limit $ \sigma\rightarrow 0 $, the Damour-Solodukin shadow's radius is similar to that of the Schwarzschild wormhole. At the other end of the parameter space, the shadow radius tends towards the  Schwarzschild BH radius.
%Thus, for this wormhole, the shadow radius varies between the shadow radius of the Schwarzschild wormhole and that of the Schwarzschild BH depending on the value of $\sigma $. For the range $ 0<\sigma<\frac{2}{3} $ there is only one circular photon orbit at the throat. However, for $ \frac{2}{3}<\sigma\leq 1 $ there are two circular photon spheres, a photon sphere at $ x_{\rm ph}= 3 \sigma $ and an anti-photon sphere at the throat $ x_{\rm ph}=2 $.

We now move on to study the effect of plasma on the shadow of this wormhole. For the homogeneous plasma profile, the variation of the shadow radius as a function of $\sigma$ and plasma parameter $k_0$ is shown in Fig.(\ref{fig:DSWH_homo_shadow}). Large values of $x_{\rm sh}$ is seen for higher values of both $\sigma$ and $k_0$. To highlight the effect of plasma, we also plot the difference in shadow radius between the with and without plasma cases denoted by $\Delta x_{\rm sh} = x_{\rm sh} - x_{\rm sh}\vert_{k_0=0}$
as shown in Fig.(\ref{fig:DSWH_homo_diff}). For the case of non-homogeneous plasma, the $x_{\rm sh}$ and $\Delta x_{\rm sh}$ are shown in Fig.(\ref{fig:DSWH_nonhom_shadow}) and Fig(\ref{fig:DSWH_nonhom_diff}), respectively. Unlike the homogeneous case, we observe that the shadow radius reduces as the plasma parameter $k_x$ increases. 

% \FloatBarrier

\subsection{Hayward Wormhole: $ \sigma =0\;,\; 0<\kappa\leq\frac{4}{3\sqrt{3}}$  \label{sec:HWH}}
\begin{figure}[h]
		\makebox[0.95\paperwidth][c]{
			\hspace*{-3.5cm}
			\begin{subfigure}[t]{0.4\paperwidth}
				\centering
				\includegraphics[width=0.95 \textwidth]{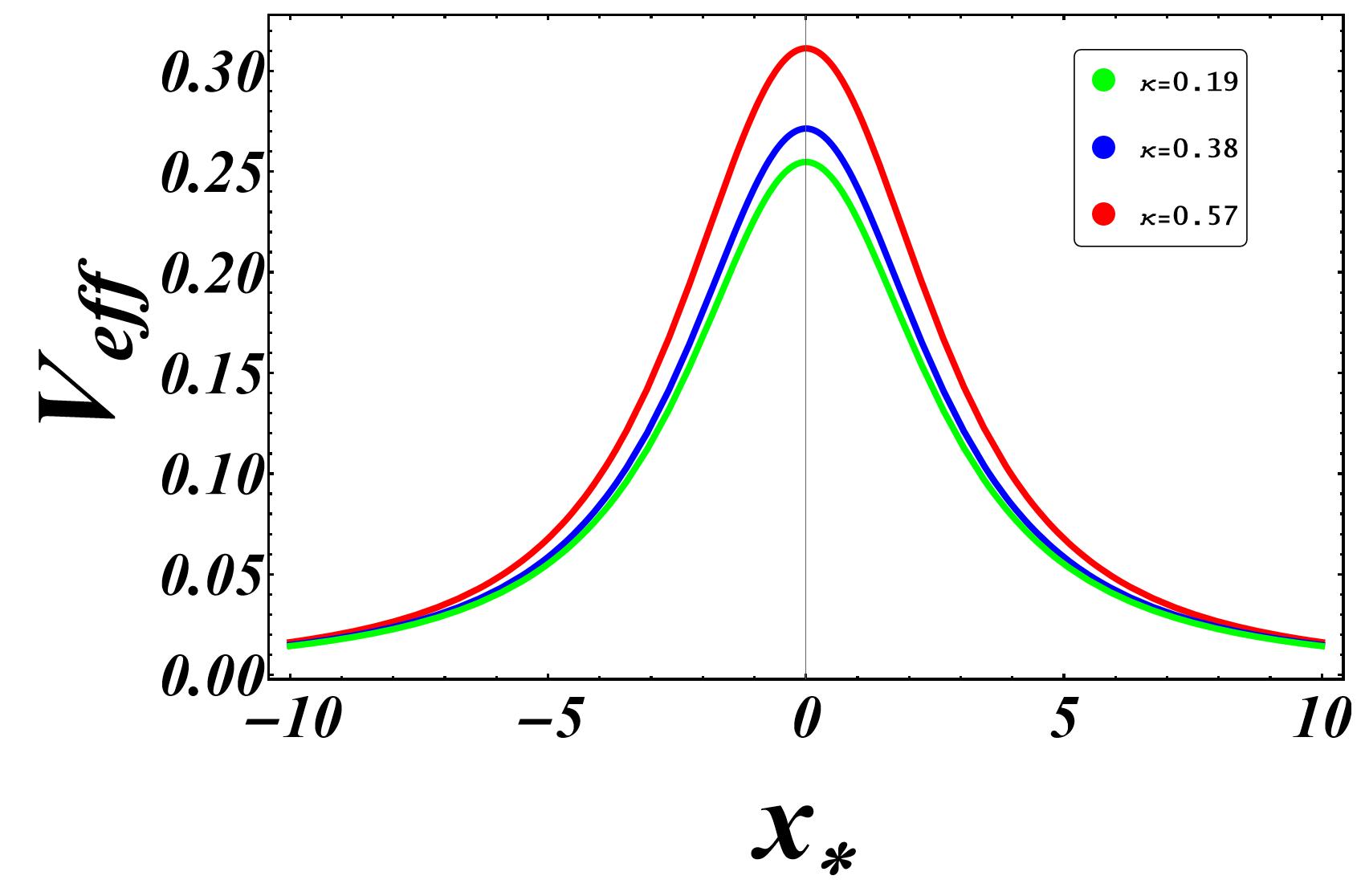}
				\caption{}
    \label{fig:HWH_pot}
			\end{subfigure}
			\begin{subfigure}[t]{0.47\paperwidth}
				\centering
				\includegraphics[width=0.95 \textwidth]{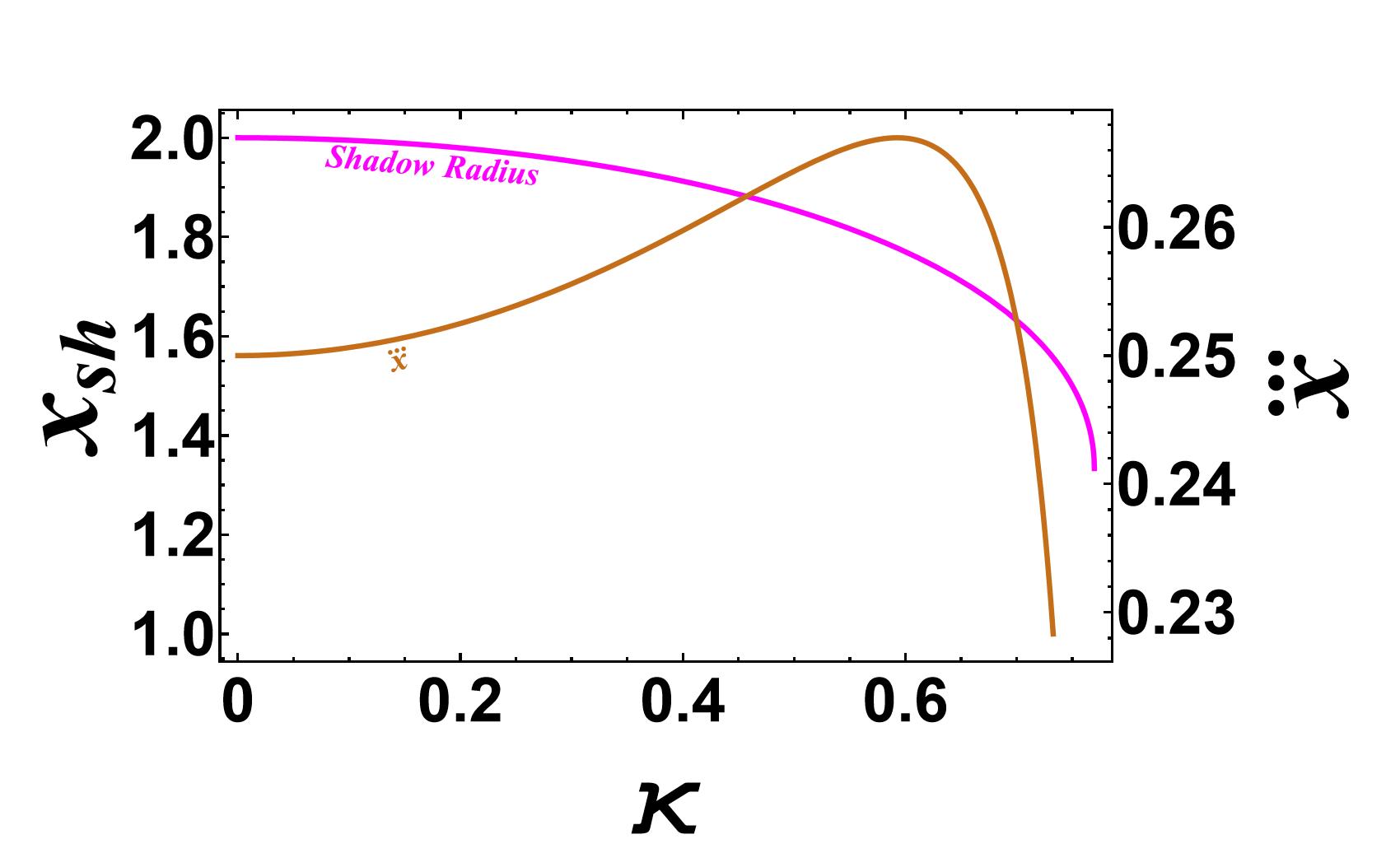}
				\caption{}
    \label{fig:HWH_shadow}
			\end{subfigure}
%			\begin{subfigure}[t]{0.3\paperwidth}
%				\centering
%				\includegraphics[width=0.95 \textwidth]{plots/nonplasma//haywardWH_stability.jpg}
%				\caption{Plot showing the instability of photon sphere}
%			\end{subfigure}
		}
	\caption{(a)The plot shows the single peak effective potential for a  few representative values of $\kappa$ for the Hayward wormhole. This feature holds for the entire parameter range of $\kappa$. (b) The plot shows the variation of the shadow radius (magenta curve) as a function of $\kappa$ while the brown curve indicates the stability of the photon spheres denoted by $\dddot{x}$, the positive value of which, for the entire parameter range, indicates the photon spheres to be unstable.}
	\end{figure}
The metric functions are of the form $f_{1}(x)=1$ and $f(x)=1-\frac{2 x^{2}}{x^{3}+2 \kappa^{2}}$ for the Hayward wormhole spacetime. The only possible solution to eq.\eqref{eq:V} and eq.\eqref{eq:V2}, for this wormhole, is $ f(x)=0 $ and $ b^2 = x_{\rm ph}^{2}$. Hence for this case, the circular photon orbits are located at the throat, which is defined as the largest root to the cubic polynomial equation  $f(x)=0$ i.e. $x^{3}+2\kappa^{2}-2x^{2}=0$. We plot the effective potential as a function of the tortoise coordinates shown in Fig.(\ref{fig:HWH_pot}), which is characterized by a single peak for all values of $\kappa$. To understand the stability of photon orbits, we compute  $\dddot{x}$ numerically and conclude that these orbits are unstable for the complete range of  $ \kappa $ as shown by the brown curve in Fig.(\ref{fig:HWH_shadow}). The instability of the photon spheres is evident from the effective potentials as well since for the complete range of $\kappa$ have a single peak around the throat. Finally, we observe that the shadow radius reduces monotonically with increasing $\kappa$ as shown by the magenta curve in Fig.(\ref{fig:HWH_shadow}).
  \begin{figure}[h]
	\makebox[0.95\paperwidth][c]{
		\hspace*{-3.5cm}
		\begin{subfigure}[t]{0.4\paperwidth}
			\includegraphics[width=0.75\textwidth]{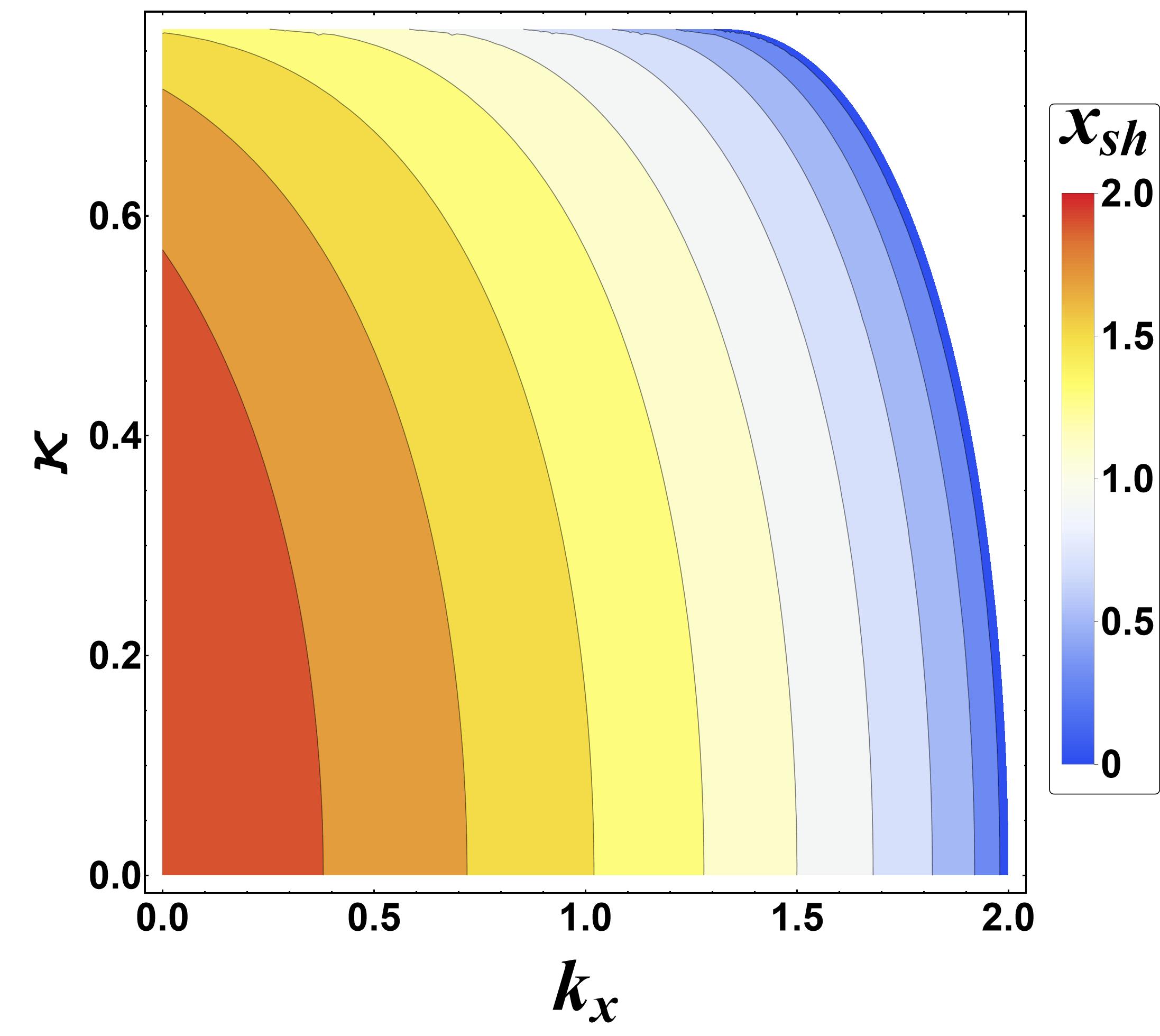}
			\caption{}%hayward Wormhole shadow plot }
             \label{fig:HWH_nonhomo_shadow}
		\end{subfigure}
		\begin{subfigure}[t]{0.4\paperwidth}
			\centering
			\includegraphics[width=0.75\textwidth]{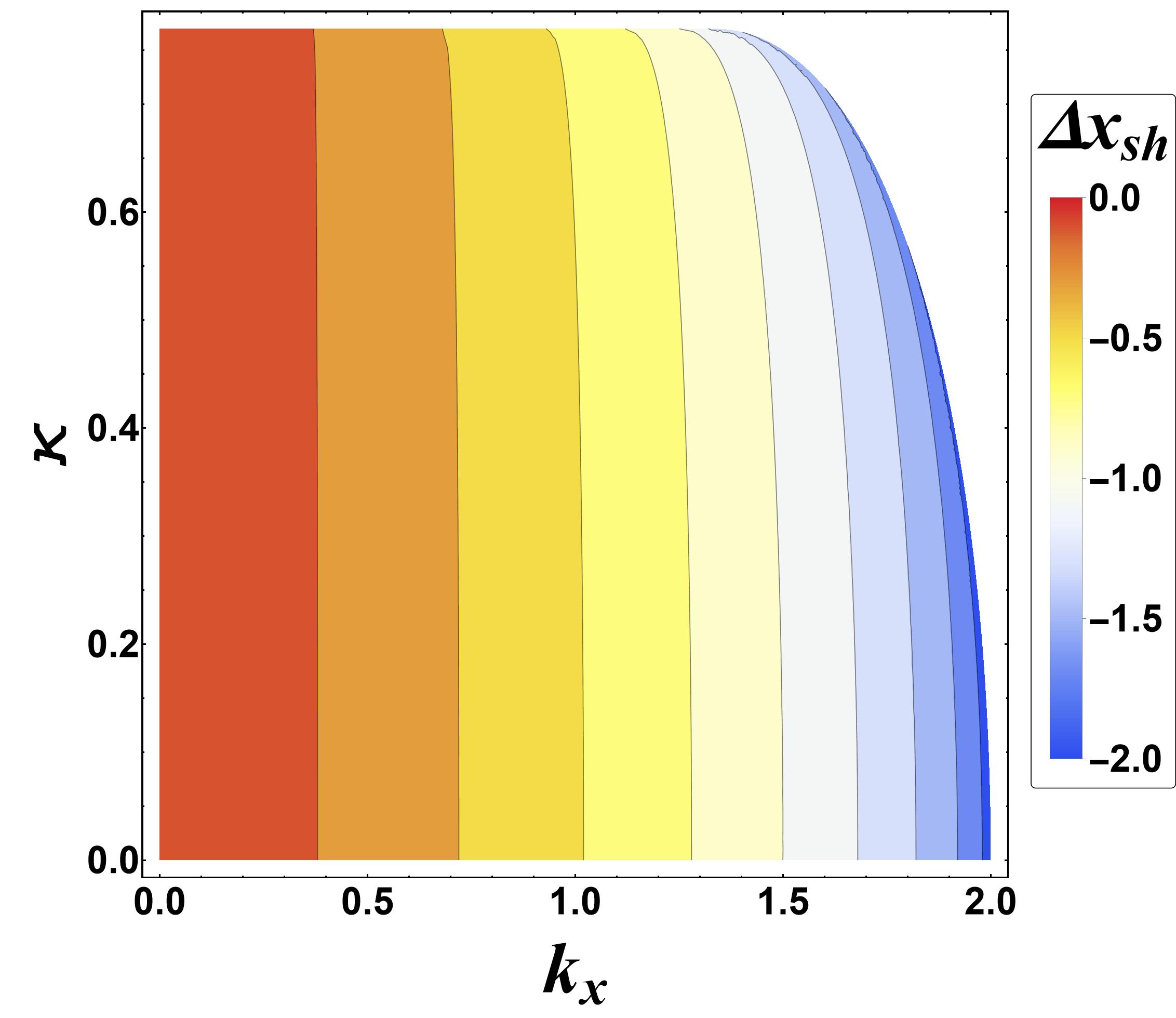}
			\caption{}%$\Delta x_{sh} = x_{sh} -x_{sh}\vert_{k_{r}=0}$}
   \label{fig:HWH_nonhomo_difference}
		\end{subfigure}
	}
 \caption{The plots show the effect of the non-homogeneous plasma profile given by $\Omega= \frac{k_x}{x}$, on the shadow of the Hayward wormhole. (a) The variation of $x_{\rm sh}$ is shown as a function of $\kappa$ and $k_x$ which reduces as $k_x$ increases for a particular $\kappa$. (b) The difference in shadow radius $\Delta x_{\rm sh} = x_{\rm sh} - x_{\rm sh}\vert_{k_x =0}$, with and without plasma, is shown as function of $\kappa$ and $k_x$. }
\end{figure}

For the case of homogeneous plasma, it can be checked from eq.\eqref{eq:homogeneousplasmashadow} that the shadow radius is independent of the plasma parameter and is identical to the vacuum case. However, for the case of non-homogeneous plasma, the shadow radius decreases as the plasma parameter $k_x$ increases as shown in Fig.(\ref{fig:HWH_nonhomo_shadow}). We also show the variation of $\Delta x_{\rm sh} = x_{\rm sh} - x_{\rm sh}\vert_{k_x =0}$ in Fig.(\ref{fig:HWH_nonhomo_difference}) with varying $\kappa$ and $k_x$ that highlights the deviation of shadow radius from the vacuum case in presence of non-homogeneous plasma.
\begin{figure}[h!]
		\makebox[0.95\paperwidth][c]{
			\hspace*{-3.5cm}
			\begin{subfigure}[t]{0.4\paperwidth}
				\centering
				\includegraphics[width=0.95 \textwidth]{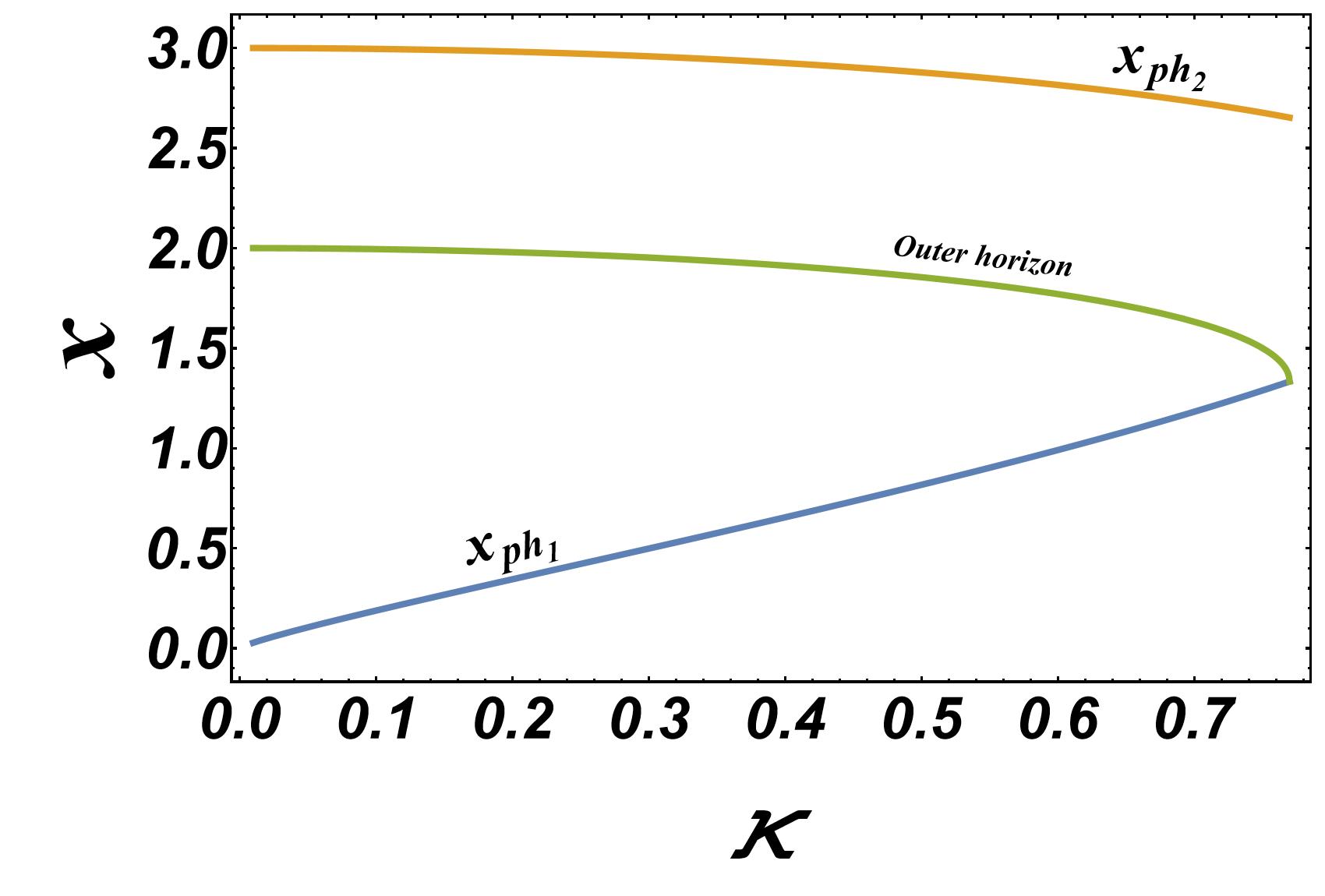}
				\caption{}
    \label{fig:HBH_roots}
			\end{subfigure}
			\begin{subfigure}[t]{0.5\paperwidth}
				\centering
				\includegraphics[width=0.95 \textwidth]{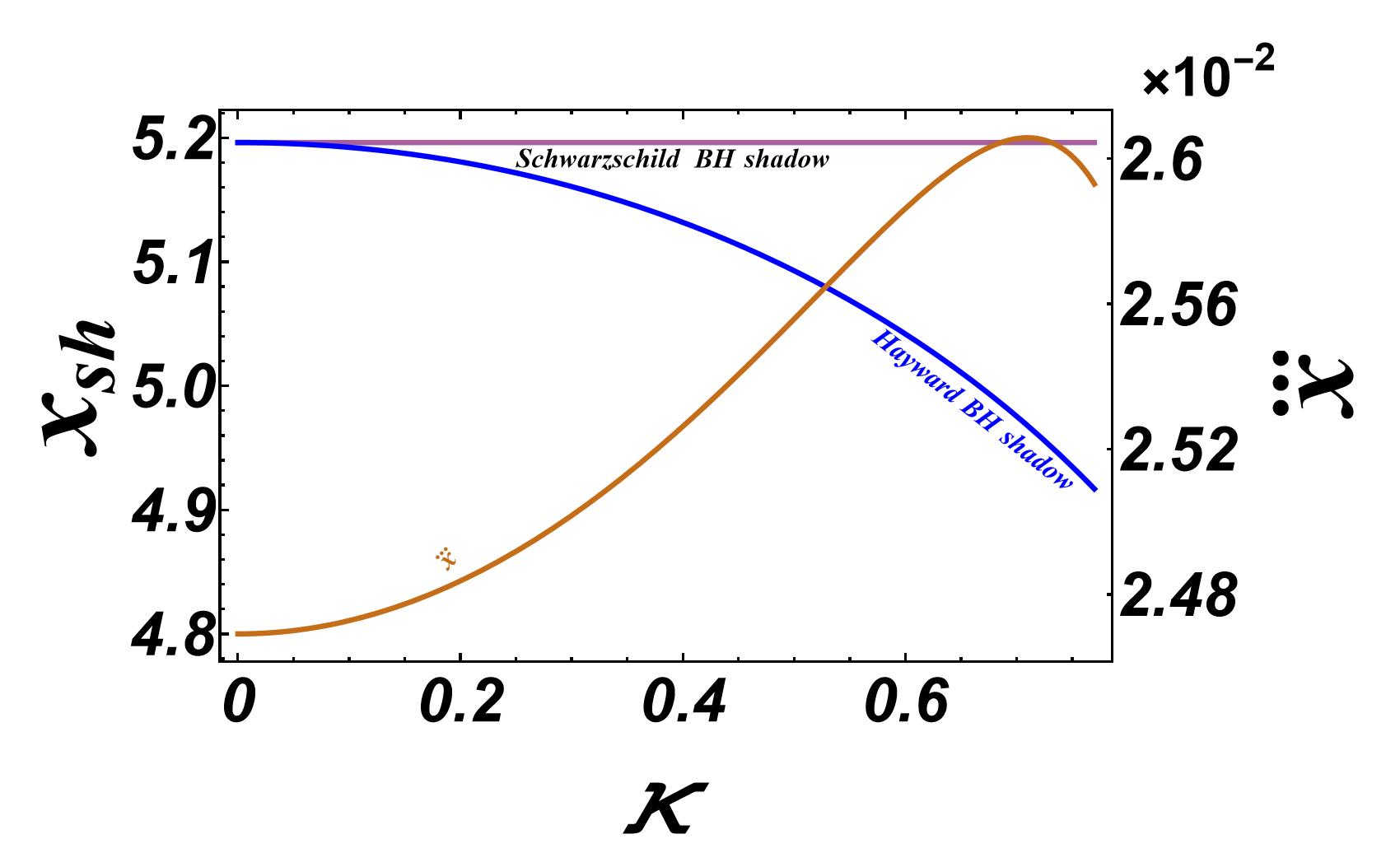}
				\caption{}
    \label{fig:HBH_shadow}
			\end{subfigure}
		}
   \caption{(a) The plot shows the relative location of the two positive photon spheres with respect to the outer horizon. Only the larger of the two photon spheres lies outside the horizon and, hence, the only physically relevant solution for generating a shadow. (b)The plot shows the shadow radius of the Hayward BH in blue along with the Schwarzschild BH shadow radius in violet for comparison. The brown curve represents $\dddot{x}$, which is positive for the complete range of parameters, indicating that the photon sphere is indeed unstable.}
	\end{figure}

%\FloatBarrier

\subsection{ Regular Hayward black hole: $ \sigma =1,\; 0<\kappa\leq\frac{4}{3\sqrt{3}}$}

Setting $ \sigma =1,\; 0<\kappa\leq\frac{4}{3\sqrt{3}}$ reduces the metric in eq.\eqref{eq:dimensionless_metric} to that of the regular Hayward BH case which was the original metric to be deformed to give rise to the generalized metric.  For this spacetime, all the curvature scalars remain finite for the complete range of the manifold and hence it is considered as a regular BH solution.   The metric functions reduce to	$f_{1}(x)=f(x)= 1- \frac{2 x^{2}}{x^{3}+ 2\kappa^{2}}$ and is characterized with two horizons corresponding to the roots of $g_{tt}=0$. Solving eq.\eqref{eq:V} for the photon sphere radius results in a sixth-order polynomial equation, which has been solved numerically. The polynomial equation has only two positive roots ($x_{\rm ph1}, x_{\rm ph2}$), of which, we are interested in the one outside the outer horizon. Fig.(\ref{fig:HBH_roots}) shows the position of ($x_{\rm ph1}, x_{\rm ph2}$) with respect to the outer horizon. The larger of the two photon spheres, i.e. $x_{\rm ph2}$, denoted by the brown curve in Fig.(\ref{fig:HBH_roots}), lies outside the outer horizon and hence gives rise to the shadow. We then calculate the $\dddot{x}$ for the photon orbit corresponding to $x_{\rm ph2}$ shown in brown curve in Fig.(\ref{fig:HBH_shadow}) and conclude that they are all unstable. The corresponding shadow radius is shown as the blue curve in  Fig.(\ref{fig:HBH_shadow}) where the Schwarzschild shadow radius (magenta curve) is also shown for comparison.

\begin{figure}[h]
	\makebox[0.95\paperwidth][c]{
		\hspace*{-3.5cm}
		\begin{subfigure}[t]{0.4\paperwidth}
			\includegraphics[width=0.75\textwidth]{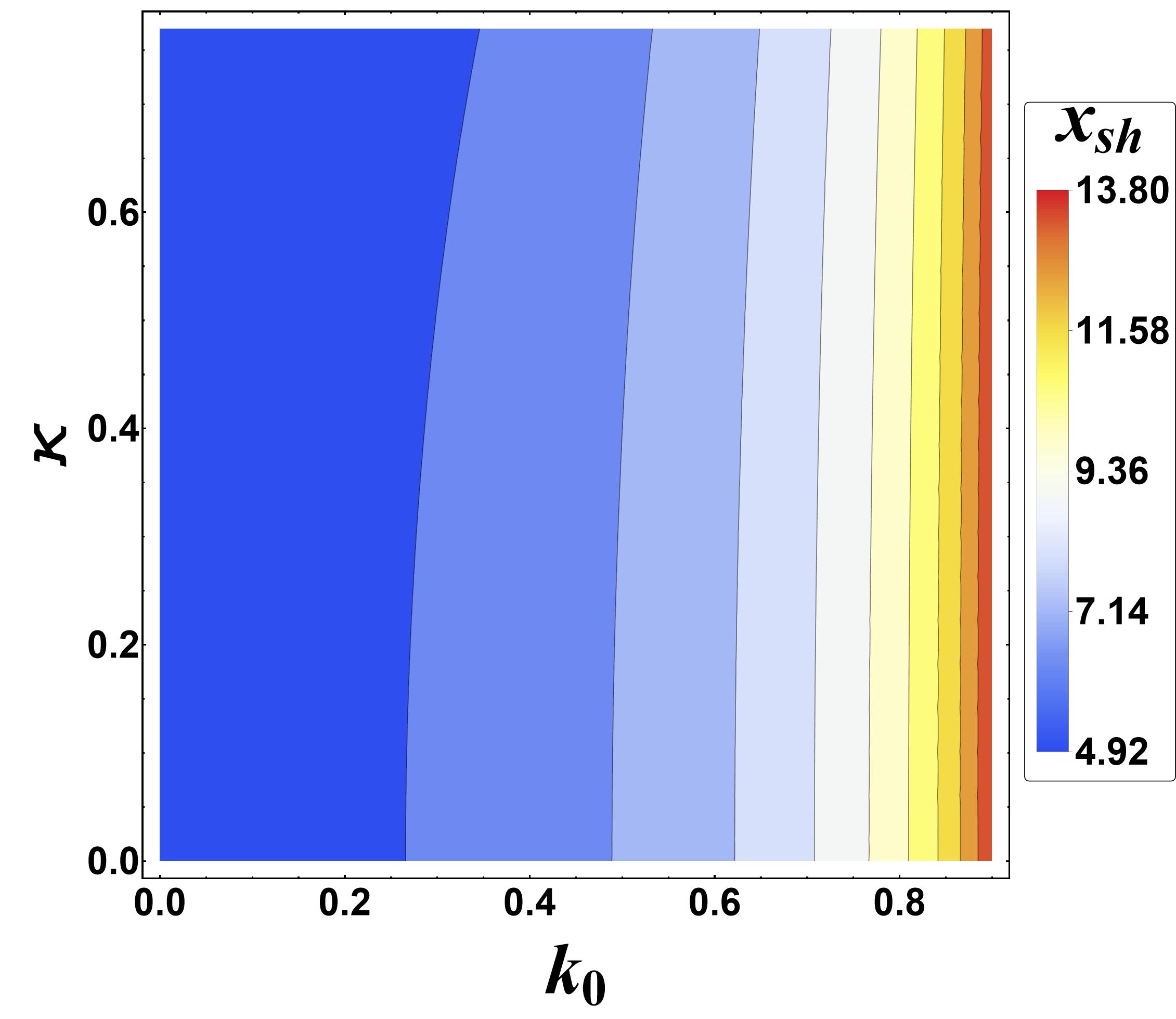}
			\caption{}%Hayward BH shadow }
   \label{fig:HBH_homo_shadow}
		\end{subfigure}
		\begin{subfigure}[t]{0.4\paperwidth}
			\centering
			\includegraphics[width=0.75\textwidth]{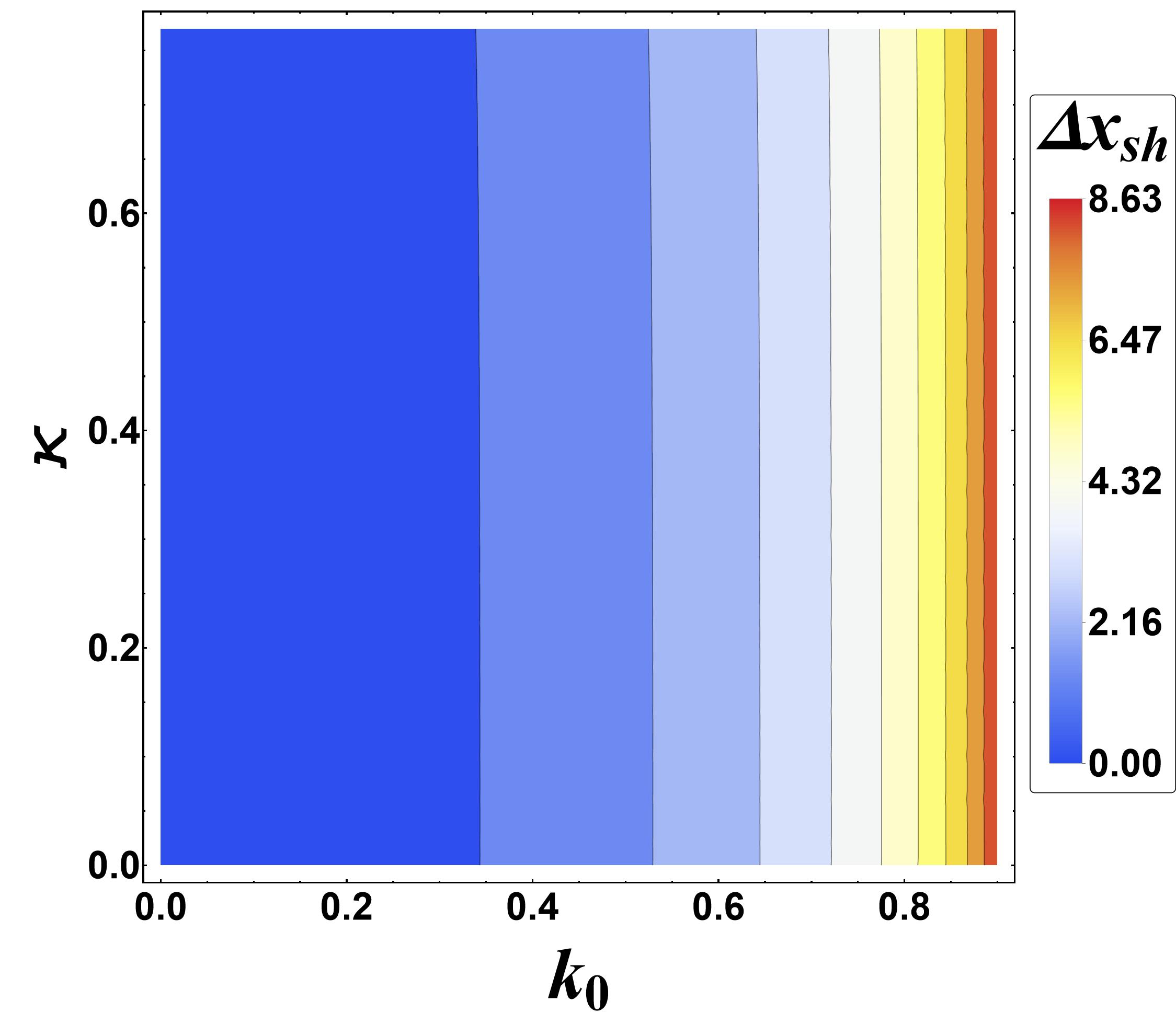}
			\caption{}%$\Delta x_{sh} = x_{sh} -x_{sh}\vert_{k_{0}=0}$}
   \label{fig:HBH_homo_diffrence}
		\end{subfigure}
	}
 \caption{The plots correspond to the shadow radius of the Hayward regular BH in presence of the homogeneous plasma profile given by $ \Omega=k_{0} $. (a) The plot shows the variation of shadow radius as a function of $\kappa$ and $k_0$. (b) The plot shows the difference in shadow radius with and without plasma i.e. $\Delta x_{\rm sh} = x_{\rm sh} - x_{\rm sh}\vert_{k_0=0}$, as a function of $\kappa$ and $k_0$.  }
\end{figure}
\begin{figure}[h]
	\makebox[0.95\paperwidth][c]{
		\hspace*{-3.5cm}
		\begin{subfigure}[t]{0.4\paperwidth}
			\includegraphics[width=0.75\textwidth]{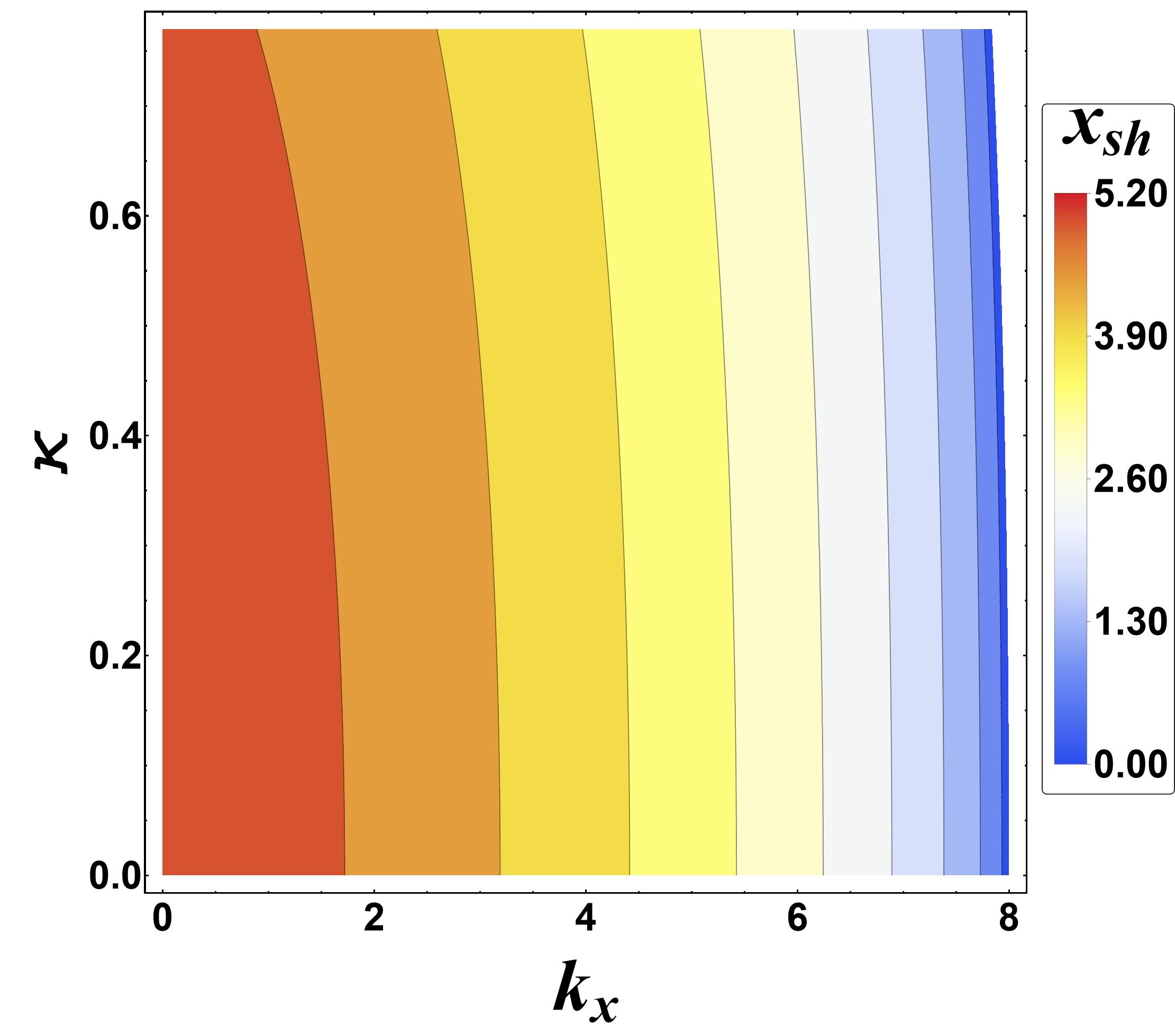}
			\caption{}%hayward BH shadow radius }
      \label{fig:HBH_nonhomo_shadow}
		\end{subfigure}
		\begin{subfigure}[t]{0.4\paperwidth}
			\centering
			\includegraphics[width=0.75\textwidth]{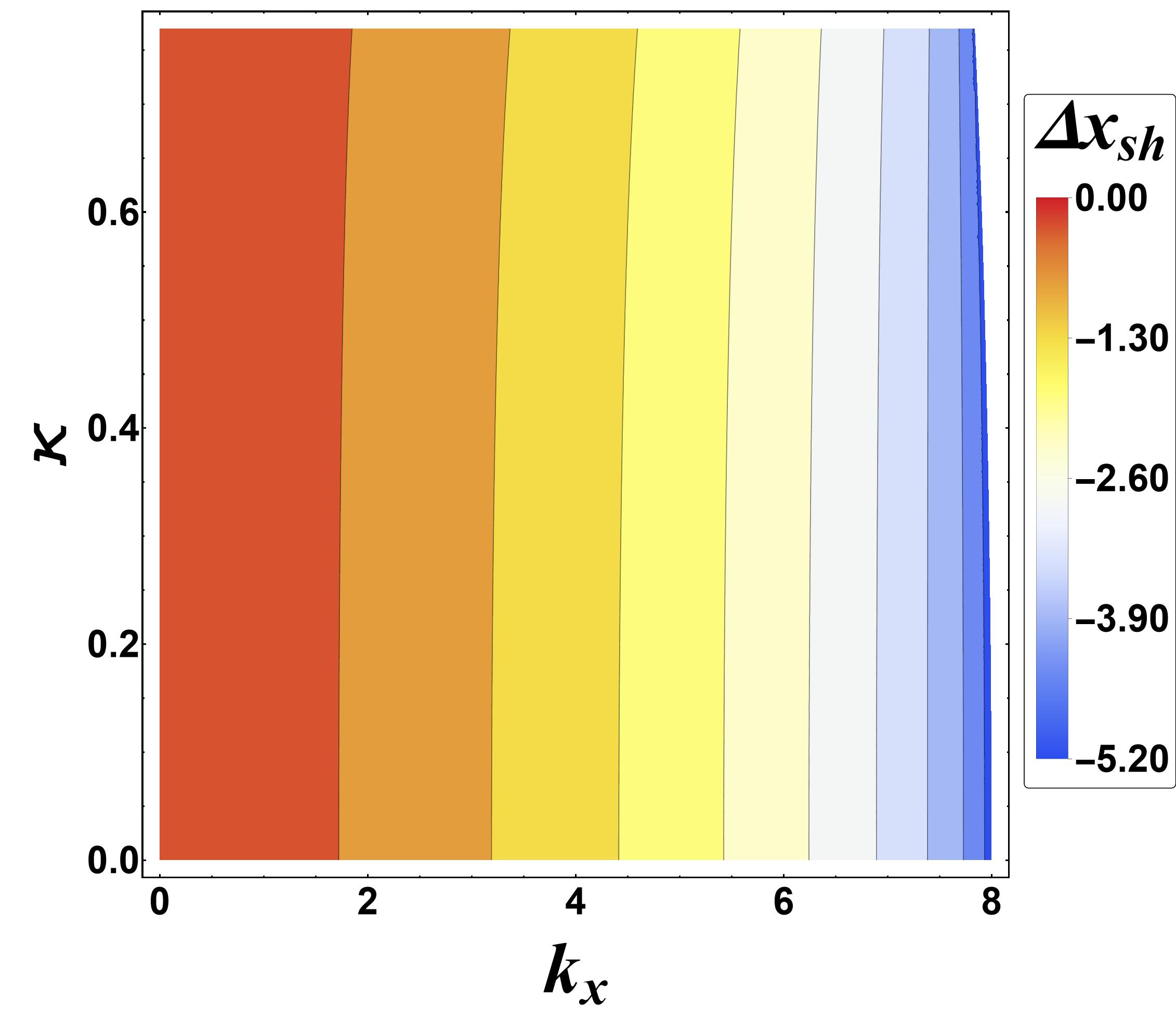}
			\caption{}%$ \Delta x_{sh} = x_{sh} -x_{sh}\vert_{k_{r}=0} $}
      \label{fig:HBH_nonhomo_diffrence}
		\end{subfigure}
	}
 \caption{The plots show the effect of non-homogeneous plasma profile given by $ \Omega=\frac{k_{x}}{x}$ on the shadow of the regular Hayward BH. (a) The plot shows the variation in shadow radius as a function of $\kappa$ and $k_x$. (b) The plot shows the difference in shadow radius in presence and absence of plasma, i.e. $\Delta x_{\rm sh} = x_{\rm sh} - x_{\rm sh}\vert_{k_x=0}$, as a function of $\kappa$ and $k_x$.  }
\end{figure}
The introduction of homogeneous plasma profile leads to increase in shadow radius with increasing $k_0$ as shown in Fig.(\ref{fig:HBH_homo_shadow}) while the reduction in the shadow radius is seen for the non-homogeneous plasma as the plasma parameter $k_x$ increases, shown in Fig.(\ref{fig:HBH_nonhomo_shadow}). The difference in the shadow radius in the presence and absence of plasma is shown as a function of metric and plasma parameters in Fig.(\ref{fig:HBH_homo_diffrence}) for the homogeneous plasma and in Fig.(\ref{fig:HBH_nonhomo_diffrence}) for the non-homogeneous case.

%\FloatBarrier
\subsection{Hayward-Damour-Solodukhin wormhole: $ 0<\sigma <1$, $0<\kappa\leq\frac{4}{3\sqrt{3}}$  }
\begin{figure}[h]
		\makebox[0.95\paperwidth][c]{
			\hspace*{-3.5cm}
			\begin{subfigure}[t]{0.4\paperwidth}
				\includegraphics[width=0.95\textwidth]{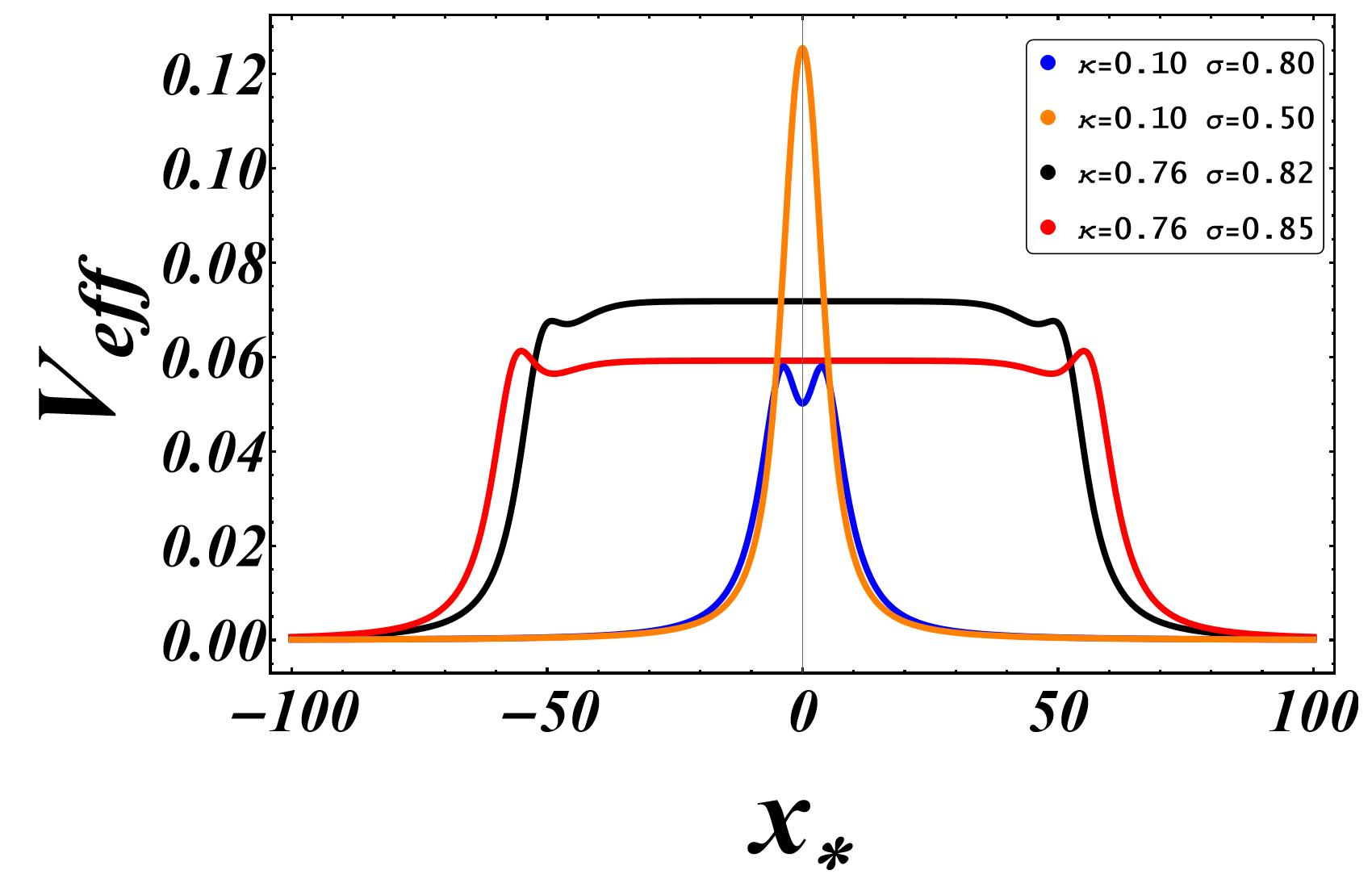}
					\caption{}%Potential Plots} 
     \label{fig:HDSWH_pot}
			\end{subfigure}
			\begin{subfigure}[t]{0.4\paperwidth}
				\centering
				\includegraphics[width=0.75\textwidth]{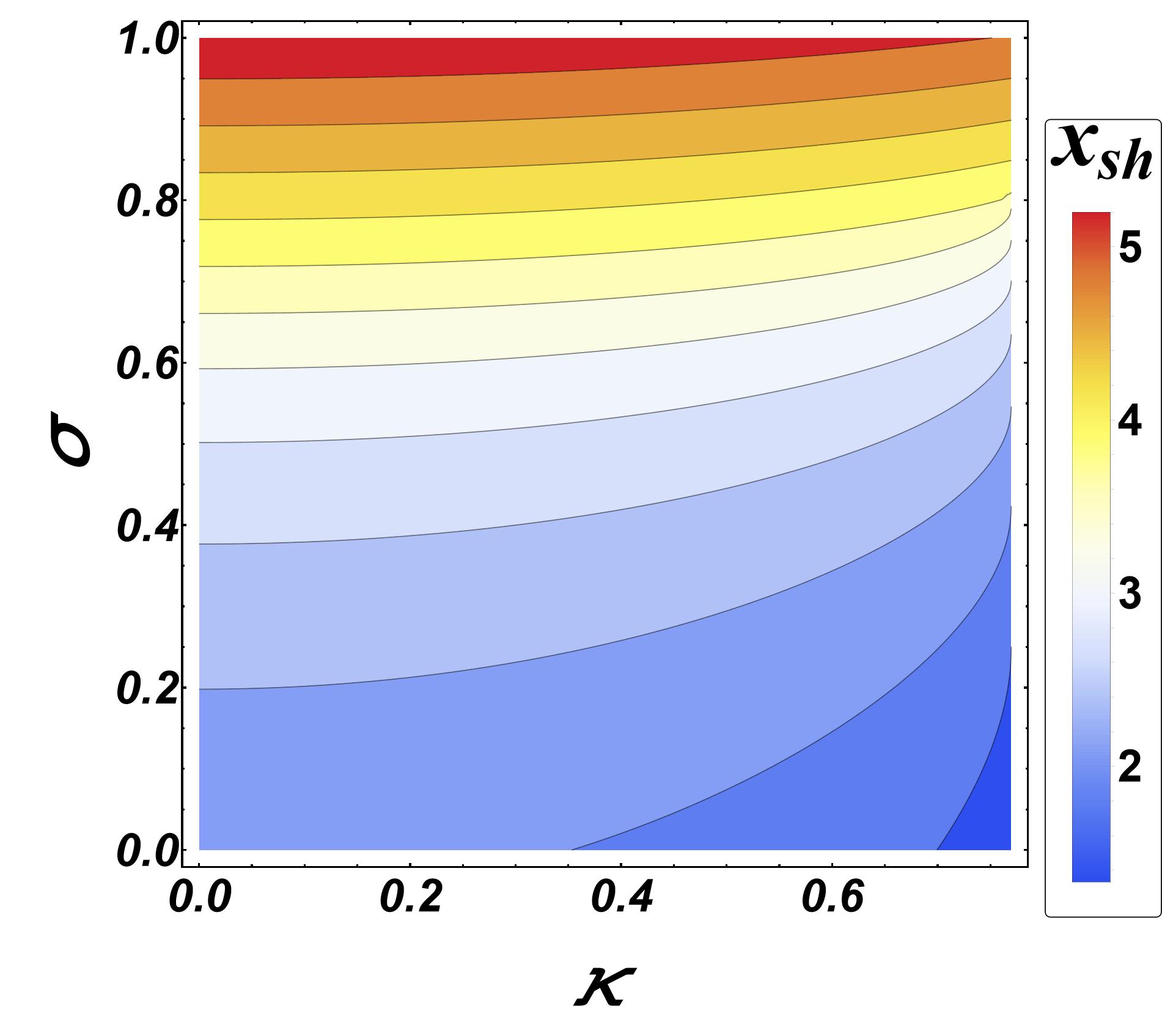}
			\caption{}%$ Shadow Plot  $}
   \label{fig:HDSWH_shadow}
			\end{subfigure}
		}
\caption{(a) The plot shows the effective potentials for different choices of ($\sigma,\kappa$) for the Hayward-Damour-Solodukhin wormhole. The orange curve represents a single barrier, the blue curve represents a double barrier, and the red and black ones represent triple barrier potentials. (b) The plot shows the shadow radius for the complete range of metric parameters.}
	\end{figure}

This wormhole spacetime is the most general case spanning the entire range of ($\sigma,\kappa$). The metric functions for this class are given by $ f_{1}(x)= 1-\frac{2 \sigma x^{2}}{x^{3}+ 2\sigma \kappa^{2}}$ and  $f(x)= 1-\frac{2 x^{2}}{x^{3}+ 2\kappa^{2}} $. There are two branches of solutions for eq.\eqref{eq:V} given by $f(x)=0$ and $-f_{1}'(x)+ \frac{2f_{1}(x)}{x}=0$. These solutions determine the radius of the circular photon orbits and the impact parameter is given by eq.\eqref{eq:V2} which reduces to $b^2 = \frac{x_{\rm ph}^{2}}{f_{1}(x_{\rm ph})}$ for this case.
\begin{figure}[h]
 \centering
  \includegraphics[width=0.7\textwidth]{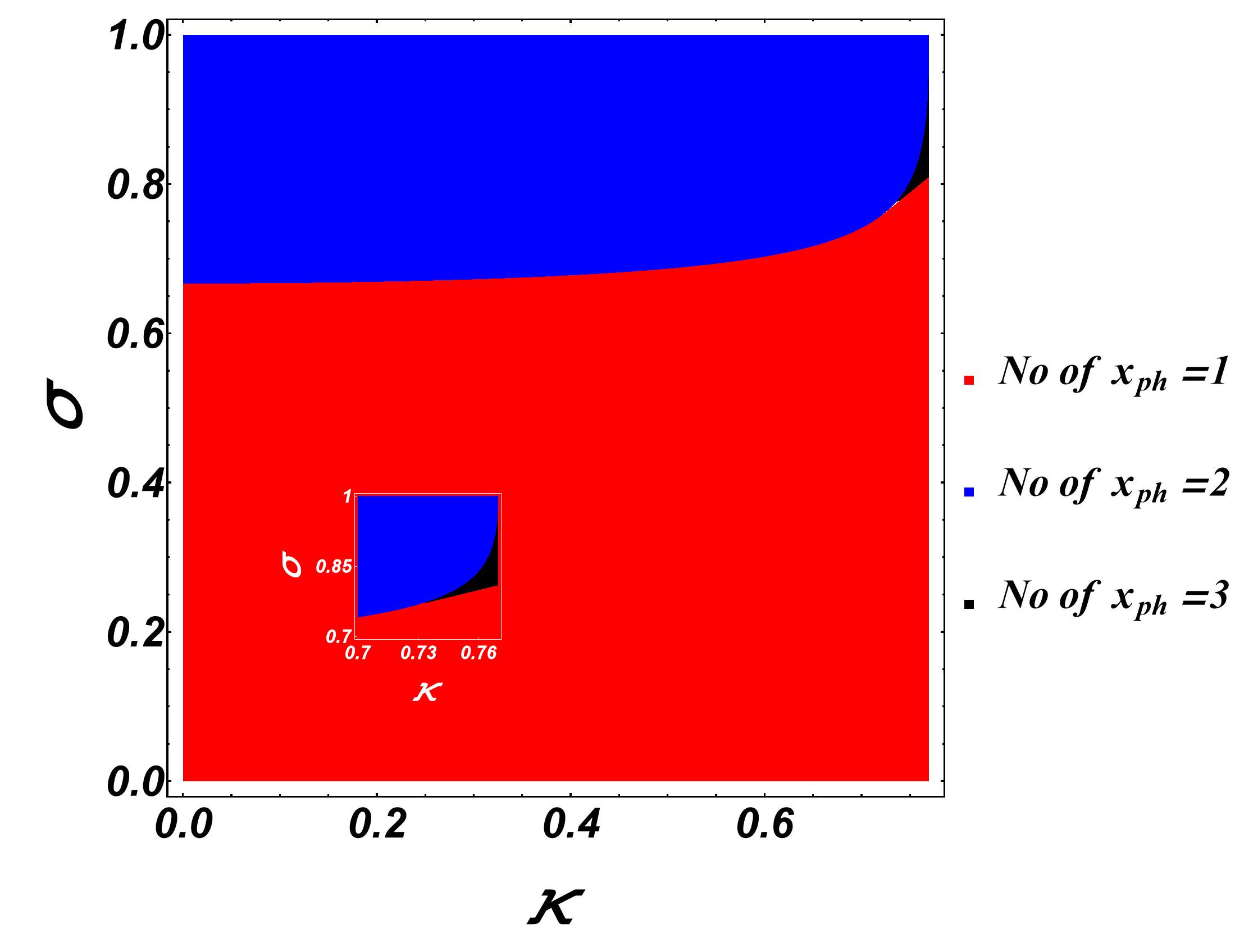}
  \caption{The plot shows the number of photon spheres occurring for different choices of $(\sigma,\kappa)$ in case of the  Hayward-Damour-Solodukhin wormhole. In case of multiple photon spheres, only the largest one is considered to be casting the observed shadow. The points $\sigma=\kappa=0$ and $\sigma=1$ are not part of the spacetime and the plot starts just after these values. }
  \label{fig:HDSWH_parameter}
\end{figure}
These equations have been solved numerically, and we observe that in this wormhole geometry, there are either one, two, or three photon spheres depending on the value of metric parameters. We have plotted the number of photon orbits as a function of parameters in Fig.(\ref{fig:HDSWH_parameter}). Note that for certain choices of parameters there are no photon spheres denoted by the black region of Fig.(\ref{fig:HDSWH_parameter}). In the case of single-peak potential, the photon sphere is located near the throat, and for the double-peak potential, the photon orbits at the throat is an anti-photon sphere. For the triple peak potential, there are two photon spheres out of which one is located at the throat, and the two photon spheres are separated by an anti-photon sphere. To get a better visual representation of these scenarios, we have plotted the effective potential for specific parameters to cover all the possible cases in Fig.(\ref{fig:HDSWH_pot}). We compute the shadow radius corresponding to the largest of the photon spheres in case there are more than one photon sphere. The shadow radius for the complete parameter range of $\sigma$ and $\kappa$ is plotted in Fig.(\ref{fig:HDSWH_shadow}). Larger shadow radius is observed for higher $\sigma$ and for a particular choice of $\kappa$.
 
\begin{figure}[h]
		\makebox[0.95\paperwidth][c]{
			\hspace*{-3.5cm}
			\begin{subfigure}[t]{0.4\paperwidth}
				\includegraphics[width=0.95\textwidth]{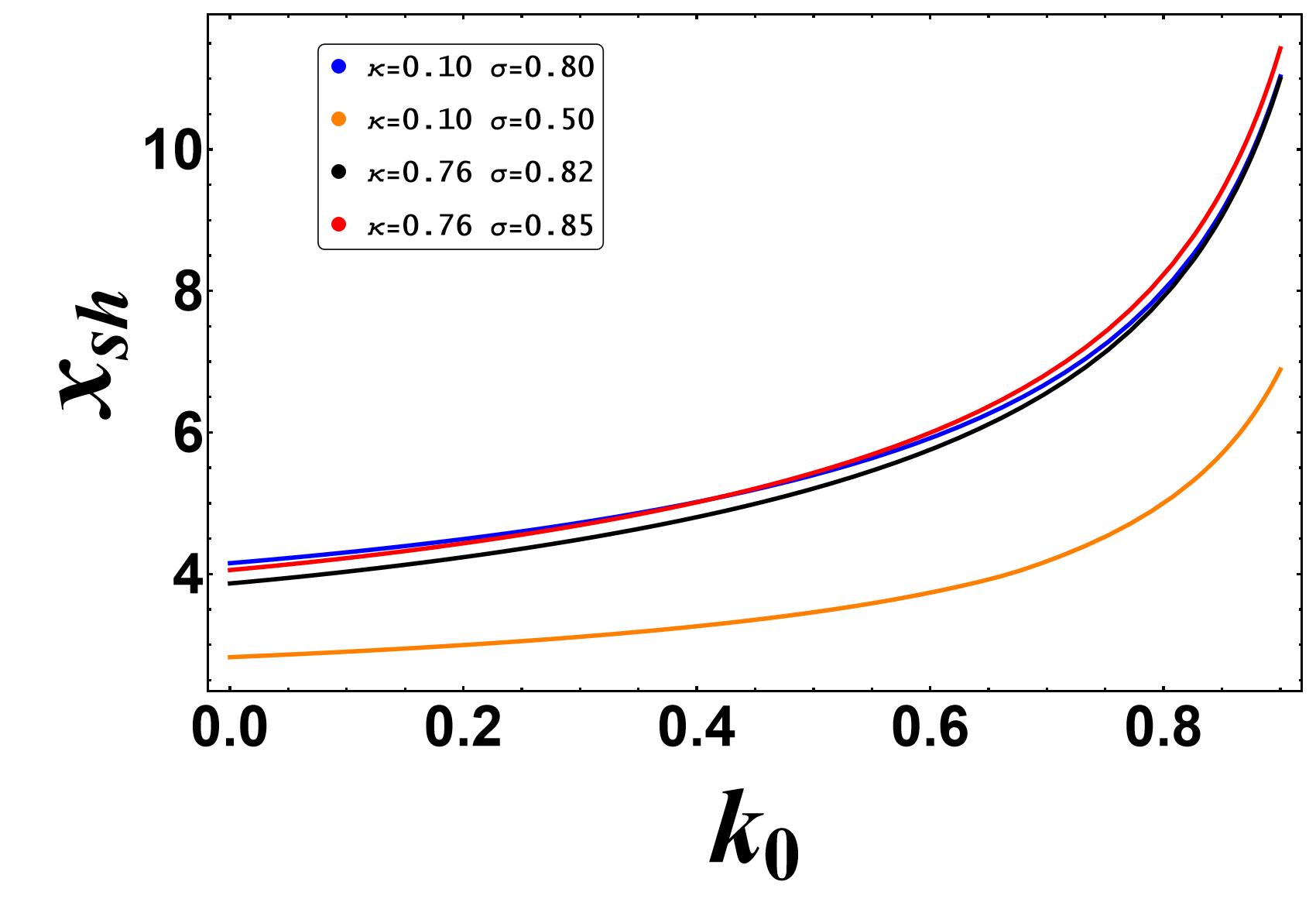}
				\caption{}%Hayward Damour Solodukin Wormhole shadow radius }
    \label{fig:HDSWH_homo_shadow}
			\end{subfigure}
			\begin{subfigure}[t]{0.4\paperwidth}
				\centering
				\includegraphics[width=0.95\textwidth]{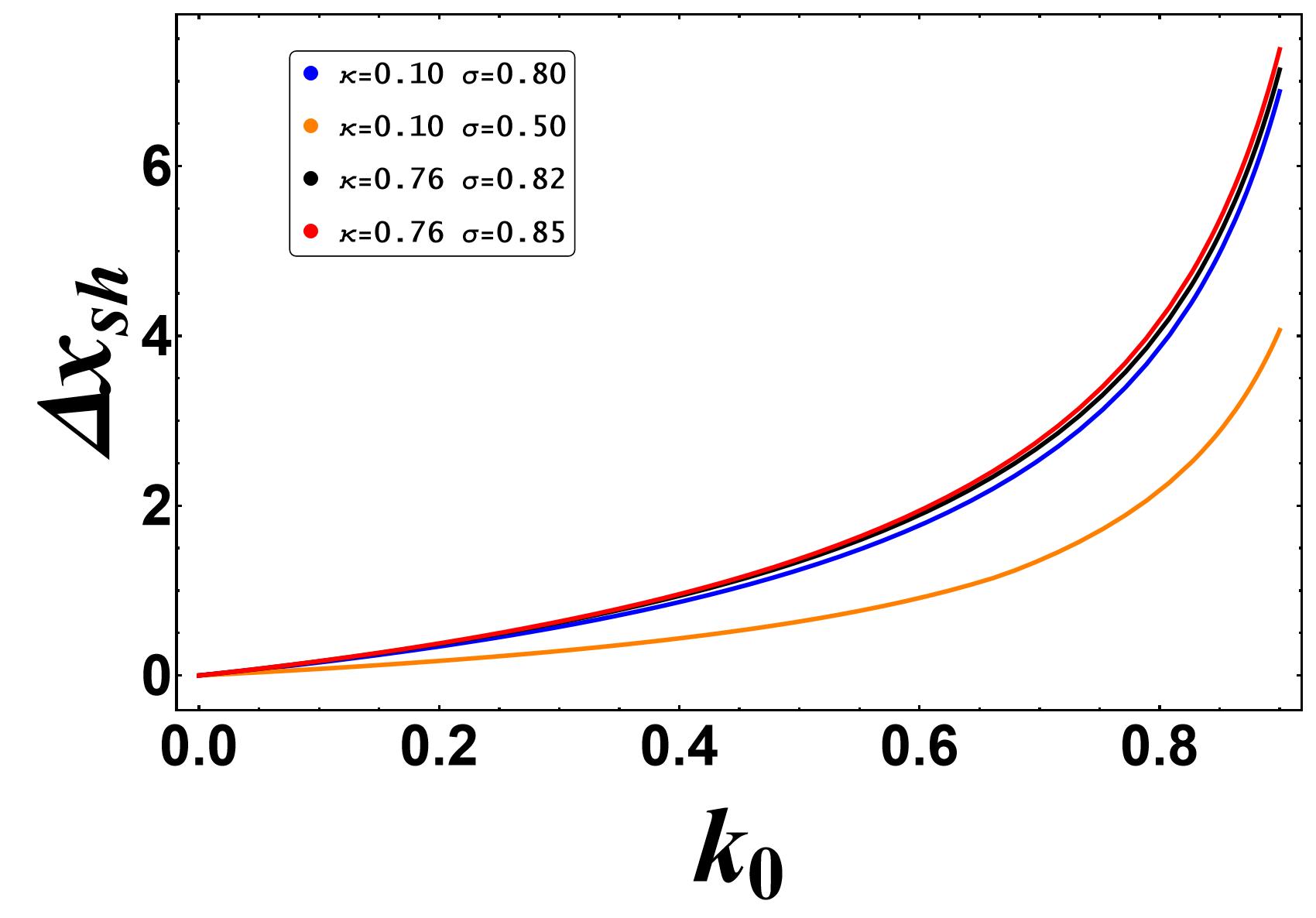}
				\caption{}%$\Delta x_{sh} = x_{sh} -x_{sh}\vert_{k_{0}=0}$}
    \label{fig:HDSWH_homo_diff}
			\end{subfigure}
		}
\caption{(a) The plot shows the variation of shadow radius for certain representative choices of $(\sigma,\kappa)$ in the presence of homogeneous plasma profile of the form $\Omega=k_{0} $ for the Hayward-Damour-Solodukhin wormhole. The effective potential for these same set of $(\sigma,\kappa)$ has been shown in Fig.\ref{fig:HDSWH_pot}. (b) The plot shows the difference in shadow radius with and without plasma i.e. $\Delta x_{\rm sh} = x_{\rm sh} - x_{\rm sh}\vert_{k_0=0}$.} 
\label{fig:HDSWH_homo}
	\end{figure}

\begin{figure}[h]
		\makebox[0.95\paperwidth][c]{
			\hspace*{-3.5cm}
			\begin{subfigure}[t]{0.4\paperwidth}
				\includegraphics[width=0.95\textwidth]{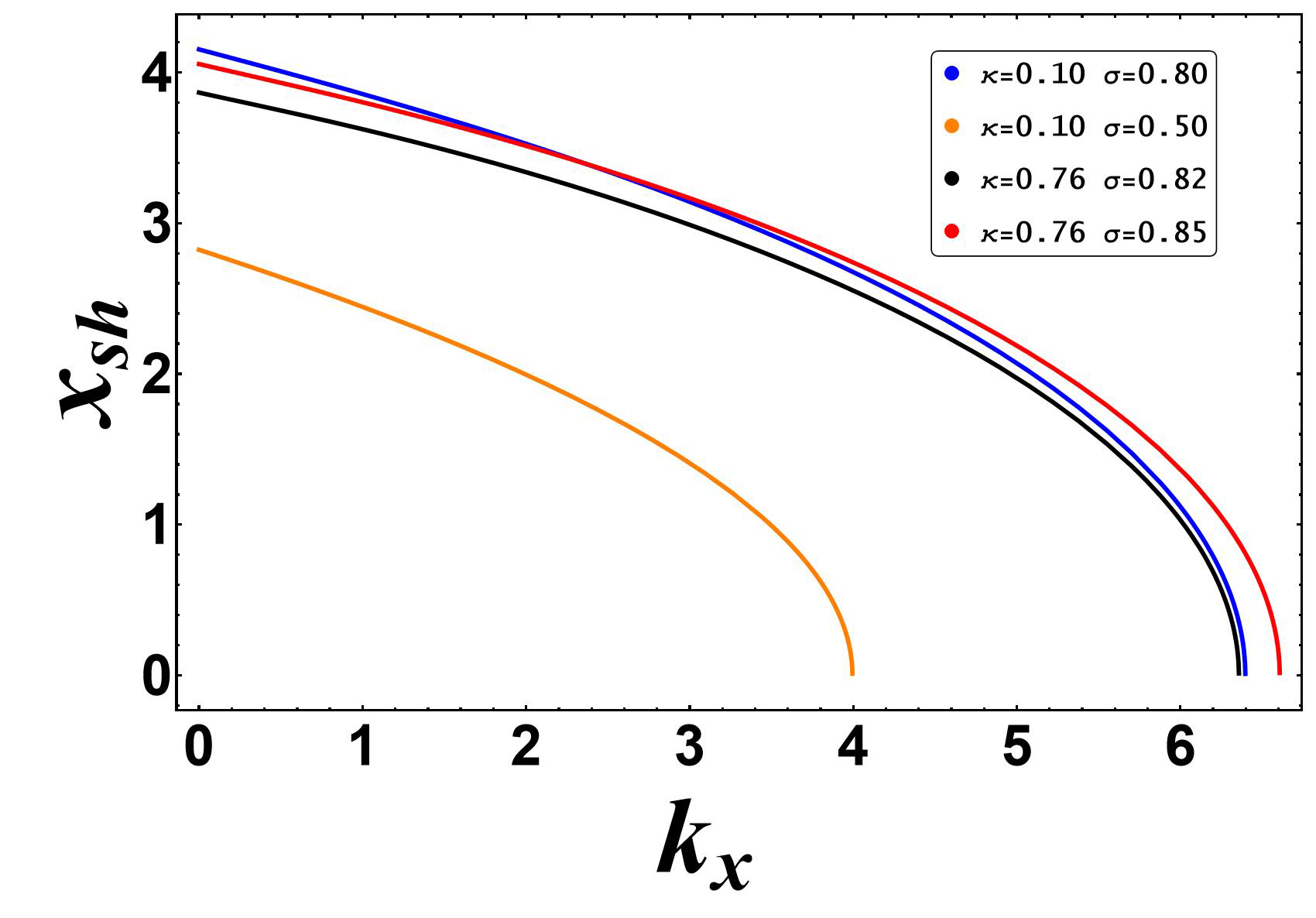}
				\caption{}%Hayward Damour Solodukin Wormhole shadow radius}
    \label{fig:HDSWH_nonhomo_shadow}
			\end{subfigure}
			\begin{subfigure}[t]{0.4\paperwidth}
				\centering
				\includegraphics[width=0.95\textwidth]{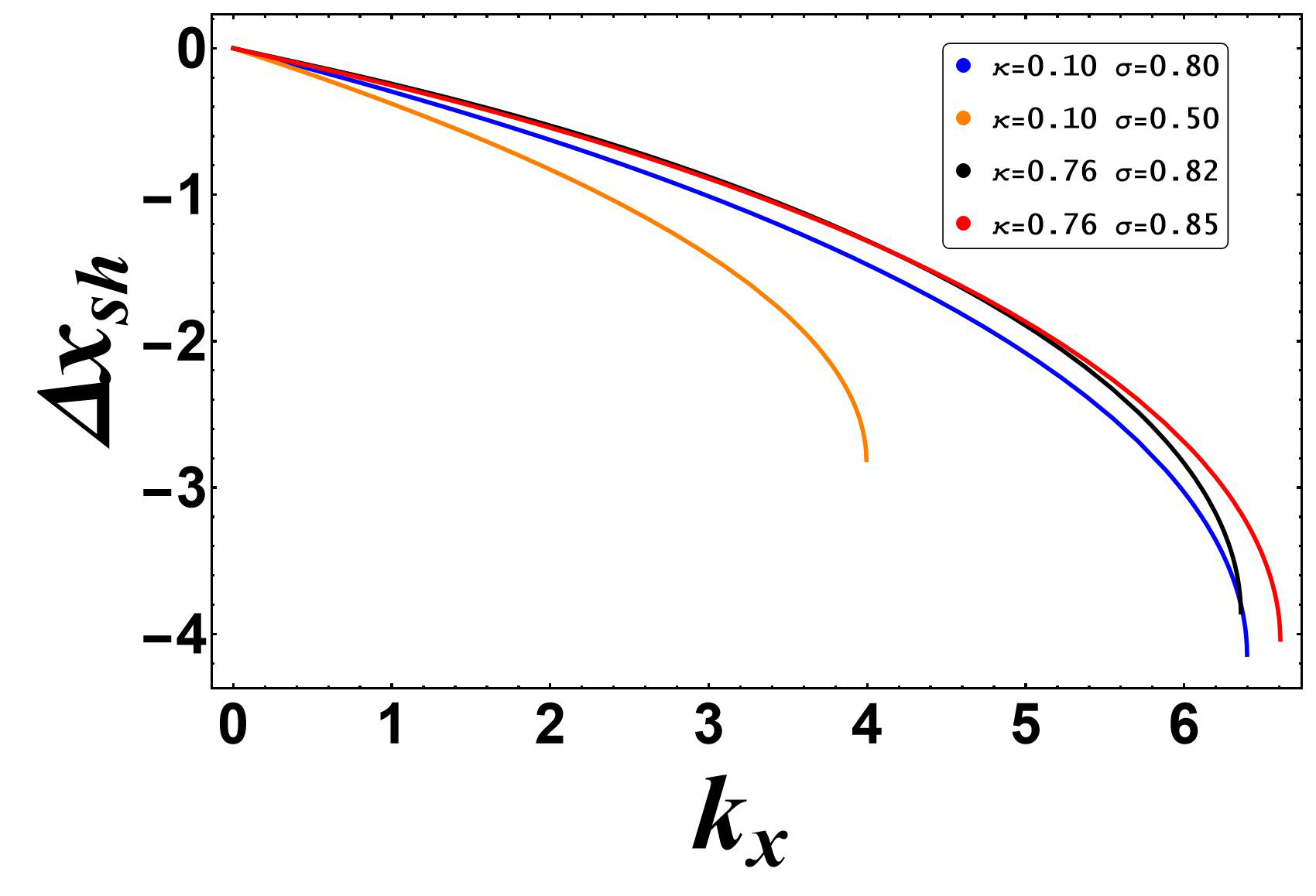}
				\caption{}%$ \Delta x_{sh} = x_{sh} -x_{sh}\vert_{k_{r}=0} $}
    \label{fig:HDSWH_nonhomo_diff}
			\end{subfigure}
		}
  \caption{The plot shows the variation of shadow radius for certain representative choices of $(\sigma,\kappa)$ in the presence of non-homogeneous plasma profile of the form $\Omega=\frac{k_{x}}{x} $ for the Hayward-Damour-Solodukhin wormhole. The effective potential for these same set of $(\sigma,\kappa)$ has been shown in Fig.\ref{fig:HDSWH_pot}. (b) The plot shows the difference in shadow radius with and without plasma i.e. $\Delta x_{\rm sh} = x_{\rm sh} - x_{\rm sh}\vert_{k_x=0}$.}
  \label{fig:HDSWH_nonhomo}
	\end{figure}

Finally, we move on to study the effect of plasma on the shadow of this wormhole metric. We plot the shadow radius and their difference with and without plasma for both homogeneous and non-homogeneous profiles in Fig.(\ref{fig:HDSWH_homo}) and Fig.(\ref{fig:HDSWH_nonhomo}) respectively. Since the variation of multiple parameters and their effect on shadow cannot be studied simultaneously, we choose certain representative parameter sets and study the effect of plasma on their shadows. For the homogeneous plasma profile, we observe that the shadow radius increases as the plasma parameter $k_0$ increases as shown in Fig.(\ref{fig:HDSWH_homo_shadow}). However, for the case of non-homogeneous plasma, the shadow radius decreases as the plasma parameter $k_x$ reduces as shown in Fig.(\ref{fig:HDSWH_nonhomo_shadow}). To complete the analysis we also plot the difference in the shadow radius with and without plasma for both homogeneous and non-homogeneous profiles as shown in Fig.(\ref{fig:HDSWH_homo_diff}) and Fig.(\ref{fig:HDSWH_nonhomo_diff}) respectively.
	
    \begin{table}[h]
		\centering
		\begin{tabular}{|c|c|c|c|c|}
			\hline
			\textbf{Spacetime} & \multicolumn{2}{|c|}{Vacuum} & Homogeneous plasma & Non-homogeneous plasma\\ 
   \hline
    & ($\sigma$,\,$x_{\rm sh}$) & ($\kappa$,\,$x_{\rm sh}$) &($k_0$,\,$x_{\rm sh}$) & ($k_x$,\,$x_{\rm sh}$)\\
             \hline 
             \hline
            Schwarzschild BH & $(1, -)$ & $(0, -)$ &  ($\uparrow$, $\uparrow$) & $(\uparrow$,$\downarrow$)\\
            \hline
           Schwarzschild WH & $(0,-)$ & $(0,-)$ & $(k_0,-)$ & ($\uparrow$,$\downarrow$) \\
           \hline
          DS WH & $(\uparrow, \uparrow)$ & $(0,-)$ &  ($\uparrow, \uparrow$) & ($\uparrow,\downarrow$) \\
          \hline
          Hayward WH & $(0,-)$ & $(\uparrow, \downarrow)$ & $(k_0, -)$ & $(\uparrow,\downarrow)$\\
          \hline
          Hayward BH & $(1,-)$ & $(\uparrow,\downarrow)$ & $(\uparrow, \uparrow)$ & $(\uparrow,\downarrow)$\\
			\hline
        Case VI WH & $(\uparrow, \uparrow)$ & $(\uparrow,\downarrow)$ &  $(\uparrow,\uparrow)$ & $(\uparrow,\downarrow)$ \\
			\hline
		\end{tabular}
		\caption{Table shows the summary of the results obtained in our study. For each spacetime, the dependence of the shadow radius ($x_{\rm sh}$) on the metric and plasma parameters is shown for vacuum and in presence of homogeneous and non-homogeneous plasma profiles. When studying the dependence of $x_{\rm sh}$ on a particular parameter, the other parameters are kept fixed. The $(\uparrow,\downarrow)$ indicates increasing and decreasing values respectively. In some cases, the $x_{\rm sh}$ is constant and does not depend on the parameters and are denoted by `$-$'.}
		\label{tab:summary}
	\end{table}

\section{Constraints from Sgr $\rm A^{*}$}\label{sec:4}

In the previous section, we conducted a general analysis of the shadows cast by the new spacetime solutions as seen by an asymptotic observer. In this section, we contextualize these results using observational data from EHT to determine if any of these spacetimes are viable candidates to represent an astrophysical compact object. Specifically, we consider the most recent EHT estimate for the angular diameter ($\Phi$) of the shadow of Sgr $\rm A^*$, given by $\Phi = 48.7 \pm 7.0\ \mu as$  \cite{EventHorizonTelescope:2022wkp} and study if any of these spacetimes can represent this object. This analysis incorporates priors on the distance to Sgr $\rm A^*$  and estimated BH mass which are obtained from: Keck and the Very Large Telescope Interferometer (VLTI) \cite{GRAVITY-1,Gravity-2}. The Keck team reported a distance of $r_o = 7959 \pm 59 \pm 32\; pc$  and a BH mass of $M = (3.975 \pm 0.058 \pm 0.026) \times 10^6 M_{\odot}$ \cite{Keck:2019txf}. In contrast, VLTI, in collaboration with GRAVITY estimates the distance to be $r_o = 8277 \pm 9 \pm 33\;  pc$ and mass $M = (4.297 \pm 0.012 \pm 0.040) \times 10^6 M_{\odot}$ \cite{GRAVITY-1, Gravity-2}.

In order to relate these results to our framework, we need to convert the angular shadow diameter to shadow radius. This is achieved by first relating the angular diameter of the shadow with shadow radius which is given by
\begin{equation}
	\frac{r_{\text{sh}}}{r_o} = \tan\left( \frac{1}{2} \Phi \right).
\end{equation}

Note that the distance $r_o$ used above is obtained from the priors considered for generating the angular diameter of the shadow. Also, the shadow radius obtained above is not dimensionless, it can be converted to a  dimensionless quantity by  normalizing it with appropriate factors of BH mass (from the mass prior), speed of light and the gravitational constant. Carrying out this computations using the VLTI and Gravity collaboration data we obtain the following constraint on dimensionless shadow radius
\begin{equation}
	4.35548 \leq x_{\text{sh}} \leq 5.81695.
\end{equation}
In principle, the same analysis can be done with the Keck data, which would lead to marginally different bounds. However, the overall conclusions remain largely unaffected. 

\begin{figure}[h]
    \centering
    \subfloat[]{\label{fig:DSWH_shadow_sgrA}%
  \includegraphics[width=0.6\textwidth]{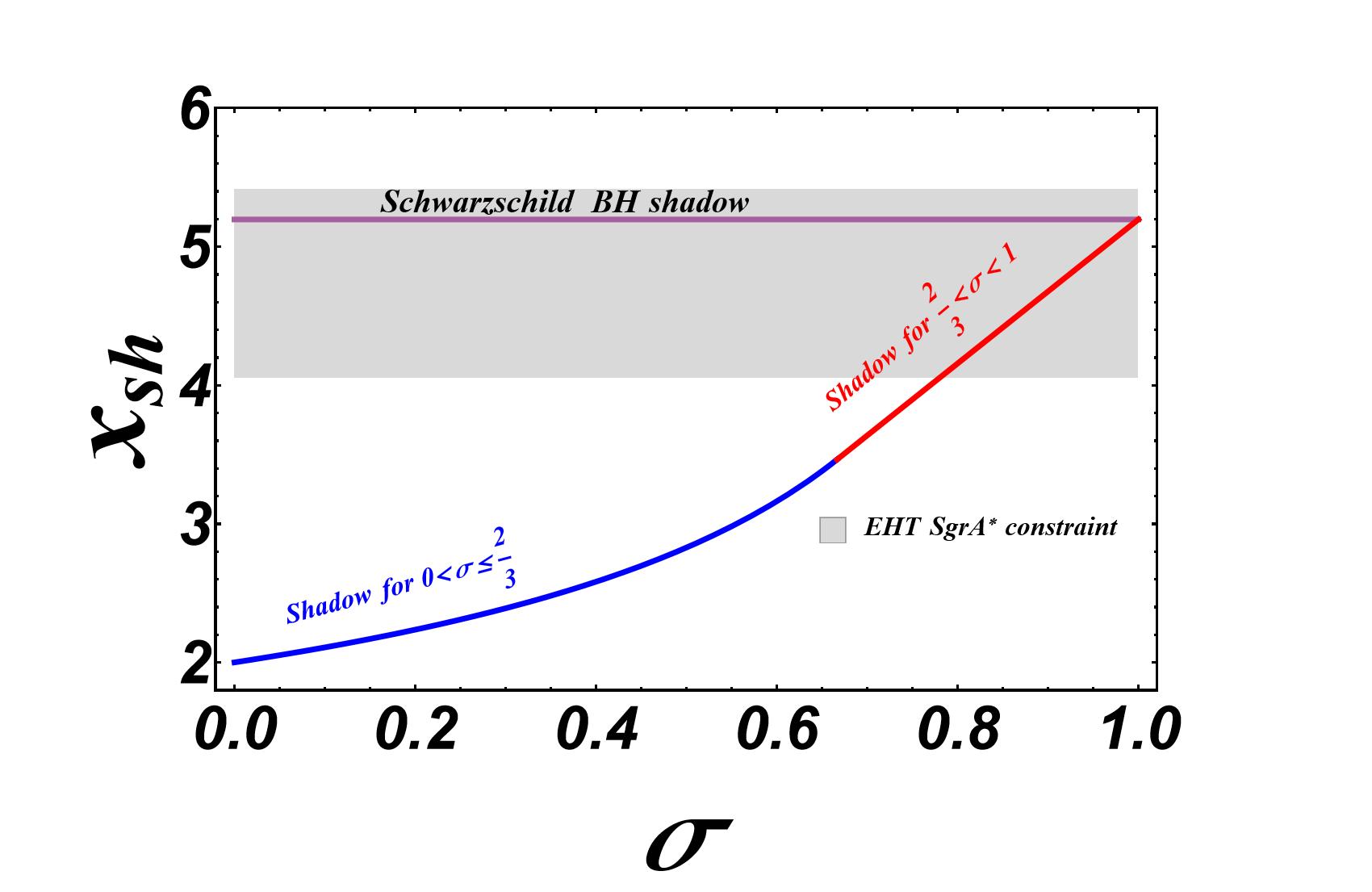}%
    }
    %\hspace{0.3in}
   \subfloat[]{\label{fig:HWH_shadow_sgrA}%
  \includegraphics[width=0.48\textwidth]{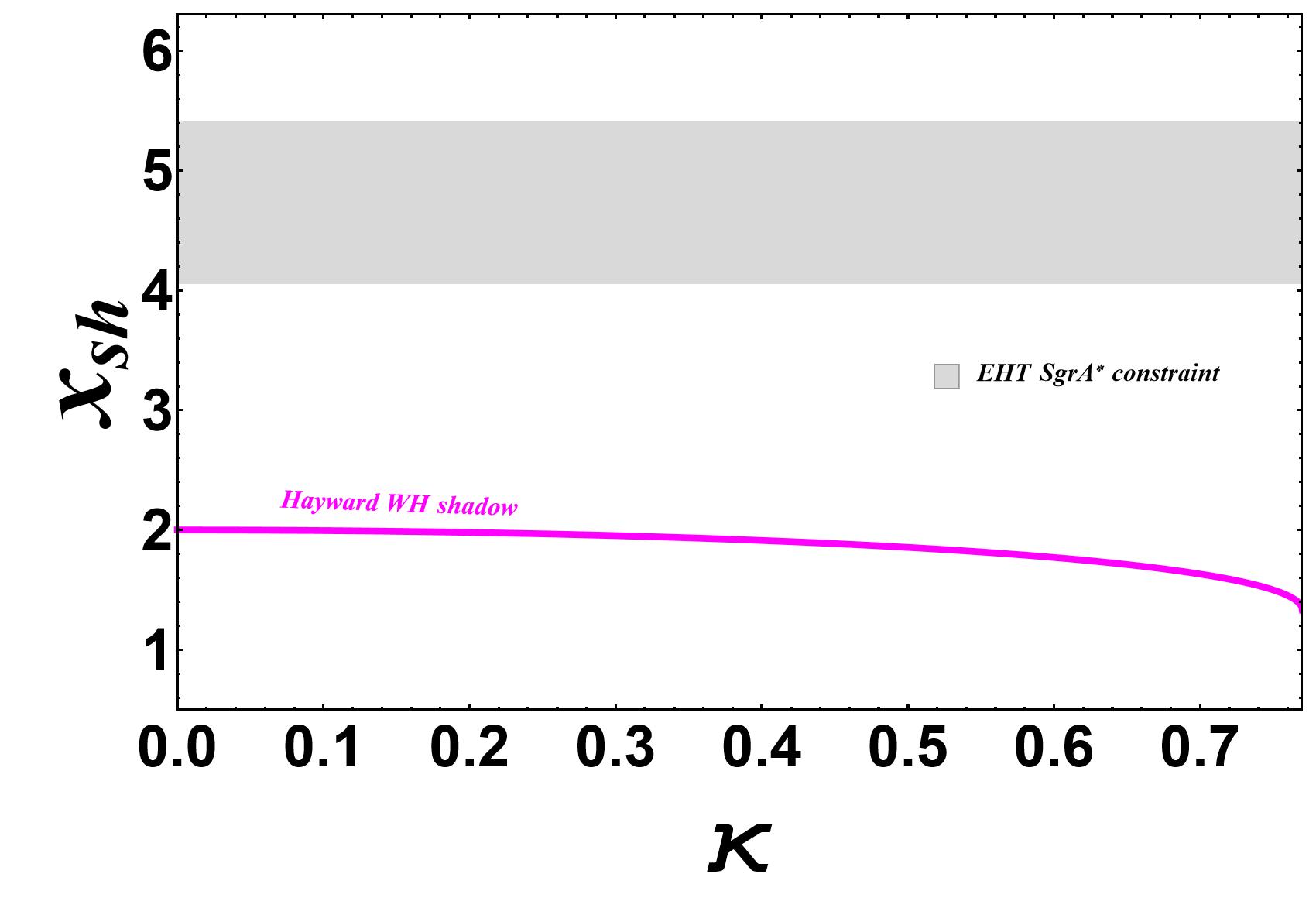}%
  } 
  \hspace{0.3in}
     \subfloat[]{\label{fig:HBH_shadow_sgrA}%
  \includegraphics[width=0.58\textwidth]{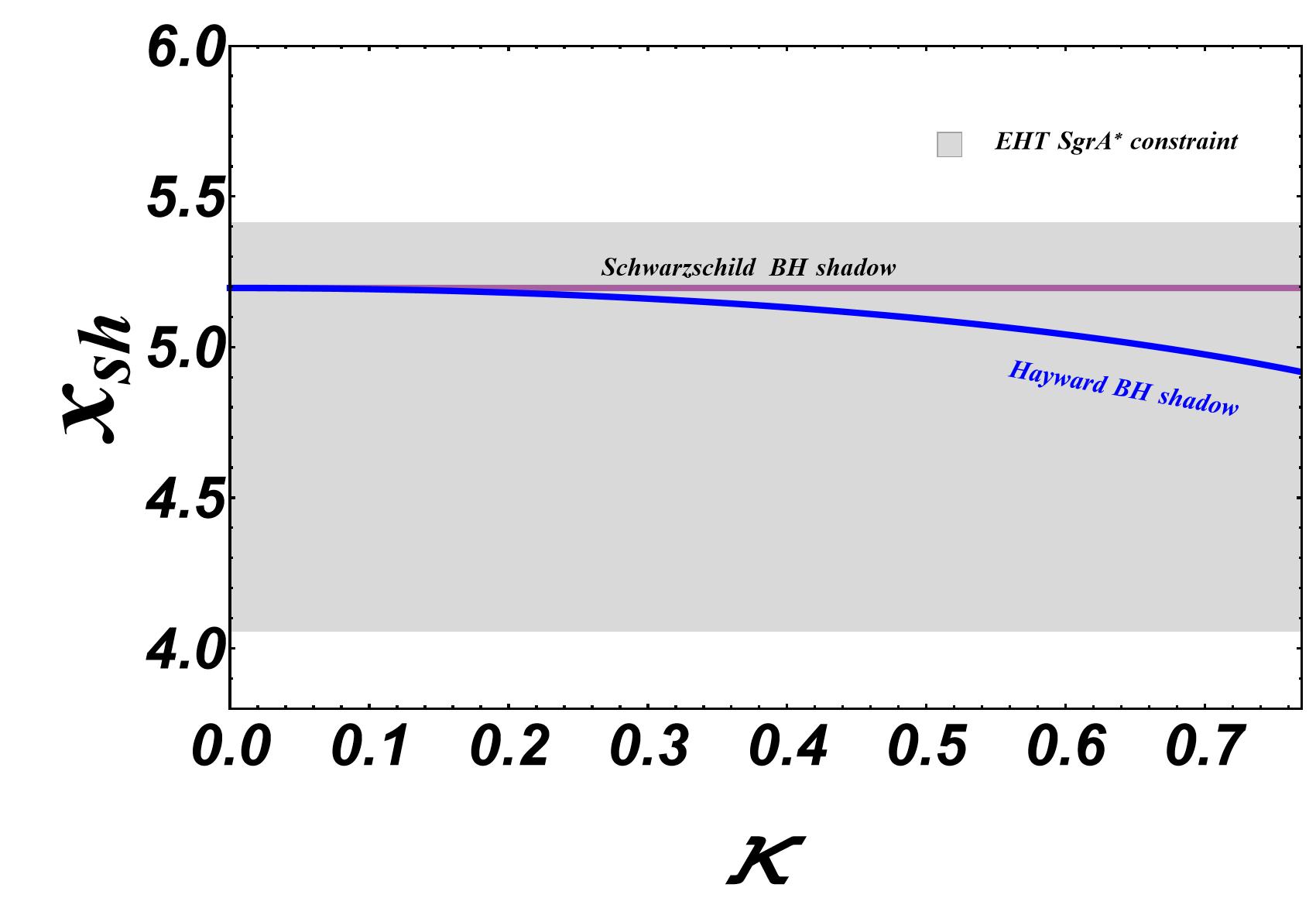}%
    }
    \caption{(a) The figure shows the shadow radius of Damour-Solodukin WH as a function of the parameter $\sigma$. Values of $\sigma$ close to 1 lie within the EHT constraint whereas the Schwarzschild BH shadow lies well within the bound. (b) The figure shows the shadow radius of Hayward WH as a function of the parameter $\kappa$ where for no values of $\kappa$ the shadow of the WH falls within EHT bound. (c) The figure shows the shadow radius of Damour-Solodukin BH as a function of the parameter $\kappa$, where for all $\kappa$ values the shadow lies within the shaded region.}
    %\label{fig:bound_density}
\end{figure}

\subsection{In vacuum}
Using the above constraint, we compare the shadow results obtained for various subclasses of spacetimes in the previous section in the absence of plasma and comment on which of these can be possible candidates to represent Sgr $\rm A^*$. For a Schwarzschild BH, the dimensionless shadow radius is $x_{\text{sh}} = 3\sqrt{3}$, which lies comfortably within the derived bounds. This aligns with the EHT analysis, where a deviation parameter of $\delta = -0.08 \pm 0.09$ was reported \cite{EventHorizonTelescope:2022xqj}. On the other hand, the Schwarzschild wormhole, with $x_{\text{sh}} = 2$, falls well outside the allowed range, disqualifying it as a viable model for Sgr A\*.

For the Damour-Solodukhin wormhole, the permitted range of the parameter $\sigma$ predominantly corresponds to double-barrier potentials i.e. with $\sigma$ close to 1, as illustrated in Fig.(\ref{fig:DSWH_shadow_sgrA}). In case of the Hayward wormhole, the shadow radius lies significantly below the observational constraints, as shown in Fig.(\ref{fig:HWH_shadow_sgrA}). Conversely, for the regular Hayward BH, the shadow radius across the complete parameter space remains within the observational bounds, as depicted in Fig.(\ref{fig:HBH_shadow_sgrA}).

\begin{figure}
    \centering
    \includegraphics[width=0.5\linewidth]{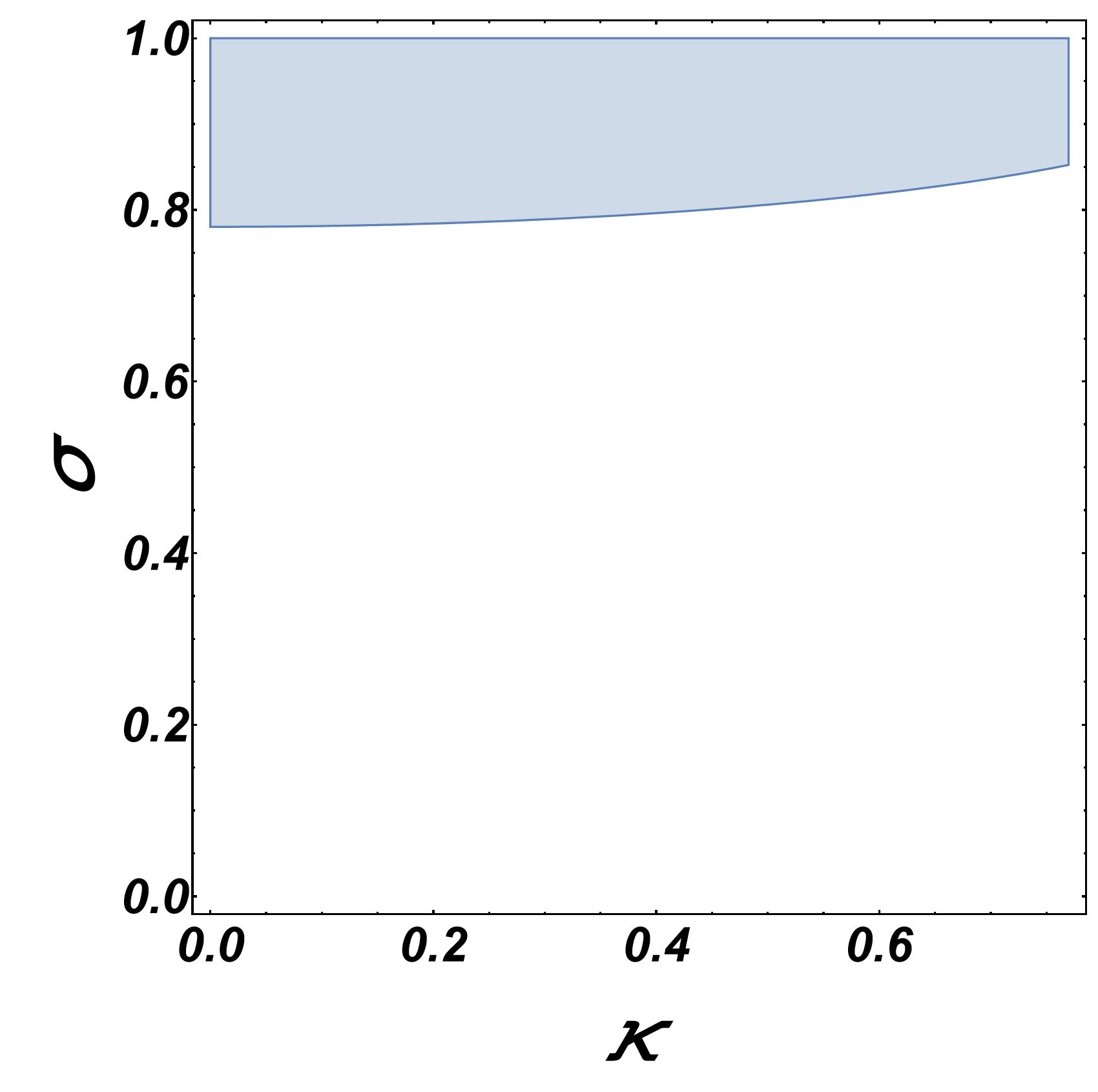}
    \caption{The allowed range of parameters of Hayward-Damour-Solodukin WH for which the shadow radius lies within the bounds obtained from Sgr $\rm A^*$}
    \label{fig:HDSWH_shadow_sgrA}
\end{figure}

Finally, for the most general case of Hayward-Damour-Solodukhin wormhole, we identify the region in the parameter space where the shadow radius complies with the EHT constraints as shown in Fig.(\ref{fig:HDSWH_shadow_sgrA}).

In summary, our analysis of the shadow radius indicates that among wormhole models, those exhibiting double- or triple-barrier effective potentials are more favored over their single-barrier counterparts. For regular BH models, current EHT observations of Sgr $\rm A^*$ do not impose additional constraints beyond those already provided by theoretical considerations. These conclusions may be revisited as more observational data become available in the future.

\subsection{With plasma} 
\begin{figure}[h]
    \centering
    \subfloat[]{\label{fig:SBH_homo_shadow_SgrA}%
  \includegraphics[width=0.52\textwidth]{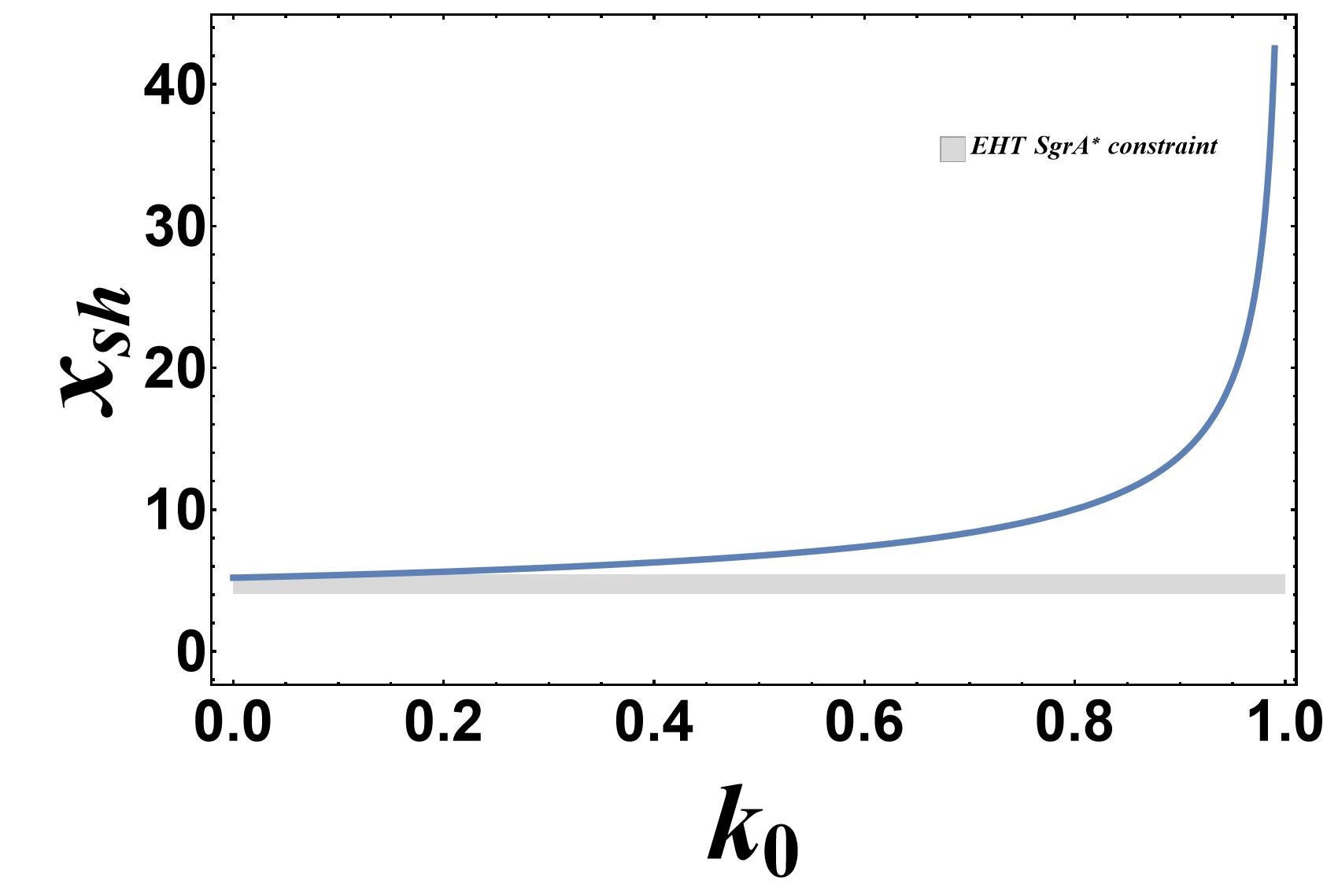}%
    }
    %\hspace{0.3in}
   \subfloat[]{\label{fig:SBH_nonhomo_shadow_SgrA}%
  \includegraphics[width=0.5\textwidth]{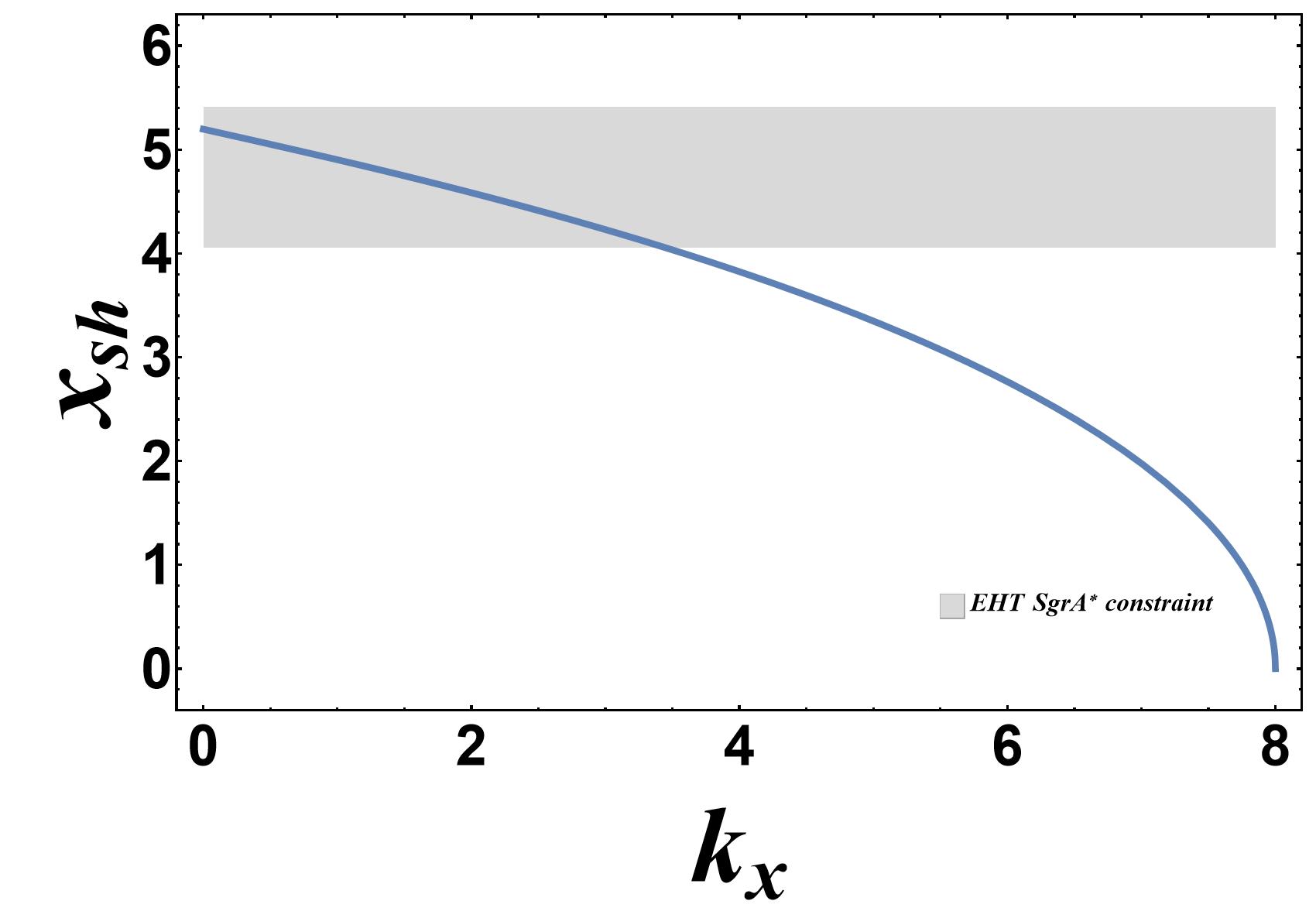}%
  } 
    \caption{The figure shows the shadow radius of Schwarzschild BH in presence of (a) homogeneous and (b) non-homogeneous plasma.}
    %\label{fig:bound_density}
\end{figure}

We now introduce a plasma medium and examine how the bounds on both spacetime parameters and plasma parameters are affected. In the case of a Schwarzschild BH, the allowed homogeneous plasma parameter ($k_0$) range remains relatively small, with support only for small values of the plasma parameter as shown in Fig.(\ref{fig:SBH_homo_shadow_SgrA}). For the non-homogeneous plasma case, the permissible parameter range is larger compared to the homogeneous scenario, however, the plasma parameter $k_x$ is still constrained to small values as depicted in Fig.(\ref{fig:SBH_nonhomo_shadow_SgrA}).

Turning to the Schwarzschild wormhole, as discussed in Sec.(\ref{sec:SWH}), the shadow radius in the homogeneous plasma is independent of the plasma parameter. Consequently, the shadow remains below the observational bounds and, thus, is inconsistent with EHT results. For the non-homogeneous plasma case, Fig.(\ref{fig:SWH_nonhomo_shadow}) clearly shows that the shadow radius decreases with increasing plasma profile parameter, further deviating from the allowed range. Therefore, both plasma scenarios for the Schwarzschild wormhole are ruled out.

\begin{figure}[h]
    \centering
    \subfloat[]{\label{fig:DSWH_homo_shadow_SgrA}%
  \includegraphics[width=0.5\textwidth]{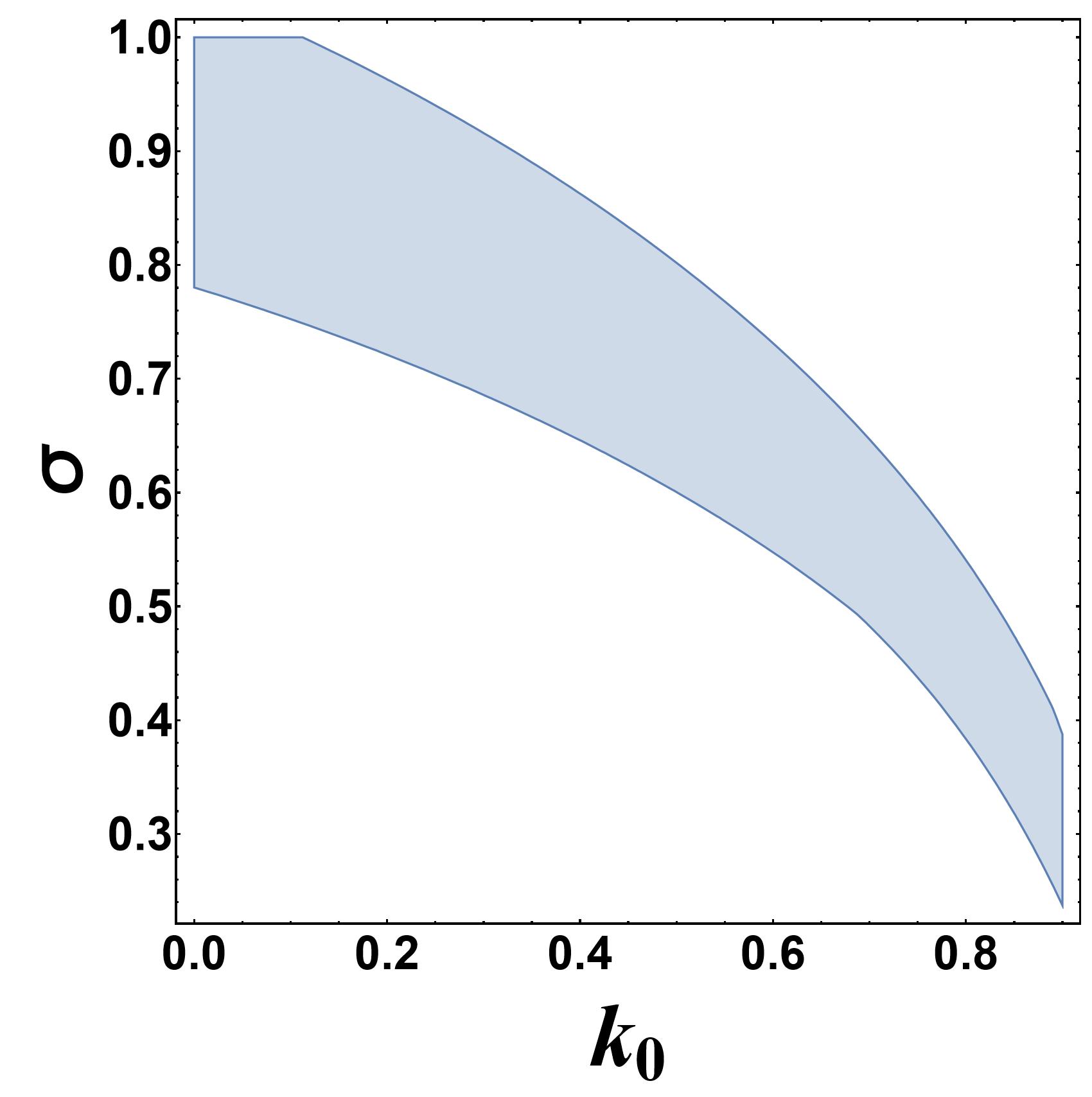}}
    %\hspace{0.3in}
   \subfloat[]{\label{fig:DSWH_nonhomo_shadow_SgrA}%
  \includegraphics[width=0.52\textwidth]{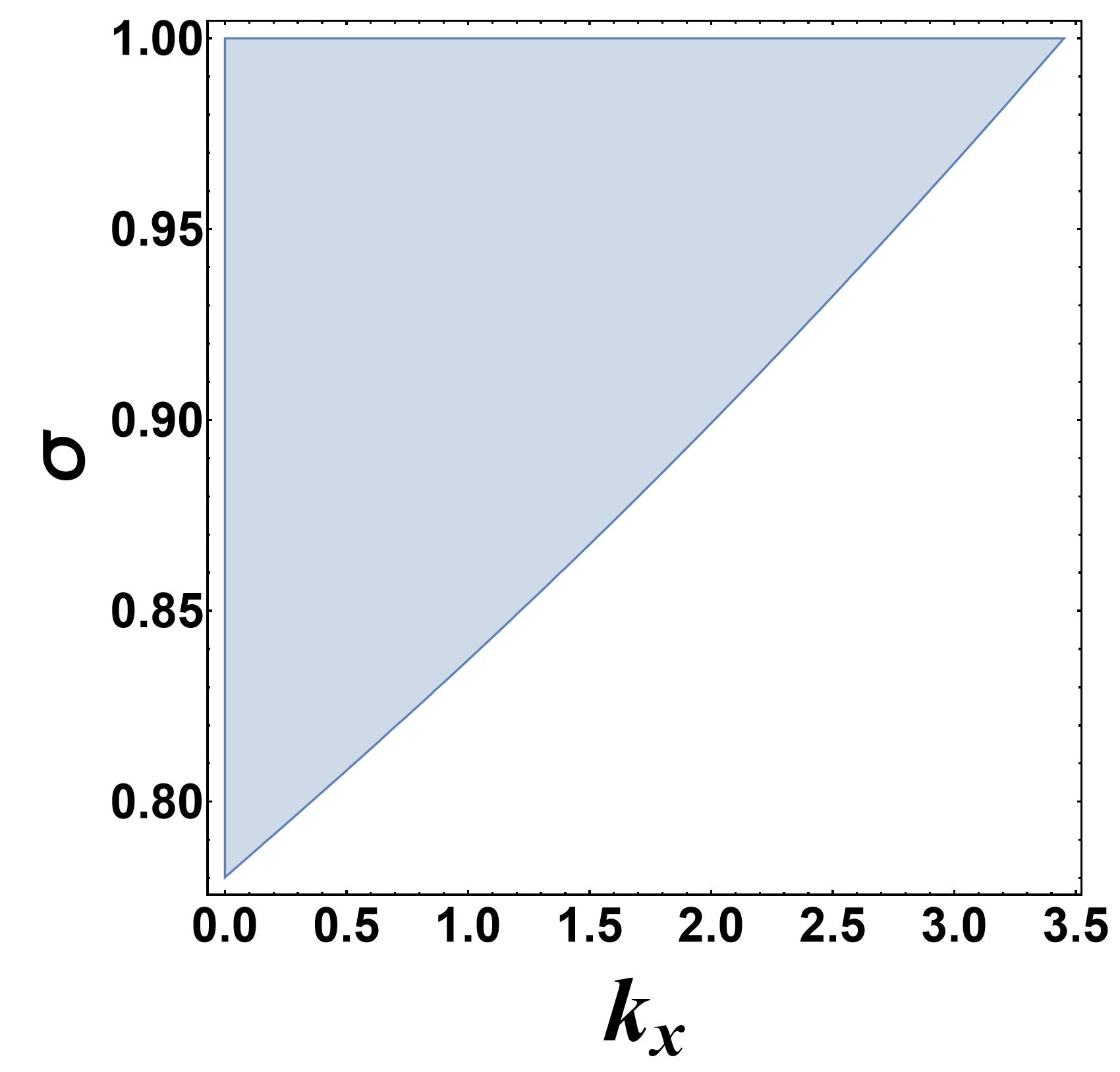}%
  } 
    \caption{The figure shows the allowed parameter space of Damour-Solodukin WH with (a) homogeneous plasma and (b) non-homogeneous plasma for which the shadow radius falls within the range of Sgr $\rm A^*$ shadow as observed by EHT .}
    %\label{fig:bound_density}
\end{figure}

For the Damour-Solodukhin wormhole, the permissible range of the parameter $\sigma$ is sensitive to the plasma  parameter in both homogeneous and non-homogeneous cases. In the homogeneous case, single-barrier potentials are supported for relatively large plasma parameters, whereas double-barrier potentials are favored at smaller values, as illustrated in Fig.(\ref{fig:DSWH_homo_shadow_SgrA}). In contrast, as shown in Fig.(\ref{fig:DSWH_nonhomo_shadow_SgrA}) under a non-homogeneous plasma profile, the parameter space corresponding to single-barrier potentials is entirely excluded. However, double-barrier potentials are allowed.

Similar to the Schwarzschild wormhole, the Hayward wormhole's shadow radius remains unaffected when homogeneous plasma is introduced (as discussed in Sec.(\ref{sec:HWH})), and thus still lies below the observationally allowed range. For the non-homogeneous case, the shadow radius decreases with increasing plasma parameter, as shown in Fig.(\ref{fig:HWH_nonhomo_shadow}), which further excludes this model from compatibility with EHT constraints.

\begin{figure}[h]
    \centering
    \subfloat[]{\label{fig:HBH_homo_shadow_SgrA}%
  \includegraphics[width=0.52\textwidth]{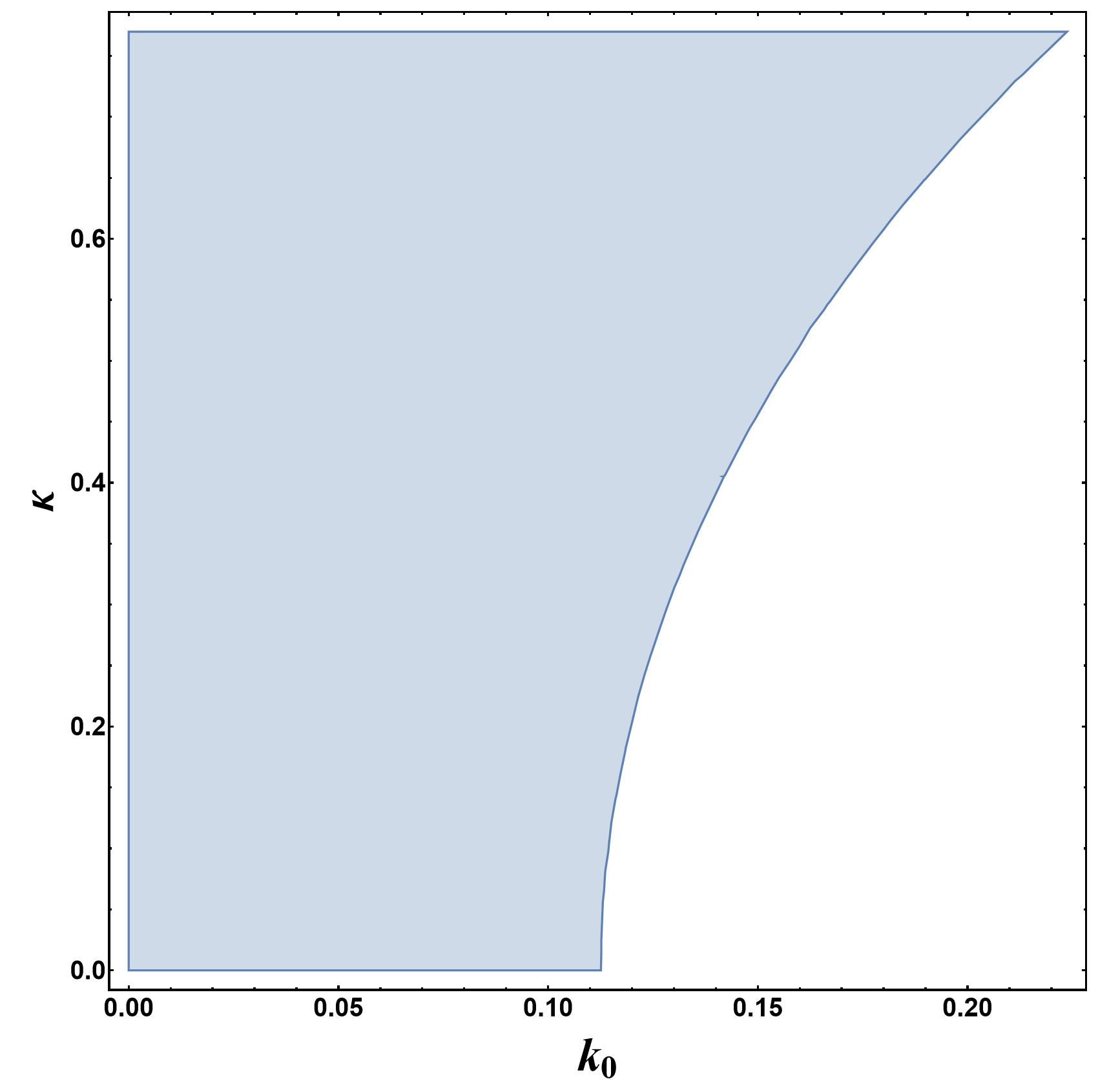}}
    %\hspace{0.3in}
   \subfloat[]{\label{fig:HBH_nonhomo_shadow_SgrA}%
  \includegraphics[width=0.52\textwidth]{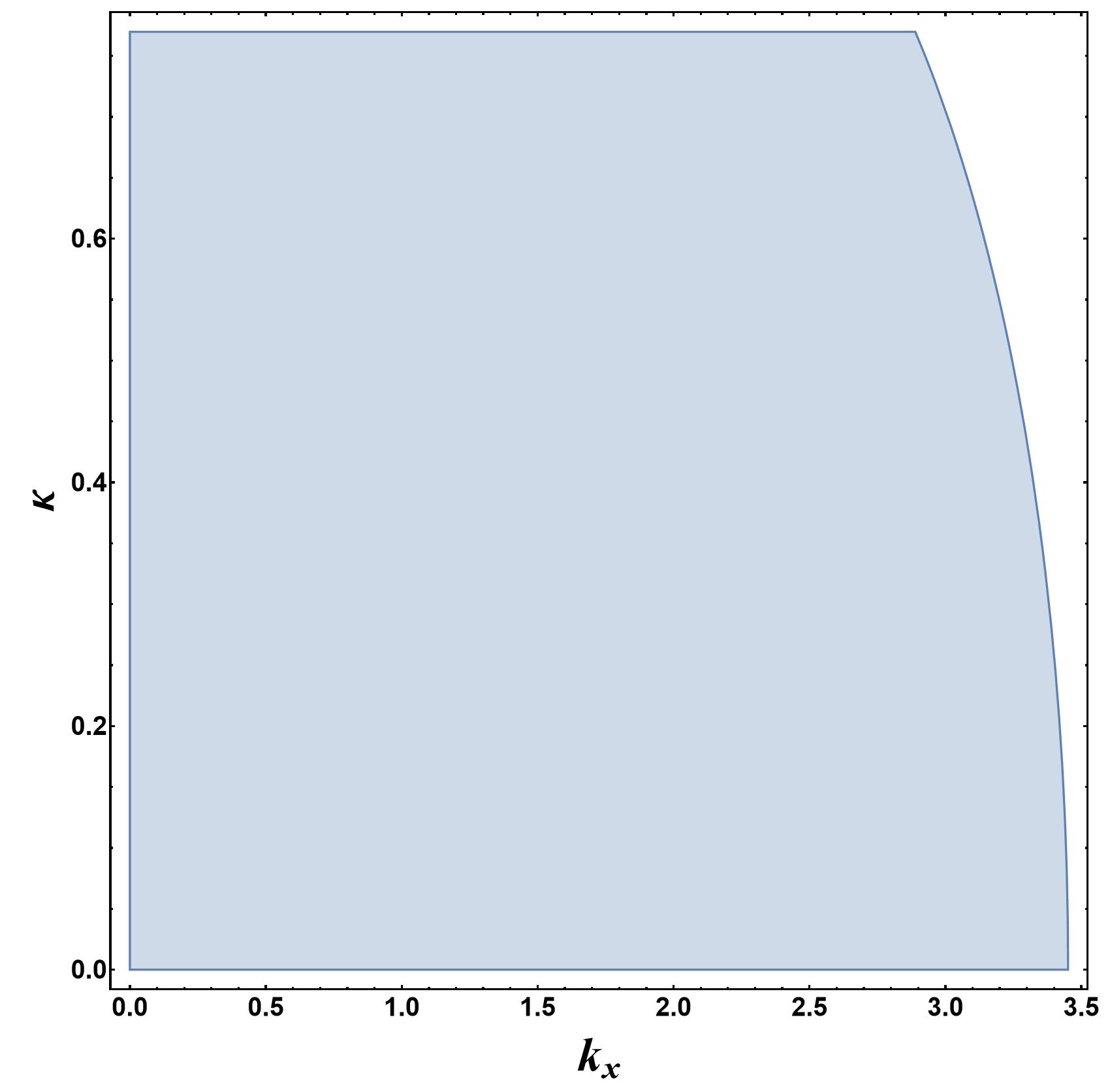}%
  } 
    \caption{The figure shows the allowed parameter space of Hayward BH with (a) homogeneous plasma and (b) non-homogeneous plasma for which the shadow radius falls within the range of Sgr $\rm A^*$ shadow as observed by EHT .
    }
    %\label{fig:bound_density}
\end{figure}

We now consider the regular Hayward BH. The allowed parameter space for both homogeneous and non-homogeneous plasma is shown in Fig.(\ref{fig:HBH_homo_shadow_SgrA}) and Fig.(\ref{fig:HBH_nonhomo_shadow_SgrA}), respectively. These figures indicate that the inclusion of observational constraints significantly narrows the viable range of plasma parameters initially obtained from theoretical limitations.

\begin{figure}[h]
    \centering
    \subfloat[]{\label{fig:HDSWH_homo_shadow_sgrA}%
  \includegraphics[width=0.52\textwidth]{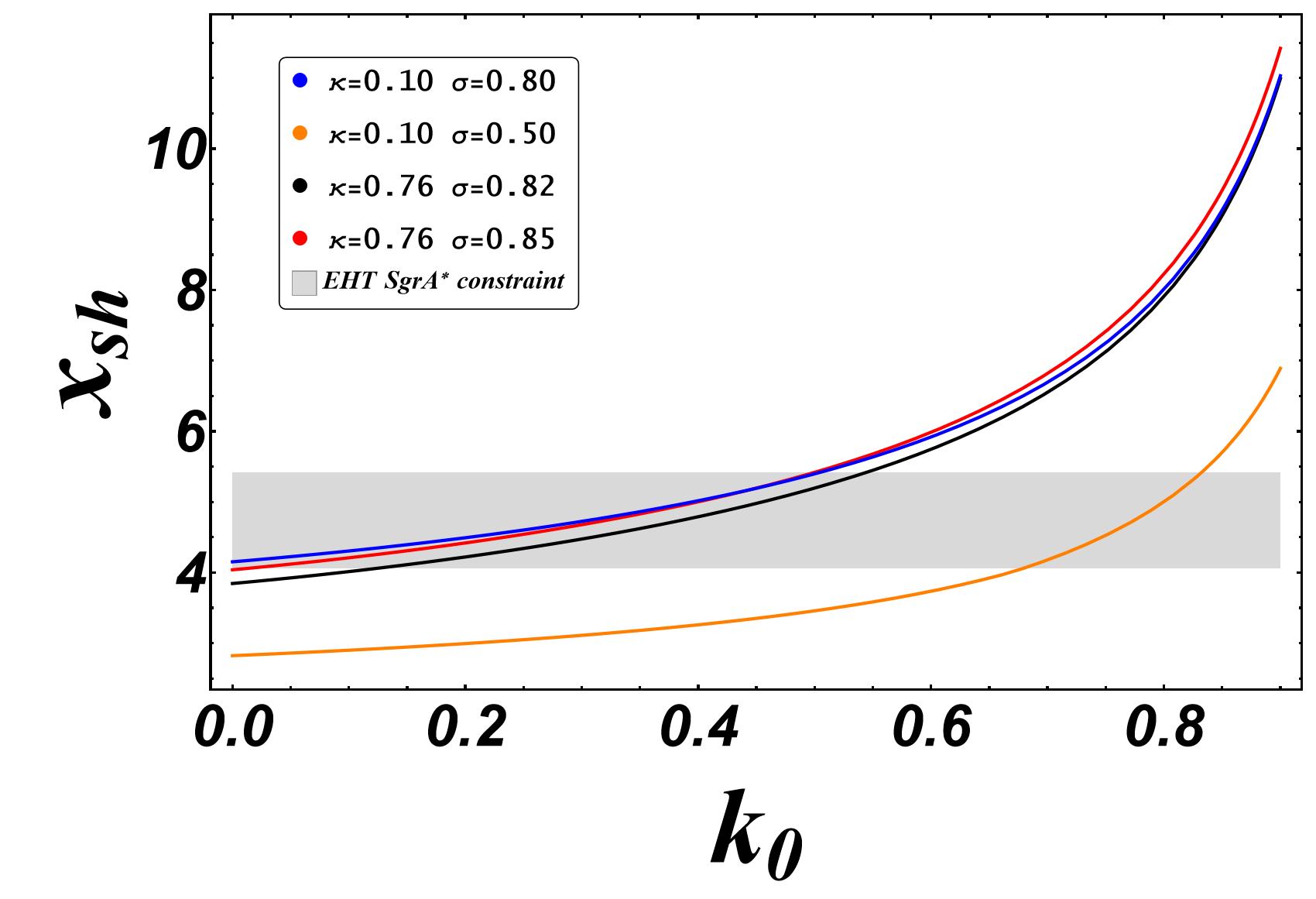}}
    %\hspace{0.3in}
   \subfloat[]{\label{fig:HDSWH_nonhomo_shadow_sgrA}%
  \includegraphics[width=0.52\textwidth]{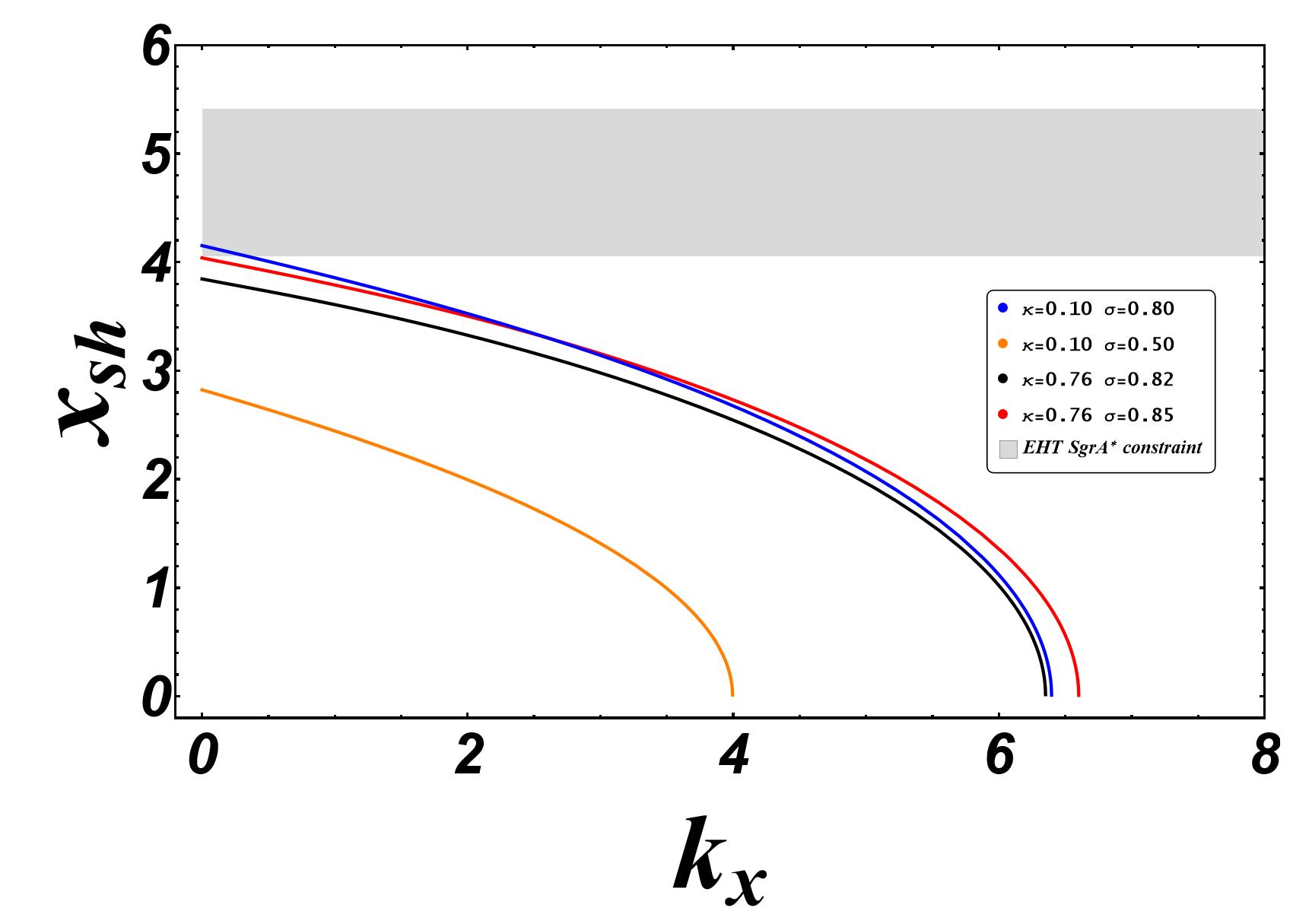}%
  } 
    \caption{The figure shows  represent the variation of shadow radius as a function of plasma parameters for the Hayward-Damour-Solodukin wormhole in the presence of (a) homogeneous and (b) non-homogeneous plasma. The shaded regions represents the constraints on shadow radius obtained from Sgr $\rm A^*$.}
    %\label{fig:bound_density}
\end{figure}
Finally, while examining the most general case of Hayward-Damour-Solodukhin wormhole, we focused on select parameter configurations that represent distinct wormhole structures characterized by the number of peaks in their effective potential. The wormholes with values of $\sigma$ closer to 1 are characterized by multi-peak potentials as shown in Fig.(\ref{fig:HDSWH_pot}). The corresponding results for homogeneous and non-homogeneous plasma profiles are presented in Fig.(\ref{fig:HDSWH_homo_shadow_sgrA}) and Fig.~~(\ref{fig:HDSWH_nonhomo_shadow_sgrA}). As observed earlier, for the homogeneous plasma case, single-peak potentials are favored at higher plasma parameter values, while double- and triple-peak configurations are supported at lower values. In the non-homogeneous case, single-peak configurations are disallowed altogether, and even higher-order peak solutions are only valid within a very narrow window at small plasma parameter values.

To summarise, the Schwarzschild BH remains consistent with EHT data in both vacuum and plasma scenarios, though only for small values of the plasma parameter. In retrospect, this is expected given the small deviation parameter as mentioned in EHT results \cite{EventHorizonTelescope:2022wkp}. In contrast, the Schwarzschild wormhole is ruled out entirely, as its shadow radius lies below the observational lower bound and decreases further in the presence of plasma. Among wormhole spacetimes, the Damour-Solodukhin wormhole remains a viable solution. In homogeneous plasma, both single- and double-peak potential configurations can be accommodated, while in the non-homogeneous case, only double-peak configurations remain possible. A similar trend is observed for the Hayward-Damour-Solodukhin wormhole, where multi-peak potentials are favored at lower plasma parameter values, and single-peak solutions are excluded in the presence of non-homogeneous plasma. The Hayward wormhole is not supported by the EHT observations, as its shadow radius consistently falls below the allowed range regardless of plasma presence. On the other hand, the regular Hayward BH remains compatible with observations, although the inclusion of plasma significantly restricts the viable range of parameters.

 \FloatBarrier
 
\section{Conclusion}\label{sec:5}
	
In this work, we have performed a detailed investigation of the shadows cast by various spacetime geometries arising from different parameter regimes of the generalized Hayward metric~\cite{DuttaRoy:2022ytr}. The metric in Eq.\eqref{eq:dimensionless_metric} depends on parameters $(\sigma,\kappa)$ and is obtained by introducing distinct mass parameters in the $g_{tt}$ and $g_{rr}$ components of the original Hayward regular black hole metric. This metric gives rise to four distinct kinds of wormholes, along with a regular and a singular BH solution as illustrated  in Fig.(\ref{fig:parameter_space}) and summarized in Table \ref{tab:classification}. While the spacetime properties and quasinormal mode spectra of these solutions have been analyzed in~\cite{DuttaRoy:2022ytr}, here we focus on their shadow characteristics. 

We have studied the dependence of shadow radius on the metric parameters for all spacetimes and plotted their deviation from the Schwarzschild BH shadow, corresponding to $(\sigma = 1, \kappa = 0)$. This analysis is especially relevant in the context of black hole mimickers, where particular parameter choices yield non-singular spacetimes whose shadow radius is nearly indistinguishable from that of a Schwarzschild BH. For instance, we observe that the Hayward regular BH yields a shadow radius nearly identical to Schwarzschild's for small $\kappa$ values (see Fig.~\ref{fig:HBH_shadow}). Similarly, the Damour-Solodukhin wormhole approaches the Schwarzschild shadow as $\sigma \rightarrow 1$ (Fig.~\ref{fig:DSWH_shadow}). However, significant deviations appear as $\kappa$ increases or $\sigma$ decreases, reflecting the growing difference in the underlying geometry from the Schwarzschild case. We also observe a notable feature  in the most general subclass the Hayward-Damour-Solodukhin (HDS) wormhole where, for certain values of $(\sigma, \kappa)$, the effective potential admits two or three photon spheres. Depending on the metric parameters, this potential can exhibit single, double, or even triple peaks. We find that the shadow radius of the HDS wormhole increases with $\sigma$ (for fixed $\kappa$) and decreases with increasing $\kappa$ (for fixed $\sigma$). In contrast, all remaining wormhole and BH solutions possess a single photon sphere.

The above conclusions pertain to vacuum spacetimes, where no surrounding medium influences the light propagation. However, realistic astrophysical BHs exist within matter-rich environments. As a first step towards incorporating such effects, we considered the impact of plasma on the shadow structure. For analytical simplicity we focused on two plasma density profiles: a homogeneous plasma $\Omega(x) = k_0$ and a non-homogeneous profile $\Omega(x) = \frac{k_x}{x}$. We computed the shadow radii for all classes of spacetimes under these conditions and compared the results to the vacuum case. A consolidated summary of the shadow's dependence on both the metric and plasma parameters is provided in Table~\ref{tab:summary}.

Finally, we compare our results with the observational constraints from the EHT image of the shadow of Sgr $\rm A^*$. Upon imposing these constraints, regular black hole solutions do not see any significant change in the allowed parameter space.  We also find that for wormhole solutions, the allowed parameter ranges for the spacetimes under consideration become significantly restricted. Our analysis indicates that a subclass of wormhole solutions exhibiting multi-peak effective potentials—particularly in the presence of plasma media—are more consistent with the observational data than those characterized by single-peak potentials.
 Interestingly, this finding stands in contrast to the conclusions drawn from quasinormal mode analyses, where single-barrier effective potentials were preferred as viable black hole mimickers \cite{DuttaRoy:2022ytr}. If the shadow-based analysis is accurate, this tension implies that compact objects resembling wormholes might exhibit characteristic late-time echoes in their gravitational wave signals offering a complementary observational signature to be explored in future detections.Current gravitational wave detectors have limited sensitivity when it comes to detecting the ringdown from the merger of compact objects and possible echoes if present. Thus the possibility of detecting echoes through improved future gravitational wave detectors keeps the wormhole geometries characterised with multi-peak potentials a viable spacetime. Regular BH solutions also remain viable, though with a significantly narrow choice of homogeneous plasma parameter when plasma is introduced. This makes it difficult to distinguish them observationally from standard GR black holes. Future high-resolution observations, both through shadows and gravitational waves, may enhance this discriminatory power, potentially uncovering subtle features unique to exotic compact objects.

It is important to note that more realistic astrophysical scenarios involve additional complexities such as magnetic fields, accretion flows, and nontrivial plasma distributions, which cannot be treated analytically and need to be addressed. However, addressing these aspects will require a fully numerical approach and is beyond the scope of the present work. Nonetheless, the generalized Hayward metric offers a versatile framework for systematically studying the shadow characteristics of a wide spectrum of regular and singular compact objects.

\section*{Acknowledgements}
 SG acknowledges Rajibul Shaikh for constructive discussions. PDR acknowledges the support from Infosys foundation. SC would like to acknowledge Mathematical Research Impact Centric Support (MATRICS) grant (MTR/2022/000318) from the Science and Engineering Research Board (SERB) of India.

	\bibliographystyle{JHEP}
%\bibliographystyle{apsrev4-2}
%\bibliographystyle{unsrt}
%	\bibliography{reff}
\bibliography{ref}

\providecommand{\href}[2]{#2}\begingroup\raggedright\begin{thebibliography}{10}

\bibitem{LIGOScientific:2016aoc}
{\scshape LIGO Scientific, Virgo} collaboration, \emph{{Observation of
  Gravitational Waves from a Binary Black Hole Merger}},
  \href{https://doi.org/10.1103/PhysRevLett.116.061102}{\emph{Phys. Rev. Lett.}
  {\bfseries 116} (2016) 061102}
  [\href{https://arxiv.org/abs/1602.03837}{{\ttfamily 1602.03837}}].

\bibitem{EventHorizonTelescope:2019dse}
{\scshape Event Horizon Telescope} collaboration, \emph{{First M87 Event
  Horizon Telescope Results. I. The Shadow of the Supermassive Black Hole}},
  \href{https://doi.org/10.3847/2041-8213/ab0ec7}{\emph{Astrophys. J. Lett.}
  {\bfseries 875} (2019) L1}
  [\href{https://arxiv.org/abs/1906.11238}{{\ttfamily 1906.11238}}].

\bibitem{EventHorizonTelescope:2022wkp}
{\scshape Event Horizon Telescope} collaboration, \emph{{First Sagittarius A*
  Event Horizon Telescope Results. I. The Shadow of the Supermassive Black Hole
  in the Center of the Milky Way}},
  \href{https://doi.org/10.3847/2041-8213/ac6674}{\emph{Astrophys. J. Lett.}
  {\bfseries 930} (2022) L12}
  [\href{https://arxiv.org/abs/2311.08680}{{\ttfamily 2311.08680}}].

\bibitem{bardeen_1968}
J.M.~Bardeen{\emph{presented at GR5, Tiflis, U.S.S.R., and published in the
  conference proceedings in the U.S.S.R.} (1968) }.

\bibitem{Ayon-Beato:1998hmi}
E.~Ayon-Beato and A.~Garcia, \emph{{Regular black hole in general relativity
  coupled to nonlinear electrodynamics}},
  \href{https://doi.org/10.1103/PhysRevLett.80.5056}{\emph{Phys. Rev. Lett.}
  {\bfseries 80} (1998) 5056}
  [\href{https://arxiv.org/abs/gr-qc/9911046}{{\ttfamily gr-qc/9911046}}].

\bibitem{Ayon-Beato:1999qin}
E.~Ayon-Beato and A.~Garcia, \emph{{Nonsingular charged black hole solution for
  nonlinear source}}, \href{https://doi.org/10.1023/A:1026640911319}{\emph{Gen.
  Rel. Grav.} {\bfseries 31} (1999) 629}
  [\href{https://arxiv.org/abs/gr-qc/9911084}{{\ttfamily gr-qc/9911084}}].

\bibitem{Ayon-Beato:1999kuh}
E.~Ayon-Beato and A.~Garcia, \emph{{New regular black hole solution from
  nonlinear electrodynamics}},
  \href{https://doi.org/10.1016/S0370-2693(99)01038-2}{\emph{Phys. Lett. B}
  {\bfseries 464} (1999) 25}
  [\href{https://arxiv.org/abs/hep-th/9911174}{{\ttfamily hep-th/9911174}}].

\bibitem{ayon-Beato_2000}
E.~Ayon-Beato and A.~Garcia, \emph{{The Bardeen model as a nonlinear magnetic
  monopole}}, \href{https://doi.org/10.1016/S0370-2693(00)01125-4}{\emph{Phys.
  Lett. B} {\bfseries 493} (2000) 149}
  [\href{https://arxiv.org/abs/gr-qc/0009077}{{\ttfamily gr-qc/0009077}}].

\bibitem{ayon-Beato_2004}
E.~Ayon-Beato and A.~Garcia, \emph{{Four parametric regular black hole
  solution}}, \href{https://doi.org/10.1007/s10714-005-0050-y}{\emph{Gen. Rel.
  Grav.} {\bfseries 37} (2005) 635}
  [\href{https://arxiv.org/abs/hep-th/0403229}{{\ttfamily hep-th/0403229}}].

\bibitem{Dymnikova20044417}
I.~Dymnikova, \emph{Regular electrically charged vacuum structures with de
  sitter centre in nonlinear electrodynamics coupled to general relativity},
  \href{https://doi.org/10.1088/0264-9381/21/18/009}{\emph{Classical and
  Quantum Gravity} {\bfseries 21} (2004) 4417 – 4428}.

\bibitem{Bronnikov2001}
K.~Bronnikov, \emph{Regular magnetic black holes and monopoles from nonlinear
  electrodynamics},
  \href{https://doi.org/10.1103/PhysRevD.63.044005}{\emph{Physical Review D}
  {\bfseries 63} (2001) }.

\bibitem{Shankaranarayanan20041095}
S.~Shankaranarayanan and N.~Dadhich, \emph{Non-singular black-holes on the
  brane}, \href{https://doi.org/10.1142/S0218271804005109}{\emph{International
  Journal of Modern Physics D} {\bfseries 13} (2004) 1095 – 1103}.

\bibitem{hayward_2006}
S.A.~Hayward, \emph{Formation and evaporation of nonsingular black holes},
  \href{https://doi.org/10.1103/PhysRevLett.96.031103}{\emph{Phys. Rev. Lett.}
  {\bfseries 96} (2006) 031103}.

\bibitem{morris_1988}
M.S.~Morris, K.S.~Thorne and U.~Yurtsever, \emph{Wormholes, time machines, and
  the weak energy condition},
  \href{https://doi.org/10.1103/PhysRevLett.61.1446}{\emph{Phys. Rev. Lett.}
  {\bfseries 61} (1988) 1446}.

\bibitem{thorne_1988}
M.S.~Morris and K.S.~Thorne, \emph{Wormholes in spacetime and their use for
  interstellar travel: A tool for teaching general relativity},
  \href{https://doi.org/10.1119/1.15620}{\emph{American Journal of Physics}
  {\bfseries 56} (1988) 395}
  [\href{https://arxiv.org/abs/https://doi.org/10.1119/1.15620}{{\ttfamily
  https://doi.org/10.1119/1.15620}}].

\bibitem{ori_1991}
A.~Ori, \emph{Inner structure of a charged black hole: An exact mass-inflation
  solution}, \href{https://doi.org/10.1103/PhysRevLett.67.789}{\emph{Phys. Rev.
  Lett.} {\bfseries 67} (1991) 789}.

\bibitem{poisson_1989}
E.~Poisson and W.~Israel, \emph{Inner-horizon instability and mass inflation in
  black holes}, \href{https://doi.org/10.1103/PhysRevLett.63.1663}{\emph{Phys.
  Rev. Lett.} {\bfseries 63} (1989) 1663}.

\bibitem{carballo_2018}
R.~Carballo-Rubio, F.~Di~Filippo, S.~Liberati, C.~Pacilio and M.~Visser,
  \emph{{On the viability of regular black holes}},
  \href{https://doi.org/10.1007/JHEP07(2018)023}{\emph{JHEP} {\bfseries 07}
  (2018) 023} [\href{https://arxiv.org/abs/1805.02675}{{\ttfamily
  1805.02675}}].

\bibitem{carballo_2021}
R.~Carballo-Rubio, F.~Di~Filippo, S.~Liberati, C.~Pacilio and M.~Visser,
  \emph{{Inner horizon instability and the unstable cores of regular black
  holes}}, \href{https://doi.org/10.1007/JHEP05(2021)132}{\emph{JHEP}
  {\bfseries 05} (2021) 132}
  [\href{https://arxiv.org/abs/2101.05006}{{\ttfamily 2101.05006}}].

\bibitem{lobo_2008}
F.S.N.~Lobo, \emph{{General class of wormhole geometries in conformal Weyl
  gravity}}, \href{https://doi.org/10.1088/0264-9381/25/17/175006}{\emph{Class.
  Quant. Grav.} {\bfseries 25} (2008) 175006}
  [\href{https://arxiv.org/abs/0801.4401}{{\ttfamily 0801.4401}}].

\bibitem{varieschi_2015}
G.U.~Varieschi and K.L.~Ault, \emph{{Wormhole geometries in fourth-order
  conformal Weyl gravity}},
  \href{https://doi.org/10.1142/S0218271816500644}{\emph{Int. J. Mod. Phys. D}
  {\bfseries 25} (2016) 1650064}
  [\href{https://arxiv.org/abs/1510.05054}{{\ttfamily 1510.05054}}].

\bibitem{kord_2015}
M.~Kord~Zangeneh, F.S.N.~Lobo and M.H.~Dehghani, \emph{{Traversable wormholes
  satisfying the weak energy condition in third-order Lovelock gravity}},
  \href{https://doi.org/10.1103/PhysRevD.92.124049}{\emph{Phys. Rev. D}
  {\bfseries 92} (2015) 124049}
  [\href{https://arxiv.org/abs/1510.07089}{{\ttfamily 1510.07089}}].

\bibitem{ovgun_2019}
A.~\"Ovg\"un, K.~Jusufi and I.~Sakalli, \emph{Exact traversable wormhole
  solution in bumblebee gravity},
  \href{https://doi.org/10.1103/PhysRevD.99.024042}{\emph{Phys. Rev. D}
  {\bfseries 99} (2019) 024042}.

\bibitem{zubair_2017}
M.~Zubair, F.~Kousar and S.~Bahamonde, \emph{{Static spherically symmetric
  wormholes in generalized $f(R,\phi)$ gravity}},
  \href{https://doi.org/10.1140/epjp/i2018-12344-y}{\emph{Eur. Phys. J. Plus}
  {\bfseries 133} (2018) 523}
  [\href{https://arxiv.org/abs/1712.05699}{{\ttfamily 1712.05699}}].

\bibitem{lobo_2009}
F.S.N.~Lobo and M.A.~Oliveira, \emph{Wormhole geometries in $f(r)$ modified
  theories of gravity},
  \href{https://doi.org/10.1103/PhysRevD.80.104012}{\emph{Phys. Rev. D}
  {\bfseries 80} (2009) 104012}.

\bibitem{boehmer_2012}
C.~Boehmer, T.~Harko and F.S.~Lobo, \emph{Wormhole geometries in modified
  teleparallel gravity and the energy conditions}, {\emph{Physical Review D}
  {\bfseries 85} (2012) 044033}.

\bibitem{shaikh_2016}
R.~Shaikh and S.~Kar, \emph{Wormholes, the weak energy condition, and
  scalar-tensor gravity},
  \href{https://doi.org/10.1103/PhysRevD.94.024011}{\emph{Phys. Rev. D}
  {\bfseries 94} (2016) 024011}.

\bibitem{kanti_2012}
P.~Kanti, B.~Kleihaus and J.~Kunz, \emph{Stable lorentzian wormholes in
  dilatonic einstein-gauss-bonnet theory},
  \href{https://doi.org/10.1103/PhysRevD.85.044007}{\emph{Phys. Rev. D}
  {\bfseries 85} (2012) 044007}.

\bibitem{mehdizadeh_2015}
M.R.~Mehdizadeh, M.K.~Zangeneh and F.S.N.~Lobo, \emph{Einstein-gauss-bonnet
  traversable wormholes satisfying the weak energy condition},
  \href{https://doi.org/10.1103/PhysRevD.91.084004}{\emph{Phys. Rev. D}
  {\bfseries 91} (2015) 084004}.

\bibitem{maeda_2008}
H.~Maeda and M.~Nozawa, \emph{Static and symmetric wormholes respecting energy
  conditions in einstein-gauss-bonnet gravity},
  \href{https://doi.org/10.1103/PhysRevD.78.024005}{\emph{Phys. Rev. D}
  {\bfseries 78} (2008) 024005}.

\bibitem{kanti_2011}
P.~Kanti, B.~Kleihaus and J.~Kunz, \emph{Wormholes in dilatonic
  einstein-gauss-bonnet theory},
  \href{https://doi.org/10.1103/PhysRevLett.107.271101}{\emph{Phys. Rev. Lett.}
  {\bfseries 107} (2011) 271101}.

\bibitem{shaikh_2015}
R.~Shaikh, \emph{Lorentzian wormholes in eddington-inspired born-infeld
  gravity}, \href{https://doi.org/10.1103/PhysRevD.92.024015}{\emph{Phys. Rev.
  D} {\bfseries 92} (2015) 024015}.

\bibitem{Carballo-Rubio:2022kad}
R.~Carballo-Rubio, F.~Di~Filippo, S.~Liberati, C.~Pacilio and M.~Visser,
  \emph{{Regular black holes without mass inflation instability}},
  \href{https://doi.org/10.1007/JHEP09(2022)118}{\emph{JHEP} {\bfseries 09}
  (2022) 118} [\href{https://arxiv.org/abs/2205.13556}{{\ttfamily
  2205.13556}}].

\bibitem{bonanno_2021}
A.~Bonanno, A.-P.~Khosravi and F.~Saueressig, \emph{Regular black holes with
  stable cores}, \href{https://doi.org/10.1103/PhysRevD.103.124027}{\emph{Phys.
  Rev. D} {\bfseries 103} (2021) 124027}.

\bibitem{Synge:1966mon}
J.L.~Synge, \emph{{The Escape of Photons from Gravitationally Intense Stars}},
  \href{https://doi.org/10.1093/mnras/131.3.463}{\emph{Mon. Not. Roy. Astron.
  Soc.} {\bfseries 131} (1966) 463}.

\bibitem{Luminet:1979nyg}
J.P.~Luminet, \emph{{Image of a spherical black hole with thin accretion
  disk}}, {\emph{Astron. Astrophys.} {\bfseries 75} (1979) 228}.

\bibitem{bardeen_1973}
J.M.~Bardeen{\emph{in Proceedings of the Ecole d’Eté De Physique Theorique:
  Les Astres Occlus: Les Houches 1972} (1973) 215–240}.

\bibitem{deVries:1999tiy}
A.~de~Vries, \emph{{The apparent shape of a rotating charged black hole, closed
  photon orbits and the bifurcation set $A_4$}},
  \href{https://doi.org/10.1088/0264-9381/17/1/309}{\emph{Class. Quant. Grav.}
  {\bfseries 17} (1999) 123}.

\bibitem{PhysRevD.80.024042}
K.~Hioki and K.-i.~Maeda, \emph{Measurement of the kerr spin parameter by
  observation of a compact object's shadow},
  \href{https://doi.org/10.1103/PhysRevD.80.024042}{\emph{Phys. Rev. D}
  {\bfseries 80} (2009) 024042}.

\bibitem{Shao-Wen_Wei_2013}
S.-W.~Wei and Y.-X.~Liu, \emph{Observing the shadow of
  einstein-maxwell-dilaton-axion black hole},
  \href{https://doi.org/10.1088/1475-7516/2013/11/063}{\emph{Journal of
  Cosmology and Astroparticle Physics} {\bfseries 2013} (2013) 063}.

\bibitem{Abdujabbarov:2012bn}
A.~Abdujabbarov, F.~Atamurotov, Y.~Kucukakca, B.~Ahmedov and U.~Camci,
  \emph{{Shadow of Kerr-Taub-NUT black hole}},
  \href{https://doi.org/10.1007/s10509-012-1337-6}{\emph{Astrophys. Space Sci.}
  {\bfseries 344} (2013) 429}
  [\href{https://arxiv.org/abs/1212.4949}{{\ttfamily 1212.4949}}].

\bibitem{Moffat:2015kva}
J.W.~Moffat, \emph{{Modified Gravity Black Holes and their Observable
  Shadows}}, \href{https://doi.org/10.1140/epjc/s10052-015-3352-6}{\emph{Eur.
  Phys. J. C} {\bfseries 75} (2015) 130}
  [\href{https://arxiv.org/abs/1502.01677}{{\ttfamily 1502.01677}}].

\bibitem{PhysRevD.85.064019}
L.~Amarilla and E.F.~Eiroa, \emph{Shadow of a rotating braneworld black hole},
  \href{https://doi.org/10.1103/PhysRevD.85.064019}{\emph{Phys. Rev. D}
  {\bfseries 85} (2012) 064019}.

\bibitem{PhysRevD.88.064004}
F.~Atamurotov, A.~Abdujabbarov and B.~Ahmedov, \emph{Shadow of rotating
  non-kerr black hole},
  \href{https://doi.org/10.1103/PhysRevD.88.064004}{\emph{Phys. Rev. D}
  {\bfseries 88} (2013) 064004}.

\bibitem{Roy:2020dyy}
R.~Roy and S.~Chakrabarti, \emph{{Study on black hole shadows in asymptotically
  de Sitter spacetimes}},
  \href{https://doi.org/10.1103/PhysRevD.102.024059}{\emph{Phys. Rev. D}
  {\bfseries 102} (2020) 024059}
  [\href{https://arxiv.org/abs/2003.14107}{{\ttfamily 2003.14107}}].

\bibitem{Rodriguez:2024ijx}
B.~Rodr\'\i{}guez, J.~Chagoya and C.~Ortiz, \emph{{Shadows of black holes in
  dynamical Chern-Simons modified gravity}},
  \href{https://arxiv.org/abs/2403.13062}{{\ttfamily 2403.13062}}.

\bibitem{Cunha:2018acu}
P.V.P.~Cunha and C.A.R.~Herdeiro, \emph{{Shadows and strong gravitational
  lensing: a brief review}},
  \href{https://doi.org/10.1007/s10714-018-2361-9}{\emph{Gen. Rel. Grav.}
  {\bfseries 50} (2018) 42} [\href{https://arxiv.org/abs/1801.00860}{{\ttfamily
  1801.00860}}].

\bibitem{Perlick:2021aok}
V.~Perlick and O.Y.~Tsupko, \emph{{Calculating black hole shadows: Review of
  analytical studies}},
  \href{https://doi.org/10.1016/j.physrep.2021.10.004}{\emph{Phys. Rept.}
  {\bfseries 947} (2022) 1} [\href{https://arxiv.org/abs/2105.07101}{{\ttfamily
  2105.07101}}].

\bibitem{Lupsasca:2024wkp}
A.~Lupsasca, D.R.~Mayerson, B.~Ripperda and S.~Staelens, \emph{{A
  Beginner\textquoteright{}s Guide to~Black Hole Imaging and~Associated Tests
  of~General Relativity}},  in \emph{{Recent Progress on Gravity Tests.
  Challenges and Future Perspectives}}, C.~Bambi and A.~Cardenas-Avendano,
  eds., pp.~183--237 (2024),
  \href{https://doi.org/10.1007/978-981-97-2871-8_6}{DOI}
  [\href{https://arxiv.org/abs/2402.01290}{{\ttfamily 2402.01290}}].

\bibitem{Li:2013jra}
Z.~Li and C.~Bambi, \emph{{Measuring the Kerr spin parameter of regular black
  holes from their shadow}},
  \href{https://doi.org/10.1088/1475-7516/2014/01/041}{\emph{JCAP} {\bfseries
  01} (2014) 041} [\href{https://arxiv.org/abs/1309.1606}{{\ttfamily
  1309.1606}}].

\bibitem{Abdujabbarov:2016hnw}
A.~Abdujabbarov, M.~Amir, B.~Ahmedov and S.G.~Ghosh, \emph{{Shadow of rotating
  regular black holes}},
  \href{https://doi.org/10.1103/PhysRevD.93.104004}{\emph{Phys. Rev. D}
  {\bfseries 93} (2016) 104004}
  [\href{https://arxiv.org/abs/1604.03809}{{\ttfamily 1604.03809}}].

\bibitem{Stuchlik:2019uvf}
Z.~Stuchl\'\i{}k and J.~Schee, \emph{{Shadow of the regular Bardeen black holes
  and comparison of the motion of photons and neutrinos}},
  \href{https://doi.org/10.1140/epjc/s10052-019-6543-8}{\emph{Eur. Phys. J. C}
  {\bfseries 79} (2019) 44}.

\bibitem{Dymnikova:2019vuz}
I.~Dymnikova and K.~Kraav, \emph{{Identification of a Regular Black Hole by Its
  Shadow}}, \href{https://doi.org/10.3390/universe5070163}{\emph{Universe}
  {\bfseries 5} (2019) 163}.

\bibitem{GHOSH2020115088}
S.G.~Ghosh, M.~Amir and S.D.~Maharaj, \emph{Ergosphere and shadow of a rotating
  regular black hole},
  \href{https://doi.org/https://doi.org/10.1016/j.nuclphysb.2020.115088}{\emph{Nuclear
  Physics B} {\bfseries 957} (2020) 115088}.

\bibitem{Uniyal:2023ahv}
A.~Uniyal, S.~Chakrabarti, M.~Fathi and A.~\"Ovg\"un, \emph{{Observational
  signatures: Shadow cast by the effective metric of photons for black holes
  with rational non-linear electrodynamics}},
  \href{https://doi.org/10.1016/j.aop.2024.169614}{\emph{Annals Phys.}
  {\bfseries 462} (2024) 169614}
  [\href{https://arxiv.org/abs/2309.13680}{{\ttfamily 2309.13680}}].

\bibitem{KumarWalia:2024yxn}
R.~Kumar~Walia, \emph{{Exploring nonlinear electrodynamics theories: Shadows of
  regular black holes and horizonless ultracompact objects}},
  \href{https://doi.org/10.1103/PhysRevD.110.064058}{\emph{Phys. Rev. D}
  {\bfseries 110} (2024) 064058}
  [\href{https://arxiv.org/abs/2409.13290}{{\ttfamily 2409.13290}}].

\bibitem{DuttaRoy:2022ytr}
P.~Dutta~Roy and S.~Kar, \emph{{Generalized Hayward spacetimes: Geometry,
  matter, and scalar quasinormal modes}},
  \href{https://doi.org/10.1103/PhysRevD.106.044028}{\emph{Phys. Rev. D}
  {\bfseries 106} (2022) 044028}
  [\href{https://arxiv.org/abs/2206.04505}{{\ttfamily 2206.04505}}].

\bibitem{Kumar:2023wfp}
S.~Kumar, A.~Uniyal and S.~Chakrabarti, \emph{{Shadow and weak gravitational
  lensing of rotating traversable wormhole in nonhomogeneous plasma
  spacetime}}, \href{https://doi.org/10.1103/PhysRevD.109.104012}{\emph{Phys.
  Rev. D} {\bfseries 109} (2024) 104012}
  [\href{https://arxiv.org/abs/2308.05545}{{\ttfamily 2308.05545}}].

\bibitem{Perlick:2015prd}
V.~Perlick, O.Y.~Tsupko and G.S.~Bisnovatyi-Kogan, \emph{{Influence of a plasma
  on the shadow of a spherically symmetric black hole}},
  \href{https://doi.org/10.1103/PhysRevD.92.104031}{\emph{Phys. Rev. D}
  {\bfseries 92} (2015) 104031}
  [\href{https://arxiv.org/abs/1507.04217}{{\ttfamily 1507.04217}}].

\bibitem{Breuer:1980proc}
R.A.~Breuer, J.~Ehlers and R.~Penrose, \emph{Propagation of high-frequency
  electromagnetic waves through a magnetized plasma in curved space-time. i},
  \href{https://doi.org/10.1098/rspa.1980.0040}{\emph{Proceedings of the Royal
  Society of London. A. Mathematical and Physical Sciences} {\bfseries 370}
  (1980) 389}
  [\href{https://arxiv.org/abs/https://royalsocietypublishing.org/doi/pdf/10.1098/rspa.1980.0040}{{\ttfamily
  https://royalsocietypublishing.org/doi/pdf/10.1098/rspa.1980.0040}}].

\bibitem{Breuer:1981proc}
R.A.~Breuer, J.~Ehlers and R.~Penrose, \emph{Propagation of high-frequency
  electromagnetic waves through a magnetized plasma in curved space-time. ii.
  application of the asymptotic approximation},
  \href{https://doi.org/10.1098/rspa.1981.0011}{\emph{Proceedings of the Royal
  Society of London. A. Mathematical and Physical Sciences} {\bfseries 374}
  (1981) 65}
  [\href{https://arxiv.org/abs/https://royalsocietypublishing.org/doi/pdf/10.1098/rspa.1981.0011}{{\ttfamily
  https://royalsocietypublishing.org/doi/pdf/10.1098/rspa.1981.0011}}].

\bibitem{Perlick:2000book}
V.~Perlick, \emph{{Ray optics, Fermat's principle, and applications to general
  relativity}}, Lecture Notes in Physics Monographs, Springer, Berlin, Germany,
  2000~ed. (feb, 2000).

\bibitem{Synge:1960book}
J.L.~Synge, ed., \emph{{Relativity: The General theory}} (1960).

\bibitem{Bisnovatyi:2008grav}
G.S.~Bisnovatyi-Kogan and O.Y.~Tsupko, \emph{{Gravitational
  radiospectrometer}},
  \href{https://doi.org/10.1134/S020228930901006X}{\emph{Grav. Cosmol.}
  {\bfseries 15} (2009) 20} [\href{https://arxiv.org/abs/0809.1021}{{\ttfamily
  0809.1021}}].

\bibitem{Bisnovatyi:2010mon}
G.S.~Bisnovatyi-Kogan and O.Y.~Tsupko, \emph{{Gravitational lensing in a
  non-uniform plasma}},
  \href{https://doi.org/10.1111/j.1365-2966.2010.16290.x}{\emph{Mon. Not. Roy.
  Astron. Soc.} {\bfseries 404} (2010) 1790}
  [\href{https://arxiv.org/abs/1006.2321}{{\ttfamily 1006.2321}}].

\bibitem{Tsupko:2013prd}
O.Y.~Tsupko and G.S.~Bisnovatyi-Kogan, \emph{{Gravitational lensing in plasma:
  Relativistic images at homogeneous plasma}},
  \href{https://doi.org/10.1103/PhysRevD.87.124009}{\emph{Phys. Rev. D}
  {\bfseries 87} (2013) 124009}
  [\href{https://arxiv.org/abs/1305.7032}{{\ttfamily 1305.7032}}].

\bibitem{Morozova:2013}
V.S.~Morozova, B.J.~Ahmedov and A.A.~Tursunov, \emph{Gravitational lensing by a
  rotating massive object in a plasma},
  \href{https://doi.org/10.1007/s10509-013-1458-6}{\emph{Astrophysics and Space
  Science} {\bfseries 346} (2013) 513}.

\bibitem{damour_2007}
T.~Damour and S.N.~Solodukhin, \emph{Wormholes as black hole foils},
  \href{https://doi.org/10.1103/PhysRevD.76.024016}{\emph{Phys. Rev. D}
  {\bfseries 76} (2007) 024016}.

\bibitem{Gralla:2019xty}
S.E.~Gralla, D.E.~Holz and R.M.~Wald, \emph{{Black Hole Shadows, Photon Rings,
  and Lensing Rings}},
  \href{https://doi.org/10.1103/PhysRevD.100.024018}{\emph{Phys. Rev. D}
  {\bfseries 100} (2019) 024018}
  [\href{https://arxiv.org/abs/1906.00873}{{\ttfamily 1906.00873}}].

\bibitem{Vazquez:2003zm}
S.E.~Vazquez and E.P.~Esteban, \emph{{Strong field gravitational lensing by a
  Kerr black hole}},
  \href{https://doi.org/10.1393/ncb/i2004-10121-y}{\emph{Nuovo Cim. B}
  {\bfseries 119} (2004) 489}
  [\href{https://arxiv.org/abs/gr-qc/0308023}{{\ttfamily gr-qc/0308023}}].

\bibitem{Bardeen:1973tla}
J.M.~Bardeen, \emph{{Timelike and null geodesics in the Kerr metric}},
  {\emph{Proceedings, Ecole d'Et\'e de Physique Th\'eorique: Les Astres Occlus
  : Les Houches, France, August, 1972, 215-240} (1973) 215}.

\bibitem{GRAVITY-1}
{\scshape GRAVITY} collaboration, \emph{{Detection of faint stars near
  Sagittarius A* with GRAVITY}},
  \href{https://doi.org/10.1051/0004-6361/202039544}{\emph{Astron. Astrophys.}
  {\bfseries 645} (2021) A127}
  [\href{https://arxiv.org/abs/2011.03058}{{\ttfamily 2011.03058}}].

\bibitem{Gravity-2}
{GRAVITY Collaboration}, {Abuter, R.}, {Amorim, A.}, {Bauböck, M.}, {Berger,
  J. P.}, {Bonnet, H.} et~al., \emph{Improved gravity astrometric accuracy from
  modeling optical aberrations},
  \href{https://doi.org/10.1051/0004-6361/202040208}{\emph{Astron. Astrophys.}
  {\bfseries 647} (2021) A59}.

\bibitem{Keck:2019txf}
T.~Do et~al., \emph{{Relativistic redshift of the star S0-2 orbiting the
  Galactic center supermassive black hole}},
  \href{https://doi.org/10.1126/science.aav8137}{\emph{Science} {\bfseries 365}
  (2019) 664} [\href{https://arxiv.org/abs/1907.10731}{{\ttfamily
  1907.10731}}].

\bibitem{EventHorizonTelescope:2022xqj}
{\scshape Event Horizon Telescope} collaboration, \emph{{First Sagittarius A*
  Event Horizon Telescope Results. VI. Testing the Black Hole Metric}},
  \href{https://doi.org/10.3847/2041-8213/ac6756}{\emph{Astrophys. J. Lett.}
  {\bfseries 930} (2022) L17}
  [\href{https://arxiv.org/abs/2311.09484}{{\ttfamily 2311.09484}}].

\end{thebibliography}\endgroup
\end{document}